\documentclass[fleqn,10pt]{article}
\usepackage{latexsym, graphicx, epsfig, amsmath,  amssymb,amsfonts}
\usepackage{natbib,amsthm,version}
\usepackage{amsbsy,bm,multirow,enumerate}
\usepackage[titletoc,page]{appendix}
\usepackage[mathscr]{eucal}
\usepackage{mathtools}

\title{
Physical Formulation and 
Numerical Algorithm \\ 
for Simulating N 
Immiscible Incompressible Fluids Involving
General Order Parameters
} 
\author{
  S. Dong\thanks{ Email: sdong@purdue.edu} \\
  Center for Computational and Applied Mathematics \\
  Department of Mathematics \\
  Purdue University 
 } 

\date{} 
\begin{document}
\maketitle



\begin{abstract}

We present a physical formulation, and a  numerical algorithm,
based on a class of general order parameters for simulating
the motion of a mixture of $N$ ($N\geqslant 2$) 
immiscible incompressible fluids
with given densities, dynamic viscosities, and pairwise
surface tensions.
The introduction of general order 
parameters leads to a more strongly coupled system
of phase field equations, in contrast to that with
certain special choice of the order parameters.
However, the general form enables one to compute the N-phase
mixing energy density coefficients in
an {\em explicit} fashion in terms of 
the pairwise surface tensions.
From the simulation perspective, 
the increased complexity in the form of
the  phase field equations 
with general order parameters in actuality  does not induce
essential computational difficulties. 
Our numerical algorithm reformulates the ($N-1$) strongly-coupled
phase field equations for general order parameters 
into $2(N-1)$ Helmholtz-type equations that are completely de-coupled
from one another, leading to a computational complexity
essentially the same 
as that of the simpler phase field equations
associated with special choice of order parameters.
We demonstrate the capabilities of the method
developed herein 
using several test problems involving multiple fluid phases
and large  contrasts in densities and viscosities  among
the multitude of fluids.
In particular, by comparing 
simulation results with the Langmuir-de Gennes
 theory of floating liquid
lenses we show that the method produces physically accurate
results for multiple fluid phases.

\end{abstract}


\vspace{0.05cm}

\section{Introduction}
\label{sec:intro}

%
%
%
%
%

The present work focuses on
the motion of a mixture of $N$ ($N\geqslant 2$)
immiscible incompressible fluids with given densities,
dynamic viscosities, and pairwise surface tensions.
The system is assumed to contain no solid phase
(e.g. solid particles), except for possible solid-wall
boundaries.
The situation is a  generalization of incompressible
two-phase flows, 
which have been under intensive investigations
by the community for decades.
The applications and potential implications 
of N-phase problems are
enormous, from both the practical engineering perspective
and fundamental physics 
perspective \cite{Taylor1934,deGennesBQ2003,SayeS2011}.

N-phase flows have been the subject of a number of 
past research efforts in the literature.
A summary of the existing studies is provided
in the following paragraphs.
Several researchers have reviewed 
two-phase flows  comprehensively 
for different approaches and techniques
\cite{OsherS1988,UnverdiT1992,AndersonMW1998,LowengrubT1998,Jacqmin1999,ScardovelliZ1999,Tryggvasonetal2001,SethianS2003};
see also the references therein.
We will therefore restrict our attention in the following
review, and also in the main work of the 
present paper, to flows with 
three or more fluid phases ($N\geqslant 3$),
noting that the technique developed herein equally
applies to two-phase flows.
It should also be noted that our attention is limited
to incompressible fluids.

%

Since the overall approach of the current work
pertains to the phase field framework, 
we will first briefly mention the representative works 
for multiple phases
based on other related approaches such as level set or volume of
fluids,
and then will concentrate on the existing studies 
with phase fields.
The level set technique as proposed by 
\cite{OsherS1988} is extended from two to 
N components in several studies
(see e.g. \cite{MerrimanBO1994,ZhaoCMO1996,SmithSC2002,ZlotnikD2009,VillaF2010}, among others),
where different fluid components are differentiated
by using $N$ \cite{MerrimanBO1994,ZhaoCMO1996}, 
($N-1$) \cite{SmithSC2002,ZlotnikD2009}, 
or $\left\lceil\log_2^N\right\rceil$ \cite{ChanV2001}
level set functions. 
In \cite{SayeS2011} the multiple phases are 
characterized by a single unsigned level set function,
and the $\epsilon$ ($\epsilon>0$ is a small number) level sets
are convected in the usual way by the flow, while
the fluid interfaces at the new step are 
re-constructed from the 
$\epsilon$ level sets by a Voronoi tessellation.
The work of \cite{BonhommeMDP2012}
combines the experiments and numerical
simulations based on a volume-of-fluid approach,
and investigates in detail the dynamics of an air bubble
crossing an interface between two fluids. 


Let us now concentrate on the past studies of N-phase flows
with the phase field (or diffuse interface) approach,
which appears to constitute the majority of
efforts in this area.
The main developments are primarily thanks to 
the contributions of
Kim and collaborators \cite{KimL2005,Kim2009,Kim2012}, 
 Boyer and
collaborators \cite{BoyerL2006,BoyerLMPQ2010,BoyerM2011},
and Heida and collaborators \cite{HeidaMR2012}.
%
Among these, several investigations have been devoted to
the discretizations of
the three-component \cite{KimKL2004,BoyerM2011} 
or N-component \cite{LeeK2008,LeeK2012}
Cahn-Hilliard equations or Allen-Cahn equations,
where the hydrodynamic interaction is absent.
In particular, energy-stable schemes for the three-component
Cahn-Hilliard equations are discussed in
\cite{BoyerM2011}, and a nonlinear
multigrid method combined with a finite-difference
discretization is presented in
\cite{KimKL2004,LeeK2008}.
%
When the hydrodynamic interaction is
present, in \cite{KimL2005}
a thermodynamically-consistent phase field model 
for N fluid components
is derived based on the balance equations and
the second law of thermodynamics. This is 
a generalization of the two-phase model proposed 
by \cite{LowengrubT1998}.
An important feature is that,
the mixture velocity in this N-phase model 
is the mass-averaged velocity and therefore
it is not divergence free.
In \cite{BoyerL2006}
a three-component Cahn-Hilliard system is coupled
with the Navier-Stokes equation, supplemented
by a capillary force, 
to model three-phase flows. The effects of 
different forms for the bulk free energy have been
studied.
Noting a solvability difficulty in determining the
coefficients for the surface-tension forces
when more than three phases are involved,
Kim \cite{Kim2009} proposed a phenomenological
surface-tension force for multiple
fluid components;
see its applications in \cite{Kim2012,LeeK2013}.
In \cite{BoyerLMPQ2010} a combined 
Cahn-Hilliard/Navier-Stokes model
has been studied for three-phase flows,
in which the Cahn-Hilliard model uses
a particular free-energy form due to \cite{BoyerL2006}
and the Navier-Stokes equation uses
a special form for the inertia term 
due to \cite{GuermondQ2000}.
%
More recently, Heida et al \cite{HeidaMR2012}
present a Cahn-Hilliard-Navier-Stokes type 
phase field model for N fluid phases 
in which the constitutive relations
are obtained by requiring the maximization of
the rate of entropy production.
The mixture velocity in this model, similar to
that of \cite{KimL2005,LiW2014},
is also the mass-averaged velocity and is not 
divergence free.
%
Another interesting work is
\cite{MatsutaniNS2011},
in which an Euler type equation, i.e. barring
the dissipation terms, with the surface
tensions of an N-phase field has been
derived based on the variational principle.
%


Very recently, by considering the mass conservations
of the N  individual fluid phases, the momentum
conservation, Galilean invariance 
and the second law of thermodynamics,
we have derived in \cite{Dong2014} a general phase field
model (isothermal) for the mixture of N ($N\geqslant 2$) immiscible 
incompressible fluids; see also
Section \ref{sec:nphase_model_general} below.
This model is fundamentally different from
those of \cite{KimL2005,HeidaMR2012,LiW2014},
in that the mixture velocity in our model is
the {\em volume-averaged} velocity, which 
can be rigorously shown to be divergence free \cite{Dong2014}.
In contrast, the velocity in the models of
\cite{KimL2005,HeidaMR2012,LiW2014} is the 
{\em mass-averaged} velocity, and
is not divergence free.
Our N-phase model can be considered as
a generalization of the 
formulation in \cite{AbelsGG2012} for two-phase flows.


In order to provide an N-phase formulation
suitable for numerical simulations,
the general N-phase model of \cite{Dong2014}
requires the further specification of two items: 
(1) a set of ($N-1$) order parameters
or phase field variables, and (2)
the form of the N-phase free 
energy density function.
In \cite{Dong2014} we have employed 
a very special set of order parameters,
which is also given in Section \ref{sec:formulation},
because this set 
significantly simplifies the form of the resulting 
phase field equations.
Employing this particular phase field formulation,
we have further developed 
a method for computing the mixing energy density
coefficients (see Section \ref{sec:formulation} for definition)
involved in the formulation 
by solving a linear algebraic
system based on the pairwise surface tensions among the N fluids. 
We have also developed 
 an algorithm 
for solving the coupled system of
governing equations in this formulation for
 N-phase simulations. 


In this paper we generalize the N-phase formulation
to a class of general order parameters.
This gives rise to a class of N-phase physical
formulations suitable for numerical simulations.
Within this family,
by specifying a constant non-singular matrix 
and a constant vector, one will arrive at 
a specific N-phase formulation.
This class of physical formulations with
general order parameters includes
the one of \cite{Dong2014} as a 
particular case.


The introduction of the class of 
general order parameters has two major implications:
\begin{itemize}

\item
It enables us to derive an {\em explicit expression} for 
the mixing energy density coefficients in terms of
the pairwise surface tensions. Therefore, 
the mixing energy density coefficients with
general order parameters can be {\em explicitly} computed.
In contrast, the method of \cite{Dong2014} requires one to solve
a linear algebraic system 
to determine the mixing energy density coefficients.

\item
The resulting  phase field 
equations have a more complicated form than that
employing the special set of order parameters of \cite{Dong2014}.
In particular, the ($N-1$) phase field equations
with general order parameters
become much more strongly coupled with one another.
The increased complexity raises new challenges to
their numerical solutions.

\end{itemize}


We have developed an algorithm 
for solving the new phase field equations with
general order parameters, which
 overcomes the computational challenge
caused by the increased complexity.
%
Our algorithm reformulates the ($N-1$)
strongly-coupled phase field equations for general
order parameters into 
$2(N-1)$ Helmholtz type equations that
are completely de-coupled from one another.
With this algorithm
the computational complexity
for the general order parameters
is comparable to that of \cite{Dong2014}
for the simplified phase field equations
with the special set of order parameters.
%
This algorithm for the phase field equations
with general order parameters, combined with
an algorithm for the N-phase momentum equations,
provides an efficient method for simulating
N-phase flows, which has also 
overcome the computational issues associated with variable
mixture density and viscosity.


The  current work is in line
with the following view toward the order parameters 
(or phase field variables).
The order parameters or phase field variables 
serve merely as a 
set of state variables chosen to 
formulate the system, and they 
can be chosen in different ways. 
Using a different set of order parameters 
leads to a different representation of the 
N-phase system.
While the resulting phase field equations may have
varying degrees of complexity with different order parameters, 
the various representations of the N-phase system
should be equivalent to one another.


The novelties of this paper lie in three aspects:
(1) the N-phase physical formulation with general
order parameters, (2) the explicit form
of the N-phase mixing energy density coefficients
in terms of the pairwise surface tensions for general
order parameters,
and (3) the numerical algorithm for solving
the ($N-1$) strongly-coupled phase field equations
with general order parameters.
In addition, the algorithm for solving the N-phase momentum
equations in the current paper, given in the 
Appendix B, is also new in the context of N-phase flows.
Note that it is different than that of \cite{Dong2014}
for the N-phase momentum equations.
But the essential strategies for dealing with
the variable density, variable viscosity and
the pressure-velocity coupling stem from
our previous work \cite{DongS2012} for
two-phase flows.





The rest of this paper is organized as follows.
In Section \ref{sec:nphase_model_general} we provide
a summary of the general phase field model for
a mixture of N immiscible incompressible fluids
we derived in \cite{Dong2014},
which serves as the basis for the N-phase physical
formulations with general order parameters of the current paper.
In Section \ref{sec:gop_formulation} we discuss
a class of general order parameters and the N-phase formulations
with the general order parameters.
We also derive an explicit form for the
mixing energy density coefficients in terms of
the pairwise surface tensions among the
N fluids.
Section \ref{sec:method} provides an efficient algorithm
for solving the ($N-1$) strongly-coupled phase
field equations with general order parameters. 
We further combine this algorithm, with a scheme for the N-phase
momentum equations discussed in Appendix B,
to form an overall method for N-phase flow simulations.
In Section \ref{sec:tests} we look into several
numerical examples involving three and four fluid phases to
demonstrate the accuracies and capabilities
of the presented method with general order
parameters.
Section \ref{sec:summary} concludes the discussions with
a summary of the key points.
Finally, Appendix A provides a proof for the unique solvability
of the linear algebraic system about the mixing energy
density coefficients derived in \cite{Dong2014}.
The unique solvability of this system is an un-settled issue
of \cite{Dong2014}.
 Appendix B presents a
scheme for the N-phase momentum equations,
exploiting the ideas  for treating
the variable density and variable viscosity 
in \cite{DongS2012} originally developed for two-phase flows.

\subsection{A General Phase-Field 
Model for an N-Fluid Mixture}
\label{sec:nphase_model_general}


This subsection summarizes the
general phase field model we derived in \cite{Dong2014}
based on the conservations of mass and momentum,
the second law of thermodynamics,
and Galilean invariance.
We refer to \cite{Dong2014} for detailed derivations of
this system.

Let $\Omega$ denote the flow domain in two or three dimensions,
and $\partial\Omega$ denote the boundary of $\Omega$.
consider the mixture of $N$ ($N\geqslant 2$) immiscible
incompressible fluids contained in $\Omega$.
Let $\tilde{\rho}_i$ ($1\leqslant i\leqslant N$) denote 
the constant densities of these $N$ pure fluids (before mixing),
and $\tilde{\mu}_i$ ($1\leqslant i\leqslant N$) denote
their constant dynamic viscosities.
We define the auxiliary parameters
\begin{equation}
\tilde{\gamma}_i = \frac{1}{\tilde{\rho}_i} \ \text{for} \ 1\leqslant i\leqslant N,
\qquad
\Gamma = \sum_{i=1}^N \tilde{\gamma}_i,
\qquad
\Gamma_{\mu} = \sum_{i=1}^N \tilde{\gamma}_i\tilde{\mu}_i.
\end{equation}

Let $\phi_i$ ($1\leqslant i\leqslant N-1$) 
denote the ($N-1$) independent order parameters
(or interchangeably, phase field variables) that characterize
the N-phase system, and 
$
\vec{\phi} = (\phi_1, \dots, \phi_{N-1})
$
denote the vector of phase field variables.
Let $\rho_i(\vec{\phi})$ and $c_i(\vec{\phi})$ ($1\leqslant i\leqslant N$)
respectively denote the density and the volume fraction 
of the $i$-th fluid {\em within
the mixture}. 
Let $\rho(\vec{\phi})$ denote the mixture density.
We have the relations
\begin{equation}
c_i = \frac{\rho_i}{\tilde{\rho}_i} = \tilde{\gamma}_i\rho_i, \quad
\sum_{i=1}^N c_i = 1, \quad
\rho = \sum_{i=1}^N \rho_i.
\label{equ:param_relation}
\end{equation}

Let $W(\vec{\phi},\nabla\vec{\phi})$ denote
the free energy density function of the system,
which must satisfy the following condition
\begin{equation}
\sum_{i=1}^{N-1}\nabla\phi_i \otimes
\frac{\partial W}{\partial(\nabla\phi_i)}
= \sum_{i=1}^{N-1}\frac{\partial W}{\partial(\nabla\phi_i)}
\otimes \nabla\phi_i,
\label{equ:energy_condition}
\end{equation}
where $\otimes$ denotes the tensor product.
Then this N-phase system is described by 
the following equations \cite{Dong2014}:
\begin{subequations}
\begin{equation}
\rho\left(
  \frac{\partial\mathbf{u}}{\partial t}
  + \mathbf{u}\cdot\nabla\mathbf{u}
\right)
+ \tilde{\mathbf{J}}\cdot\nabla\mathbf{u}
= 
-\nabla p
+ \nabla\cdot\left[
  \mu(\vec{\phi}) \mathbf{D}(\mathbf{u})
\right]
- \sum_{i=1}^{N-1} \nabla\cdot\left(
  \nabla\phi_i \otimes \frac{\partial W}{\partial(\nabla\phi_i)}
\right),
\label{equ:nse_original}
\end{equation}
\begin{equation}
\nabla\cdot\mathbf{u} = 0,
\label{equ:continuity_original}
\end{equation}
\begin{equation}
\sum_{j=1}^{N-1}\frac{\partial\varphi_i}{\partial\phi_j}\left(
  \frac{\partial\phi_j}{\partial t} + \mathbf{u}\cdot\nabla\phi_j
\right)
=
\nabla\cdot\left[
  \tilde{m}_i(\vec{\phi}) \nabla C_i
\right],
\qquad 1 \leqslant i \leqslant N-1,
\label{equ:CH_original}
\end{equation}
\end{subequations}
where $\mathbf{u}(\mathbf{x},t)$ is velocity,
$p(\mathbf{x},t)$ is pressure, 
$
\mathbf{D}(\mathbf{u}) = \nabla\mathbf{u} + \nabla\mathbf{u}^T
$ 
(superscript $T$ denoting transpose),
$\mathbf{x}$ and $t$
are respectively the spatial and temporal coordinates.
$\tilde{m}_i(\vec{\phi})\geqslant 0$ ($1\leqslant i\leqslant N-1$)
are the mobilities associated with $\phi_i$.
$\varphi_i(\vec{\phi})$ are defined by
\begin{equation}
\varphi_i(\vec{\phi}) = \rho_i(\vec{\phi}) - \rho_N(\vec{\phi}),
\quad 1\leqslant i\leqslant N-1.
\end{equation}
The chemical potentials
$C_i(\vec{\phi},\nabla\vec{\phi})$ ($1\leqslant i\leqslant N-1$)
are given by the following linear
algebraic system
\begin{equation}
\sum_{j=1}^{N-1} \frac{\partial\varphi_j}{\partial\phi_i} C_j
=
\frac{\partial W}{\partial \phi_i}
- \nabla\cdot \frac{\partial W}{\partial(\nabla\phi_i)},
\quad 1\leqslant i\leqslant N-1,
\label{equ:chem_potential}
\end{equation}
which can be solved once $W(\vec{\phi},\nabla\vec{\phi})$
and $\varphi_i(\vec{\phi})$ are given.
$\tilde{\mathbf{J}}(\vec{\phi},\nabla\vec{\phi})$ is
given by
\begin{equation}
\tilde{\mathbf{J}} = -\sum_{i=1}^{N-1}\left(  
1 - \frac{N}{\Gamma}\tilde{\gamma}_i
\right)
\tilde{m}_i(\vec{\phi})\nabla C_i.
\end{equation}
The mixture density $\rho(\vec{\phi})$ and 
dynamic viscosity $\mu(\vec{\phi})$
are given by
\begin{equation}
\rho(\vec{\phi}) = \frac{N}{\Gamma} + \sum_{i=1}^{N-1}\left(
1 - \frac{N}{\Gamma}\tilde{\gamma}_i
\right) \varphi_i(\vec{\phi}),
\qquad
\mu(\vec{\phi}) = \frac{\Gamma_{\mu}}{\Gamma} + \sum_{i=1}^{N-1}\left(
\tilde{\mu}_i - \frac{\Gamma_{\mu}}{\Gamma}
\right) \tilde{\gamma}_i \varphi_i(\vec{\phi}).
\end{equation}

\section{Order Parameters and N-Phase Physical Formulation}
\label{sec:gop_formulation}

%
%

\subsection{N-Phase Formulations with General Order Parameters}
\label{sec:formulation}

To arrive at an N-phase physical formulation suitable for 
numerical simulations, the  phase field
model given in Section \ref{sec:nphase_model_general}
requires the specification of:
(1) the form of the free energy density function
$W(\vec{\phi},\nabla\vec{\phi})$, and
(2) the set of order parameters $\phi_i$ ($1\leqslant i\leqslant N-1$).

Following \cite{Dong2014}, we assume the following form for the free
energy density function of the N-phase system
\begin{equation}
W(\vec{\phi},\nabla\vec{\phi}) = 
\sum_{i,j=1}^{N-1} \frac{\lambda_{ij}}{2} \nabla\phi_i\cdot\nabla\phi_j
  + \frac{\beta^2}{2\eta^2}H(\vec{\phi}),
\qquad
H(\vec{\phi}) = \sum_{k=1}^N c_k^2 (1 - c_k)^2,
\label{equ:free_energy}
\end{equation}
where $\beta^2$ is a characteristic scale for the energy,
and $\eta$ is a characteristic scale for the interfacial
thickness.
$c_k(\vec{\phi})$ ($1\leqslant k\leqslant N$) is the
volume fraction of the fluid $k$ in the mixture, whose
specific form is given subsequently.
$\lambda_{ij}$ ($1\leqslant i,j\leqslant N-1$) are
referred to as the mixing energy density coefficients,
and they are assumed to be constant in the current paper.
The condition \eqref{equ:energy_condition} requires that
the matrix
\begin{equation}
\mathbf{A} = \begin{bmatrix}
\lambda_{ij}
\end{bmatrix}_{(N-1)\times (N-1)}
\label{equ:A_matrix_expr}
\end{equation}
be symmetric.
We further require that $\mathbf{A}$
be positive definite to ensure the positivity
of the first term in the 
$W(\vec{\phi},\nabla\vec{\phi})$ expression.
Therefore, the matrix $\mathbf{A}$ is required to be symmetric
 positive definite (SPD) in the current paper.


The form of the free energy density function \eqref{equ:free_energy},
in particular the cross terms 
$\nabla\phi_i\cdot\nabla\phi_j$ ($i\neq j$) therein,
give rise to a set of phase-field equations 
that are very different from  those of 
the existing N-phase studies 
\cite{KimL2005,BoyerL2006,Kim2009,BoyerLMPQ2010,Kim2012,LeeK2013}.
It is the key that enables us to determine
the mixing energy density coefficients
$\lambda_{ij}$ uniquely, and to provide
their explicit expressions, based on the 
pairwise surface tensions among the
$N$ fluids. This will be
discussed subsequently in Section
\ref{sec:lambda_ij}.

We now focus on the order parameters $\vec{\phi}$,
and this is the departure point of the current work. 
Let 
\begin{equation}
\mathbf{A}_1 = \begin{bmatrix}
a_{ij}
\end{bmatrix}_{(N-1)\times(N-1)},
\qquad
\mathbf{b}_1 = \begin{bmatrix}
b_i
\end{bmatrix}_{(N-1)\times 1}
\label{equ:A1_expr}
\end{equation}
respectively denote a prescribed non-singular constant matrix
and a prescribed constant vector.
We define the ($N-1$)  order parameters
$\phi_i$ as follows,
\begin{equation}
\varphi_i(\vec{\phi}) = \rho_i(\vec{\phi}) - \rho_N(\vec{\phi}) 
= \sum_{j=1}^{N-1} a_{ij} \phi_j + b_i,
\qquad 1\leqslant i \leqslant N-1.
\label{equ:order_param}
\end{equation}
Note that $\phi_i$ ($1\leqslant i\leqslant N-1$) as 
defined above
can in general be dimensional or 
non-dimensional variables. 
However, in the current paper
we will require that $\phi_i$ be non-dimensional
in the simulations.
Equation \eqref{equ:order_param} defines a family of order parameters.
Given a specific set of $\mathbf{A}_1$ and $\mathbf{b}_1$ in
\eqref{equ:A1_expr},
equation \eqref{equ:order_param} will
define a unique set of order parameters
$\phi_i$ ($1\leqslant i\leqslant N-1$).
We refer to the family of order parameters
defined by \eqref{equ:order_param}
as the general order parameters.


With the set of general order parameters 
given by \eqref{equ:order_param},
and the free energy density function given 
by \eqref{equ:free_energy},
the motion of the N-phase mixture is described by
the following system of equations,
\begin{subequations}
\begin{align}
&
\rho(\vec{\phi})\left(\frac{\partial\mathbf{u}}{\partial t} + \mathbf{u}\cdot\nabla\mathbf{u} \right)
  + \tilde{\mathbf{J}}(\vec{\phi},\nabla\vec{\phi})\cdot\nabla\mathbf{u}
= -\nabla p + \nabla\cdot\left[ \mu(\vec{\phi})\mathbf{D}(\mathbf{u}) \right]
  - \sum_{i,j=1}^{N-1} \nabla\cdot\left(\lambda_{ij}\nabla\phi_i\nabla\phi_j \right)
  + \mathbf{f}
\label{equ:nse} \\
&
\nabla\cdot\mathbf{u} = 0
\label{equ:continuity} \\
&
\sum_{j=1}^{N-1} d_{ij} \left( \frac{\partial\phi_j}{\partial t} 
  + \mathbf{u}\cdot\nabla\phi_j \right) 
= 
\nabla^2\left[
  -\sum_{j=1}^{N-1}\lambda_{ij}\nabla^2\phi_j + h_i(\vec{\phi})
\right] + g_i(\mathbf{x},t),
\qquad 1\leqslant i \leqslant N-1, 
\label{equ:CH}
\end{align}
\end{subequations}
where 
we have taken into account an external body force
$\mathbf{f}(\mathbf{x},t)$ in the momentum equation \eqref{equ:nse},
and included a prescribed source term $g_i$ in
the phase field equations \eqref{equ:CH}.
$g_i$ ($1\leqslant i\leqslant N-1$) 
are for the purpose of numerical testing only, and will be set
to $g_i=0$ in actual simulations.
%
The constants $d_{ij}$ ($1\leqslant i, j\leqslant N-1$)
are defined by 
\begin{equation}
\left\{
\begin{split}
&
\mathbf{A}_2 = \begin{bmatrix} d_{ij} \end{bmatrix}_{(N-1)\times(N-1)}
= \mathbf{A}_1^T \mathbf{M}^{-1} \mathbf{A}_1, \\
&
\mathbf{M} = \text{diag}\left(
   m_1\left(\frac{\tilde{\rho}_1+\tilde{\rho}_N}{2} \right)^2, 
   m_2\left(\frac{\tilde{\rho}_2+\tilde{\rho}_N}{2} \right)^2, 
   \dots, 
   m_{N-1}\left(\frac{\tilde{\rho}_{N-1}+\tilde{\rho}_N}{2} \right)^2
\right),
\end{split}
\right.
\label{equ:A2_expr}
\end{equation}
where $m_i>0$ ($1\leqslant i\leqslant N-1$)
are the interfacial mobility coefficients associated with $\phi_i$
and are assumed to be positive constants.
Note that the matrix $\mathbf{A}_2$ is symmetric positive definite
(SPD) based on the assumptions about the non-singularity of
$\mathbf{A}_1$ and the positivity of $m_i$.

The function $h_i(\vec{\phi})$ in \eqref{equ:CH} is given by
\begin{equation}
h_i(\vec{\phi}) = \frac{\beta^2}{2\eta^2} \frac{\partial H}{\partial\phi_i},
\qquad 1\leqslant i\leqslant N-1,
\end{equation}
where $H(\vec{\phi})$ is defined in \eqref{equ:free_energy}.
The volume fractions
$c_k(\vec{\phi})$ ($1\leqslant k\leqslant N$) with the general
order parameters are
given by
\begin{equation}
c_k(\vec{\phi}) = \tilde{\gamma}_k \rho_k(\vec{\phi}),
\qquad
\rho_k(\vec{\phi}) = \left\{
\begin{array}{ll}
\frac{1}{\Gamma} - \sum_{i=1}^{N-1}\frac{\tilde{\gamma}_i}{\Gamma}
      \left(\sum_{j=1}^{N-1} a_{ij}\phi_j + b_i \right),
& \text{if} \ k=N, \\
\rho_N(\vec{\phi}) + \left(\sum_{j=1}^{N-1} a_{ij}\phi_j + b_i \right),
& \text{if} \ 1\leqslant k\leqslant N-1.
\end{array}
\right.
\label{equ:volfrac_expr}
\end{equation}
These expressions for $c_k$ and $\rho_k$
are obtained
based on the mass balance relations 
for the N-phase mixture
(see \cite{Dong2014} for details), 
and the definition of
the order parameters in \eqref{equ:order_param}.


The mixture density $\rho(\vec{\phi})$ is given by
\begin{equation}
\rho(\vec{\phi}) = \sum_{i=1}^N \rho_i(\vec{\phi})
 = \frac{N}{\Gamma}
+ \sum_{i=1}^{N-1}\left( 1 - \frac{N}{\Gamma}\tilde{\gamma}_i  \right)
  \left(\sum_{j=1}^{N-1} a_{ij}\phi_j + b_i  \right).
\label{equ:rho_expr}
\end{equation}
The mixture dynamic viscosity $\mu(\vec{\phi})$
is given by
\begin{equation}
\mu(\vec{\phi}) = \sum_{k=1}^N \tilde{\mu}_k c_k(\vec{\phi})
=\frac{\Gamma_{\mu}}{\Gamma} + \sum_{i=1}^{N-1}\left(
\tilde{\mu}_i - \frac{\Gamma_{\mu}}{\Gamma}
\right) \tilde{\gamma}_i \left(\sum_{j=1}^{N-1} a_{ij}\phi_j + b_i  \right).
\label{equ:mu_expr}
\end{equation}
The term $\tilde{\mathbf{J}}(\vec{\phi},\nabla\vec{\phi})$
is given by
\begin{equation}
\tilde{\mathbf{J}}(\vec{\phi},\nabla\vec{\phi}) = 
-\sum_{i=1}^{N-1}\left(1 - \frac{N}{\Gamma}\tilde{\gamma}_i \right)
  \left(\frac{\tilde{\rho}_i+\tilde{\rho}_N}{2}  \right)^2 m_i
  \sum_{j=1}^{N-1}R_{ij}\nabla\left[
    -\sum_{k=1}^{N-1}\lambda_{jk}\nabla^2\phi_k 
    + h_j(\vec{\phi})
  \right],
\label{equ:J_tilde_expr}
\end{equation}
where the constants $R_{ij}$ ($1\leqslant i,j\leqslant N-1$)
are defined by
\begin{equation}
\left(\mathbf{A}_1^{T}  \right)^{-1} = \begin{bmatrix}
R_{ij}
\end{bmatrix}_{(N-1)\times (N-1)}.
\label{equ:A1_inverse_expr}
\end{equation}



The N-phase physical formulation 
given by
\eqref{equ:nse}--\eqref{equ:CH}, which is associated with
the general order parameters defined by \eqref{equ:order_param},
is thermodynamically
consistent, in the sense that this formulation stems from 
the phase field model given by 
\eqref{equ:nse_original}--\eqref{equ:CH_original},
which in turn is derived based on 
conservations of mass/momentum and the second law of
thermodynamics \cite{Dong2014}. 
The ($N-1$) phase field equations \eqref{equ:CH}
reflects the mass conservations for the 
individual fluid phases. Equation \eqref{equ:nse}
reflects the momentum conservation of the mixture.
The velocity $\mathbf{u}$ is the volume-averaged
mixture velocity and can be shown to 
be divergence free \cite{Dong2014}.
This is reflected by equation \eqref{equ:continuity}.
We refer the reader to the Appendix of \cite{Dong2014}
for detailed derivations of the general 
N-phase phase field model
based on the conservations of mass/momentum,
Galilean invariance, and the second law of thermodynamics.

By using equations \eqref{equ:CH} with $g_i=0$, one can show that
the $\rho(\vec{\phi})$ and $\tilde{\mathbf{J}}(\vec{\phi},\nabla\vec{\phi})$
given by \eqref{equ:rho_expr} and \eqref{equ:J_tilde_expr}
satisfy the relation
\begin{equation}
\frac{\partial\rho}{\partial t} + \mathbf{u}\cdot\nabla\rho
= - \nabla\cdot\tilde{\mathbf{J}}.
\label{equ:mixture_mass_relation}
\end{equation}
Using the above relation, one can further show that the formulation
given by \eqref{equ:nse}--\eqref{equ:CH}
admits the following energy law,
assuming that $g_i=0$ in \eqref{equ:CH}
and that all surface fluxes vanish on the domain boundary,
\begin{equation}
\frac{\partial}{\partial t} \int_{\Omega}
\left[
\frac{1}{2}\rho(\vec{\phi}) \left|\mathbf{u} \right|^2
+ W(\vec{\phi},\nabla\vec{\phi})
\right]
= 
-\int_{\Omega} \frac{\mu(\vec{\phi})}{2} \left\|\mathbf{D}(\mathbf{u}) \right\|^2
- \sum_{i=1}^{N-1} m_i \left(\frac{\tilde{\rho}_i+\tilde{\rho}_N}{2}  \right)^2
   \int_{\Omega} \left|\nabla C_i \right|^2
+ \int_{\Omega} \mathbf{f}\cdot \mathbf{u},
\end{equation}
where
\begin{equation}
C_i = \sum_{j=1}^{N-1} R_{ij} \left[
  -\sum_{k=1}^{N-1} \lambda_{jk} \nabla^2\phi_k + h_j(\vec{\phi})
\right],
\qquad
1 \leqslant i \leqslant N-1,
\end{equation}
are the chemical potentials.


It is instructive to compare the current 
N-phase formulation given by
equations \eqref{equ:nse}--\eqref{equ:CH}
with that of \cite{Dong2014}.
The formulation of \cite{Dong2014}
is based on a special set of order parameters 
for simplifying the form of the phase field equations, 
specifically as follows:
\begin{equation}
\varphi_i(\vec{\phi}) = \rho_i(\vec{\phi}) - \rho_N(\vec{\phi})
= \frac{1}{2}\left(\tilde{\rho}_i - \tilde{\rho}_N  \right)
+ \frac{1}{2}\left(\tilde{\rho}_i + \tilde{\rho}_N  \right)\phi_i,
\qquad
1\leqslant i\leqslant N-1.
\label{equ:special_order_param}
\end{equation}
This set is a special case of \eqref{equ:order_param},
corresponding to
\begin{equation}
a_{ij} = \frac{1}{2}\left(\tilde{\rho}_i + \tilde{\rho}_N  \right) \delta_{ij},
\qquad
b_i = \frac{1}{2}\left(\tilde{\rho}_i - \tilde{\rho}_N  \right),
\qquad 
1\leqslant i, j\leqslant N-1,
\end{equation}
where $\delta_{ij}$ is the Kronecker delta.
Because the matrix $\mathbf{A}_2$ defined in \eqref{equ:A2_expr}
for this case is diagonal, 
the form of the phase field equations \eqref{equ:CH}
becomes significantly simplified.

In contrast, the order parameters defined by
\eqref{equ:order_param} are in more general form.
They give rise to the $\mathbf{A}_2$ matrix ($d_{ij}$ terms)
in \eqref{equ:CH}.
This causes the ($N-1$) phase field equations \eqref{equ:CH}
to couple with one another in a much stronger fashion,
which presents new challenges to the design of numerical algorithms
for solving these equations.
Despite the increased complexity in the form of
phase field equations,
the general order parameters provide
a crucial advantage.
The general form 
\eqref{equ:order_param} encompasses a certain order-parameter set,
with which the determination of
the mixing energy density coefficients $\lambda_{ij}$
based on the pairwise surface tensions will be 
dramatically simplified.
This enables one,  
for {\em any} set of 
order parameters defined by \eqref{equ:order_param}, 
to express $\lambda_{ij}$
in {\em explicit forms}
in terms of 
the pairwise surface tensions (see Section \ref{sec:lambda_ij}).
This obviates the need for solving a linear
algebraic system for $\lambda_{ij}$, as with
the method discussed in \cite{Dong2014}.
In addition, 
we will show in Section \ref{sec:algorithm} that,
the increased complexity in the  form of the phase field equations
\eqref{equ:CH} actually does not entail essential 
computational difficulties. We will present a
numerical algorithm for \eqref{equ:CH}
that involves a computational complexity essentially the same 
as that  for 
the simpler phase field equations corresponding to
the special set of order parameters 
\eqref{equ:special_order_param} in \cite{Dong2014}.


We next briefly mention several specific sets of order
parameters as illustrations of the general form
given by \eqref{equ:order_param}:
\begin{itemize}

\item 
Volume fractions as order parameters. 
Let 
\begin{equation}
\phi_i = c_i, 
\quad \phi_i\in [0,1],
\quad 1\leqslant i\leqslant N-1
\label{equ:order_param_volfrac}
\end{equation}
be the order parameters, 
where $c_i$ is the volume fraction of fluid $i$ within 
the mixture. 
Then 
$c_N = 1-\sum_{i=1}^{N-1} \phi_i$. Consequently,
$
\varphi_i = \rho_i - \rho_N
=\tilde{\rho}_i c_i - \tilde{\rho}_N c_N
= \sum_{j=1}^{N-1} \left(\tilde{\rho}_i \delta_{ij} + \tilde{\rho}_N  \right)\phi_j
  - \tilde{\rho}_N.
$ 
The 
matrices $\mathbf{A}_1$ and $\mathbf{b}_1$ in \eqref{equ:A1_expr}
are then given by
\begin{equation}
a_{ij} = \tilde{\rho}_i\delta_{ij} + \tilde{\rho}_N,
\qquad
b_i = -\tilde{\rho}_N,
\qquad
1\leqslant i,j \leqslant N-1.
\label{equ:A1_expr_volfrac}
\end{equation}

\item 
Re-scaled volume fraction differences as order parameters.
Define the order parameters $\phi_i$ ($1\leqslant i\leqslant N-1$) by
\begin{equation}
2\phi_i -1 = c_i - c_N,
\quad \phi_i \in [0, 1].
\label{equ:id_gop_1}
\end{equation}
By noting
$ 
\sum_{i=1}^N c_i = 1,
$ 
one can obtain
$ 
c_N = 1 -  \frac{2}{N} \sum_{j=1}^{N-1} \phi_j,
$
$
c_i = 2\phi_i - \frac{2}{N} \sum_{j=1}^{N-1}\phi_j,
$
($
1\leqslant i\leqslant N-1
$). 
Consequently, $\mathbf{A}_1$ and $\mathbf{b}_1$
are given by
\begin{equation}
a_{ij} = 2\tilde{\rho}_i\delta_{ij} 
  + \frac{2}{N}\left( \tilde{\rho}_N - \tilde{\rho}_i \right),
\qquad
b_i =  - \tilde{\rho}_N,
\qquad
1\leqslant i, j\leqslant N-1.
\end{equation}

\item 
Densities as order parameters.
Let 
\begin{equation}
\phi_i = \rho_i,
\quad \phi_i \in [0, \tilde{\rho}_i],
\quad 1\leqslant i\leqslant N-1
\end{equation}
be the order parameters. This is in fact a simple re-scaling
to the case with volume fractions as order
parameters, due to equation \eqref{equ:param_relation}.
Then $\mathbf{A}_1$ and $\mathbf{b}_1$
are given by
\begin{equation}
a_{ij} = \delta_{ij} + \frac{\tilde{\rho}_N}{\tilde{\rho}_j},
\qquad
b_i = -\tilde{\rho}_N, \qquad
1\leqslant i,j \leqslant N-1.
\end{equation}

\item 
Re-scaled density differences as order parameters.
Define the order parameters $\phi_i$ ($1\leqslant i\leqslant N-1$) by
\begin{equation}
 \rho_i - \rho_{i+1} = -\tilde{\rho}_{i+1} 
   + (\tilde{\rho}_{i}+\tilde{\rho}_{i+1})\phi_i,
\quad \phi_i \in[0, 1],
\quad 1\leqslant i\leqslant N-1.
\label{equ:id_gop_4} 
\end{equation}
Then 
$
\varphi_i = \rho_i - \rho_N = -\sum_{j=i}^{N-1} \tilde{\rho}_{j+1}
  + \sum_{j=i}^{N-1}(\tilde{\rho}_j+\tilde{\rho}_{j+1}) \phi_j
$
for $1\leqslant i\leqslant N-1$.
$\mathbf{A}_1$ and $\mathbf{b}_1$ are therefore
given by
\begin{equation}
a_{ij} = \left\{
\begin{matrix}
\tilde{\rho}_j+\tilde{\rho}_{j+1}, & i\leqslant j, \\
0, & i>j,
\end{matrix}
\right.
\qquad
b_i = -\sum_{j=i}^{N-1} \tilde{\rho}_{j+1},
\qquad
1\leqslant i,j \leqslant N-1.
\end{equation}

\item 
Another set of order parameters. 
Define the order parameters $\phi_i$ by
\begin{equation}
\phi_i = \sum_{j=i}^{N-1} \rho_i - (N-i)\rho_N,
\qquad
1\leqslant i \leqslant N-1.
\end{equation}
Then 
$
\varphi_i = \rho_i - \rho_N = \phi_i - \phi_{i+1}
$
 ($1\leqslant i\leqslant N-2$),
and
$
\varphi_{N-1} = \rho_{N-1}-\rho_N = \phi_{N-1}.
$
So $\mathbf{A}_1$ and $\mathbf{b}_1$
are given by
\begin{equation}
a_{ij} = \delta_{ij} - \delta_{i+1,j},
\qquad
b_i = 0, \qquad
1 \leqslant i, j\leqslant N-1.
\end{equation}

\end{itemize}



\subsection{Mixing Energy Density 
Coefficients $\lambda_{ij}$
for General Order Parameters}
\label{sec:lambda_ij}


The physical formulation \eqref{equ:nse}--\eqref{equ:CH}
involves,
noting the symmetry
of matrix $\mathbf{A}$ in \eqref{equ:A_matrix_expr},
$\frac{1}{2}N(N-1)$ independent mixing energy
density coefficients $\lambda_{ij}$, which need to be
determined based on other known physical parameters.
In this section we derive {\em explicit
 expressions} of $\lambda_{ij}$ in terms of
the $\frac{1}{2}N(N-1)$ pairwise surface tensions 
among the $N$ fluid components 
for the general order parameters defined in \eqref{equ:order_param}.

The result of this section, incidentally,  also provides the explicit
formulas of $\lambda_{ij}$ for the set of special 
order parameters (see equation \eqref{equ:special_order_param}) 
employed in \cite{Dong2014}.
Note that in \cite{Dong2014}, 
using the special set of order parameters \eqref{equ:special_order_param},
we obtained a system of
$\frac{1}{2}N(N-1)$ linear algebraic equations 
about $\lambda_{ij}$, and then numerically solved that linear system
to obtain $\lambda_{ij}$.
Although numerical experiments indicate that that linear algebraic
system for $\lambda_{ij}$ in \cite{Dong2014} 
always has a unique solution,
the well-posedness is  an un-settled issue for general
N ($N\geqslant 4$) fluid phases.


In the following, we first obtain an explicit expression 
of $\lambda_{ij}$ 
for the formulation with volume fractions
as the order parameters. Then we generalize the result
to formulations with the general order parameters
defined by \eqref{equ:order_param}.


\subsubsection{Volume Fractions as Order Parameters}
\label{sec:form_volfrac}


Let us first focus on the formulation with volume fractions as
the order parameters. See equation \eqref{equ:order_param_volfrac}
for the definition, 
and the coefficients $a_{ij}$ and $b_i$ for this formulation are
given by \eqref{equ:A1_expr_volfrac}.
The formulation with volume fractions as the order parameters
plays a special role when computing $\lambda_{ij}$
among the general order parameters.
To distinguish this formulation from those with
the other order parameters, we use $\Lambda_{ij}$
($1\leqslant i, j\leqslant N-1$) to specifically denote
the mixing energy density coefficients $\lambda_{ij}$
for this formulation, and use
\begin{equation}
\bm{\Lambda} = \begin{bmatrix}
\Lambda_{ij}
\end{bmatrix}_{(N-1)\times(N-1)}
\label{equ:Lambda_matrix_expr}
\end{equation}
to denote the matrix of mixing energy density coefficients
of this formulation, in contrast with \eqref{equ:A_matrix_expr}.
In addition, we use $\bm{\Lambda}_1$
to specifically denote the $\mathbf{A}_1$ matrix
for this formulation, that is,
\begin{equation}
\bm{\Lambda}_1 = \begin{bmatrix}
a_{ij}
\end{bmatrix}_{(N-1)\times (N-1)},
\label{equ:A1_volfrac_spec}
\end{equation}
where $a_{ij}$ are given by equation \eqref{equ:A1_expr_volfrac}.
One can verify that $\bm{\Lambda}_1$
is symmetric positive definite.

To determine $\Lambda_{ij}$ for the N-phase system, 
we employ an idea similar to that of
 \cite{Dong2014},
namely, by imposing the following consistency requirement
on the N-phase formulation.
We recognize that in the N-phase system,
if only a pair of two fluids is present  
(for any fluid pair) while all other
fluids are absent, then the N-phase system
is equivalent to a two-phase system consisting of
these two fluids. Accordingly,
for such a situation, the N-phase formulation
should reduce to the two-phase formulation for
the equivalent two-phase system. 
In particular, the free-energy density
function for the N-phase system 
should reduce to that of the equivalent
two-phase system.

Note that 
for a two-phase system ($N=2$)
the relation between
the mixing energy density coefficient 
and the surface tension is well-known.
Let 
\begin{equation}
\phi_1 = c_1 - c_2
\label{equ:2phase_order_param}
\end{equation}
denote the sole order
parameter of the two-phase system, where
$c_1$ and $c_2$ are the volume fractions of the
two fluids. Then
$
c_1 = \frac{1}{2}\left( 1 + \phi_1 \right)
$
and 
$
c_2 = \frac{1}{2}\left( 1 - \phi_1 \right).
$
The free energy density function \eqref{equ:free_energy}
is then reduced to
\begin{equation}
W(\phi_1,\nabla\phi_1) = 
\frac{\lambda_{11}}{2}\nabla\phi_1\cdot\nabla\phi_1
+ \frac{\beta^2}{16\eta^2} (1 - \phi_1^2)^2.
\label{equ:2phase_energy}
\end{equation}
In a one-dimensional setting, by requiring that at equilibrium  
the integral of the above free-energy density  across the interface
should equal the surface tension, one can
obtain the relation (see \cite{YueFLS2004,YueF2011,Dong2014} for details)
\begin{equation}
\lambda_{11} = \frac{9}{2}\frac{\eta^2}{\beta^2}\sigma_{12}^2,
\label{equ:2phase_lambda_ij}
\end{equation}
where $\sigma_{12}$ denotes the surface tension
between fluids $1$ and $2$ of the two-phase system.
Therefore, for a two-phase system with the free energy
density function given by \eqref{equ:2phase_energy},
the mixing energy density coefficient is given
by \eqref{equ:2phase_lambda_ij}.


To determine $\Lambda_{ij}$ with
N fluid phases,
let us assume that fluids $k$ and $l$ ($1\leqslant k<l\leqslant N$) 
are the only two fluids that are present in
the N-phase system, that is,
\begin{equation}
\rho_i \equiv 0, \ \
c_i \equiv 0, \ \ \text{if} \ i\neq k \ \text{and} \ i\neq l, \ \ 
\text{for} \ 1\leqslant i \leqslant N. 
\label{equ:ci_expr}
\end{equation}
Equivalently,
this N-phase system can be
considered as a two-phase system
consisting of fluids $k$ and $l$.
Therefore, this system has only one independent
order parameter. 
Noting the form of equation \eqref{equ:2phase_order_param}, we use
\begin{equation}
\phi_a = c_k - c_l
\label{equ:def_phi_a}
\end{equation}
to denote the sole independent order parameter of
this N-phase system.
Then 
\begin{equation}
c_k = \frac{1}{2}(1+\phi_a), \quad
c_l = \frac{1}{2}(1 - \phi_a),
\label{equ:ck_cl_expr_volfrac}
\end{equation}
by noting 
$
\sum_{i=1}^{N} c_i = 1
$
and the condition \eqref{equ:ci_expr}.


We will distinguish
two cases: (1) $l=N$, and (2) $l<N$.
In the first case $1\leqslant k < l=N$,
the free energy density function \eqref{equ:free_energy} becomes
\begin{equation}
\begin{split}
W & = \sum_{i,j=1}^{N-1} \frac{\Lambda_{ij}}{2} \nabla\phi_i\cdot\nabla\phi_j
    + \frac{\beta^2}{2\eta^2}\sum_{i=1}^N c_i^2(1-c_i)^2 \\
&
= \frac{\Lambda_{kk}}{2}\nabla\phi_k\cdot\nabla\phi_k
  + \frac{\beta^2}{2\eta^2}\left[ c_k^2(1-c_k)^2 + c_l^2(1-c_l)^2 \right] \\
&
= \frac{\Lambda_{kk}}{8}\nabla\phi_a\cdot\nabla\phi_a
 + \frac{\beta^2}{16\eta^2} (1-\phi_a^2)^2,
\end{split}
\end{equation}
where we have used 
equations \eqref{equ:order_param_volfrac},
\eqref{equ:ci_expr} and \eqref{equ:ck_cl_expr_volfrac}.
Comparing the above equation with
the two-phase free energy density function \eqref{equ:2phase_energy}
and using equation \eqref{equ:2phase_lambda_ij},
we have
\begin{equation}
\Lambda_{kk} = \frac{18\eta^2}{\beta^2 } \sigma_{kN}^2,
\qquad
1\leqslant k < l=N,
\label{equ:lambda_kk_expr}
\end{equation}
where $\sigma_{ij}$ ($i\neq j$) denotes the surface tension
associated with the interface formed between fluid $i$
and fluid $j$.

For the second case $1\leqslant k < l < N$, the N-phase
free energy density function \eqref{equ:free_energy} is transformed into
\begin{equation}
\begin{split}
W & = \sum_{i,j=1}^{N-1} \frac{\Lambda_{ij}}{2} \nabla\phi_i\cdot\nabla\phi_j
    + \frac{\beta^2}{2\eta^2}\sum_{i=1}^N c_i^2(1-c_i)^2 \\
&
= \frac{\Lambda_{kk}}{2}\nabla\phi_k\cdot\nabla\phi_k
  + \frac{\Lambda_{ll}}{2}\nabla\phi_l\cdot\nabla\phi_l
  + \Lambda_{kl} \nabla\phi_k\cdot\nabla\phi_l
  + \frac{\beta^2}{2\eta^2}\left[
      c_k^2(1-c_k)^2 + c_l^2(1-c_l)^2
    \right] \\
&
= \frac{1}{8}\left(
    \Lambda_{kk} + \Lambda_{ll} - 2\Lambda_{kl}
  \right) \nabla\phi_a\cdot\nabla\phi_a
+ \frac{\beta^2}{16\eta^2} (1-\phi_a^2)^2
\end{split}
\end{equation}
where we have used the symmetry of the matrix $\mathbf{\Lambda}$ and
the equations
\eqref{equ:order_param_volfrac},
 \eqref{equ:ci_expr}, and \eqref{equ:ck_cl_expr_volfrac}.
Compare the above equation with equation
\eqref{equ:2phase_energy} and use equation \eqref{equ:2phase_lambda_ij},
and one can get
\begin{equation}
\frac{1}{4}\left( \Lambda_{kk} + \Lambda_{ll} - 2\Lambda_{kl} \right)
 = \frac{9}{2} \frac{\eta^2}{\beta^2} \sigma_{kl}^2,
\qquad
1\leqslant k < l \leqslant N-1.
\label{equ:lambda_kl_equation}
\end{equation}
By using the expression \eqref{equ:lambda_kk_expr}
we have
\begin{equation}
\Lambda_{kl} = \Lambda_{lk} = \frac{9\eta^2}{\beta^2}\left(
  \sigma_{kN}^2 + \sigma_{lN}^2 - \sigma_{kl}^2
\right),
\qquad
1\leqslant k < l \leqslant N-1.
\label{equ:lambda_kl_expr}
\end{equation}

Therefore, with the volume fractions as the
order parameters, the N-phase mixing energy density
coefficients $\Lambda_{kl}$ ($1\leqslant k,l\leqslant N-1$)
are given by the explicit expressions,
\eqref{equ:lambda_kk_expr}
and \eqref{equ:lambda_kl_expr},
in terms of the pairwise surface tensions
$\sigma_{ij}$ ($1\leqslant i < j\leqslant N$)
among the N fluids.

\subsubsection{General Order Parameters}

Let us now consider physical formulations 
with the general order parameters defined by 
\eqref{equ:order_param}. 
Based on equation \eqref{equ:volfrac_expr} we have
the following relation
\begin{equation}
\nabla c_k = \sum_{i=1}^{N-1} y_{ki} \nabla\phi_i,
\quad
1\leqslant k\leqslant N. \ \ 
\label{equ:ck_deriv_expr}
\end{equation}
The coefficients $y_{ki}$ in the above equation are given by
\begin{equation}
y_{ki} = \sum_{j=1}^{N-1} e_{kj} a_{ji}, \quad 
1\leqslant k\leqslant N, \ \
1\leqslant i \leqslant N-1,
\label{equ:yki_expr}
\end{equation}
where
\begin{equation}
e_{ki} = \tilde{\gamma}_k\delta_{ki} 
         - \frac{\tilde{\gamma}_k\tilde{\gamma}_i}{\Gamma}, \quad
1\leqslant k\leqslant N, \ \
1\leqslant i \leqslant N-1. 
\end{equation}
Let 
\begin{equation}
\mathbf{Y} = \begin{bmatrix}
y_{ij}
\end{bmatrix}_{(N-1)\times(N-1)}, \quad
\mathbf{Z} = \begin{bmatrix}
e_{ij}
\end{bmatrix}_{(N-1)\times(N-1)}
\end{equation}
respectively denote 
the {\em square} matrices formed by $y_{ij}$ ($1\leqslant i, j\leqslant N-1$) 
and $e_{ij}$ ($1\leqslant i, j\leqslant N-1$). 
Then  the following matrix form
represents a subset of the equations in \eqref{equ:yki_expr},
\begin{equation}
\mathbf{Y} = \mathbf{ZA}_{1},
\end{equation}
where $\mathbf{A}_1$ is defined in \eqref{equ:A1_expr}.
It is straightforward to verify that
\begin{equation}
\mathbf{Z} = \bm{\Lambda}_1^{-1},
\label{equ:Z_mat_expr}
\end{equation}
where $\bm{\Lambda}_1$ is defined in \eqref{equ:A1_volfrac_spec}.
The matrix $\mathbf{Y}$ is therefore non-singular.

In order to determine the mixing energy density 
coefficients $\lambda_{ij}$, we recognize
the following point about the order parameters.
The physical formulations employing different sets of
order parameters are merely different representations of
the N-phase system, and the different representations 
should be equivalent. In particular,
the N-phase free energy density function can
be represented in terms of any set of
independent order parameters, and these
representations should be equivalent. 
This is an embodiment of the representation
invariance principle \cite{MaW2014}.

In light of the above point, we can re-write 
the free energy density function  of the N-phase system 
\eqref{equ:free_energy} as 
\begin{equation}
\begin{split}
W & = \sum_{i,j=1}^{N-1} \frac{\lambda_{ij}}{2} \nabla\phi_i\cdot\nabla\phi_j
      + \frac{\beta^2}{2\eta^2}\sum_{k=1}^N c_k^2(\vec{\phi})\left[1-c_k(\vec{\phi})\right]^2 \\
&
= \sum_{i,j=1}^{N-1} \frac{\Lambda_{ij}}{2} \nabla c_i\cdot\nabla c_j
      + \frac{\beta^2}{2\eta^2}\sum_{k=1}^N c_k^2(1-c_k)^2.
\end{split}
\label{equ:W_two_rep}
\end{equation}
In the above equation we have expressed the N-phase 
free energy density function in terms of
the order parameters $\phi_i$ ($1\leqslant i\leqslant N-1$),
as well as in terms of the volume fractions 
$c_i$ ($1\leqslant i\leqslant N-1$),
and we have used the results from
Section \ref{sec:form_volfrac}.
Using the relation in \eqref{equ:ck_deriv_expr},
we can obtain $\lambda_{ij}$ from \eqref{equ:W_two_rep}, 
\begin{equation}
\lambda_{ij} = \sum_{k,l=1}^{N-1} y_{ki} y_{lj} \Lambda_{kl},
\qquad
1\leqslant i, j\leqslant N-1,
\label{equ:lambda_ij_gop}
\end{equation}
where $\Lambda_{kl}$ ($1\leqslant k,l\leqslant N-1$) are
given by \eqref{equ:lambda_kk_expr}
and \eqref{equ:lambda_kl_expr}.
Equivalently, the matrix form is
\begin{equation}
\mathbf{A} = \mathbf{Y}^T \bm{\Lambda}\mathbf{Y},
\label{equ:matrix_lambda_ij_gop}
\end{equation}
where $\mathbf{A}$ and $\bm{\Lambda}$ are
defined in \eqref{equ:A_matrix_expr} and \eqref{equ:Lambda_matrix_expr}
respectively.
Equation \eqref{equ:lambda_ij_gop} or
\eqref{equ:matrix_lambda_ij_gop}
provides the explicit forms for the mixing energy density coefficients
$\lambda_{ij}$ for general order parameters 
defined by \eqref{equ:order_param}.




Based on equation \eqref{equ:matrix_lambda_ij_gop} we have 
the following observations:
\begin{itemize}

\item
If the matrix $\bm{\Lambda}$ is SPD, then
the matrix $\mathbf{A}$ for any set of order parameters
defined by \eqref{equ:order_param} is SPD. 
More generally, if the matrix $\mathbf{A}$ is SPD with one set
of order parameters defined by \eqref{equ:order_param},
then it is SPD with all sets of  order parameters defined
by \eqref{equ:order_param}.

\item
The positive definiteness of the matrix $\mathbf{A}$ for
general order parameters is only affected by 
the pairwise surface tensions $\sigma_{ij}$ among the N fluids.
The fluid densities $\tilde{\rho}_i$ affect the values of $\mathbf{A}$,
but have no effect on its positive definiteness.
This is because the dependency on $\sigma_{ij}$ is through
$\bm{\Lambda}$ and the dependency on $\tilde{\rho}_i$ is through
$\mathbf{Y}$.

\item
Given a set of arbitrary positive values for the pairwise 
surface tensions $\sigma_{ij}>0$ ($1\leqslant i<j\leqslant N$),
the matrix $\bm{\Lambda}$ is always symmetric, but
may not be positive definite. 
What conditions on $\sigma_{ij}$ will ensure the SPD
of the matrix $\bm{\Lambda}$ is currently an open question.
Numerical experiments in \cite{Dong2014} suggest that, if 
the pairwise surface-tension values are such that total
wetting occurs among some three-tuple of fluids among
these N fluids, then the matrix $\mathbf{A}$ will have a negative
eigenvalue and therefore will not be positive definite.

\item
The mixing energy density coefficients $\lambda_{ij}$ for the N-phase 
formulation employed in \cite{Dong2014} are given by 
the following explicit expression
\begin{equation}
\mathbf{A} = \begin{bmatrix} \lambda_{ij} \end{bmatrix}_{(N-1)\times (N-1)}
= \mathbf{LZ}\bm{\Lambda}\mathbf{ZL},
\qquad
\mathbf{L} = \text{diag}\left(
\frac{\tilde{\rho}_1+\tilde{\rho}_N}{2},
\frac{\tilde{\rho}_2+\tilde{\rho}_N}{2},
\cdots,
\frac{\tilde{\rho}_{N-1}+\tilde{\rho}_N}{2}
\right).
\label{equ:lambda_ij_spec_explicit}
\end{equation}
Note that in \cite{Dong2014} $\lambda_{ij}$ are obtained
by solving a linear algebraic system.
In the Appendix A, we provide a proof that
the $\lambda_{ij}$ computed based on \eqref{equ:lambda_ij_spec_explicit}
indeed are the solution to the system of linear algebraic
equations about $\lambda_{ij}$ derived in \cite{Dong2014}.

\end{itemize}



Therefore,
once a set of order parameters is chosen, that is,
the matrix $\mathbf{A}_1$ and vector $\mathbf{b}_1$
in \eqref{equ:A1_expr} are fixed,
the governing equations for the N-phase system are
given by the equations \eqref{equ:nse}--\eqref{equ:CH},
where the mixing energy density coefficients
$\lambda_{ij}$ ($1\leqslant i,j\leqslant N-1$)
are given by equation \eqref{equ:matrix_lambda_ij_gop},
in terms of the pairwise surface tensions
among the N fluids.

\vspace{0.2in}
The governing equations \eqref{equ:nse}--\eqref{equ:CH} 
need to be supplemented by
appropriate boundary conditions and initial conditions 
for the velocity and
phase field equations. 
In the current paper we consider the Dirichlet boundary condition
for the velocity,
\begin{equation}
\left.\mathbf{u}\right|_{\partial\Omega} = \mathbf{w}(\mathbf{x},t),
\label{equ:vel_bc}
\end{equation}
where $\mathbf{w}$ is the boundary velocity,
and the following simplified
boundary conditions for the phase field functions,
\begin{subequations}
\begin{equation}
\left.\mathbf{n}\cdot\nabla\left(\nabla^2\phi_i  \right)\right|_{\partial\Omega}=0, 
\qquad
1\leqslant i\leqslant N-1,
\label{equ:phi_bc_1}
\end{equation}
\begin{equation}
\left.\mathbf{n}\cdot\nabla\phi_i \right|_{\partial\Omega}=0,
\qquad
1\leqslant i\leqslant N-1.
\label{equ:phi_bc_2}
\end{equation}
\end{subequations}
The
boundary conditions \eqref{equ:phi_bc_1} and \eqref{equ:phi_bc_2}
correspond to the requirement that, if any fluid interface
intersects the domain boundary wall, the contact angle formed
between the interface and the wall shall be $90^0$.

Finally, 
for the parameter $\beta$ in the free energy density function
\eqref{equ:free_energy}, we will follow \cite{Dong2014}
and use the following expression
\begin{equation}
\beta = \sqrt{3\sqrt{2}\sigma_{min}\eta},
\label{equ:beta_expr}
\end{equation}
where $\sigma_{min} = \min\{\sigma_{ij} \}_{1\leqslant i<j\leqslant N}$
denotes the minimum of
the $\frac{1}{2}N(N-1)$ pairwise surface tensions
among the N fluids.
With this choice of $\beta$, the parameter
$\eta$ corresponds to the characteristic interfacial
thickness of the interface associated with 
the minimum pairwise surface tension $\sigma_{min}$.


\section{Numerical Algorithm}
\label{sec:method}


In this section we present a numerical algorithm for solving
the system of governing equations \eqref{equ:nse}--\eqref{equ:CH}
for the general order parameters defined in \eqref{equ:order_param},
together with the boundary conditions \eqref{equ:vel_bc}--\eqref{equ:phi_bc_2}
for the velocity and the phase field functions.

The primary challenge  lies in the system of
($N-1$) phase field equations \eqref{equ:CH}.
This system is considerably more strongly coupled 
 for the general order parameters,
compared to that in \cite{Dong2014} for the
set of special order parameters defined by \eqref{equ:special_order_param}.
In particular, the inertia terms $\frac{\partial\phi_i}{\partial t}$
are coupled with one another due to 
the   $\mathbf{A}_2$ matrix.

We will concentrate on the numerical treatment of 
the coupled system of ($N-1$) phase field equations \eqref{equ:CH}.
Our algorithm will, after discretization, 
reduce this strongly-coupled system of fourth-order
equations into ($N-1$) {\em de-coupled individual} 
fourth-order equations, each of which  can then be
further reduced into two de-coupled Helmholtz-type
equations using a technique originally developed
for two-phase phase field
equations.

For the N-phase momentum equations, 
\eqref{equ:nse} and \eqref{equ:continuity}, 
we will present an algorithm in  Appendix B.
The main strategy of this algorithm 
for treating the numerical difficulties associated with variable density
and variable dynamic viscosity
stems from the method we developed in 
\cite{DongS2012} for two-phase Navier-Stokes equations.
This algorithm is different in formulation from that of \cite{Dong2014},
in the way how the pressure computation and velocity computation
are de-coupled from each other.

\subsection{Algorithm for Coupled  
Phase-Field Equations with General Order
Parameters}
\label{sec:algorithm}


Let us focus on how to numerically solve the system of 
($N-1$) coupled phase-field equations, \eqref{equ:CH},
together with the boundary conditions, \eqref{equ:phi_bc_1}
and \eqref{equ:phi_bc_2}.
Let $n$ denote the time step index, and
$(\cdot)^n$ denote the variable $(\cdot)$
at time step $n$. 
We assume that $\mathbf{u}^n$ 
and $\phi_i^{n}$ ($1\leqslant i\leqslant N-1$)
are known. 

We discretize the coupled phase-field equations
and the boundary conditions in time as follows,
\begin{subequations}
\begin{multline}
\sum_{j=1}^{N-1} d_{ij} \left(
    \frac{\gamma_0\phi_j^{n+1}-\hat{\phi_j}}{\Delta t}
    + \mathbf{u}^{*,n+1}\cdot\nabla\phi_j^{*,n+1}
  \right)
= 
\nabla^2\left[
  -\sum_{j=1}^{N-1}\lambda_{ij} \nabla^2\phi_j^{n+1} 
  \right. \\
  \left.
  + \frac{1}{\eta^2} \sum_{j=1}^{N-1} S_{ij}\left(
       \phi_j^{n+1} - \phi_j^{*,n+1}
     \right)
  + h_i(\vec{\phi}^{*,n+1})
\right]
+ g_i^{n+1},
\qquad
1\leqslant i\leqslant N-1,
\label{equ:phase_1}
\end{multline}
\begin{equation}
\left. \mathbf{n}\cdot\nabla\left(\nabla^2\phi_i^{n+1}  \right)\right|_{\partial\Omega} = 0, \qquad 1\leqslant i\leqslant N-1,
\label{equ:phase_2}
\end{equation}
\begin{equation}
\left. \mathbf{n}\cdot\nabla\phi_i^{n+1}\right|_{\partial\Omega} = 0,
\qquad 1\leqslant i\leqslant N-1.
\label{equ:phase_3}
\end{equation}
\end{subequations}
In the above equations,
$\Delta t$ is the time step size, $\mathbf{n}$
is an outward-pointing unit vector normal to $\partial\Omega$,
and
$S_{ij}$ ($1\leqslant i,j\leqslant N-1$) are
$(N-1)^2$ chosen constants to be determined below.
Let $J$ ($J=1$ or $2$) denote the order
of temporal accuracy, and 
$\chi$ denote a generic variable.
Then $\chi^{*,n+1}$ represents a $J$-th order explicit
approximation of $\chi^{n+1}$ given by
\begin{equation}
\chi^{*,n+1} = \left\{
\begin{array}{ll}
\chi^n, & J=1 \\
2\chi^n - \chi^{n-1}, & J=2.
\end{array}
\right.
\label{equ:def_nplus1_star}
\end{equation}
$\frac{1}{\Delta t}(\gamma_0\chi^{n+1}-\hat{\chi})$
represents an approximation of 
$\left.\frac{\partial\chi}{\partial t}\right|^{n+1}$ 
by a $J$-th order backward differentiation formula, and
$\hat{\chi}$ and $\gamma_0$ are given by 
\begin{equation}
\hat{\chi} = \left\{
\begin{array}{ll}
\chi^n, & J=1 \\
2\chi^n - \frac{1}{2}\chi^{n-1}, & J=2,
\end{array}
\right.
\qquad
\gamma_0 = \left\{
\begin{array}{ll}
1, & J=1 \\
\frac{3}{2}, & J=2.
\end{array}
\right.
\label{equ:def_hat_var}
\end{equation}
$\vec{\phi}^{*,n+1}$ denotes the vector of
$\phi_i^{*,n+1}$ ($1\leqslant i\leqslant N-1$).

Equation \eqref{equ:phase_1} represents a set of
($N-1$) fourth-order equations 
about $\phi_i^{n+1}$ ($1\leqslant i\leqslant N-1$)
that are strongly coupled with one another.
The $(N-1)^2$ extra terms 
$
\sum_{j=1}^{N-1}S_{ij}\left(\phi_j^{n+1} - \phi_j^{*,n+1}  \right)
$
in the discrete form \eqref{equ:phase_1} are critical to
the current algorithm. 
After we transform \eqref{equ:phase_1}
into ($N-1$) de-coupled individual fourth-order
equations, these extra terms make it possible
to re-formulate each individual fourth-order
equation into two de-coupled Helmholtz-type (2nd order)
equations, which can be discretized in space using
$C^0$ spectral elements or finite elements in a straightforward fashion.
 
Re-write \eqref{equ:phase_1} as follows,
\begin{multline}
\sum_{j=1}^{N-1}\lambda_{ij}\nabla^2\left(\nabla^2\phi_j^{n+1} \right)
- \frac{1}{\eta^2}\sum_{j=1}^{N-1}S_{ij}\nabla^2\phi_j^{n+1}
+\frac{\gamma_0}{\Delta t}\sum_{j=1}^{N-1}d_{ij}\phi_j^{n+1} \\
= Q_i 
= Q_i^{(1)} + \nabla^2 Q_i^{(2)},
\qquad
1\leqslant i\leqslant N-1,
\label{equ:phase_1_reform}
\end{multline}
where
\begin{equation}
\left\{
\begin{split}
&
Q_i^{(1)} = g_i^{n+1} - 
  \sum_{j=1}^{N-1} d_{ij}\left(
    -\frac{1}{\Delta t}\hat{\phi_j}
    + \mathbf{u}^{*,n+1}\cdot \nabla\phi_j^{*,n+1}
  \right) \\
&
Q_i^{(2)} = h_i(\vec{\phi}^{*,n+1})
    - \frac{1}{\eta^2} \sum_{j=1}^{N-1} S_{ij} \phi_j^{*,n+1}.
\end{split}
\right.
\label{equ:Q_expr}
\end{equation}
We introduce the following vectors and matrix
\begin{equation}
\left\{
\begin{split}
&
\bm{\Phi} = \begin{bmatrix}
\vdots \\
\phi_i^{n+1} \\
\vdots
\end{bmatrix}_{(N-1)\times 1}, \
\mathbf{Q} = \begin{bmatrix}
\vdots \\
Q_i \\
\vdots
\end{bmatrix}_{(N-1)\times 1}, \
\mathbf{Q}^{(1)} = \begin{bmatrix}
\vdots \\
Q_i^{(1)} \\
\vdots
\end{bmatrix}_{(N-1)\times 1}, \
\mathbf{Q}^{(2)} = \begin{bmatrix}
\vdots \\
Q_i^{(2)} \\
\vdots
\end{bmatrix}_{(N-1)\times 1}, \ \\
&
\mathbf{S} = \begin{bmatrix} S_{ij} \end{bmatrix}_{(N-1)\times(N-1)}.
\quad
\end{split}
\right.
\label{equ:Q_vec_expr}
\end{equation}
Then the equations in \eqref{equ:phase_1_reform}
is equivalent to the following matrix form
\begin{equation}
\mathbf{A}\nabla^2\left(\nabla^2\bm{\Phi}  \right)
- \frac{1}{\eta^2}\mathbf{S}\nabla^2\bm{\Phi}
+ \frac{\gamma_0}{\Delta t}\mathbf{A}_2\bm{\Phi}
 = \mathbf{Q},
\label{equ:phase_mat_form}
\end{equation}
where the matrices $\mathbf{A}$ and $\mathbf{A}_2$
are given by \eqref{equ:A_matrix_expr} and 
\eqref{equ:A2_expr}, and 
note that both matrices are symmetric positive
definite.


Because it is SPD, the matrix $\mathbf{A}_2$
can be diagonalized as follows,
\begin{equation}
\mathbf{T}^{T} \mathbf{A}_2 \mathbf{T} = \mathbf{E}
=\text{diag}(\hat{a}_1, \hat{a}_2, \dots, \hat{a}_{N-1}),
\qquad
\mathbf{A}_2 = \mathbf{TET}^{T},
\label{equ:A2_diagonalization}
\end{equation}
where $\hat{a}_i>0$ ($1\leqslant i\leqslant N-1$)
are the eigenvalues  of $\mathbf{A}_2$,
$\mathbf{E}$ is the diagonal matrix of $\hat{a}_i$,
$\mathbf{T}$ is the orthogonal matrix formed by
the eigenvectors of $\mathbf{A}_2$,
and note that 
$
\mathbf{T}^{-1} = \mathbf{T}^T.
$

Using the expression of $\mathbf{A}_2$ in \eqref{equ:A2_diagonalization},
we can transform \eqref{equ:phase_mat_form}
into
\begin{equation}
\mathbf{B}_{1}\nabla^2\left(\nabla^2\bm{\Phi}_{1} \right)
- \frac{1}{\eta^2}\mathbf{S}_{1}\nabla^2 \bm{\Phi}_{1}
+ \frac{\gamma_0}{\Delta t}\mathbf{E}\bm{\Phi}_{1}
= \mathbf{T}^T \mathbf{Q},
\label{equ:phase_mat_form_1}
\end{equation}
where
\begin{equation}
\mathbf{B}_1 = \mathbf{T}^T\mathbf{AT}, \qquad
\mathbf{S}_1 = \mathbf{T}^T\mathbf{ST}, \qquad
\bm{\Phi}_1 = \mathbf{T}^T\bm{\Phi}.
\label{equ:B1_expr}
\end{equation}
Let
\begin{equation}
\mathbf{D} = 
\sqrt{\frac{\gamma_0}{\Delta t}} \mathbf{E}^{\frac{1}{2}},
\qquad
\frac{\gamma_0}{\Delta t}\mathbf{E} = \mathbf{D}^2,
\label{equ:D_matrix_expr}
\end{equation}
where the exponential in $\mathbf{E}^{\frac{1}{2}}$
is understood to be element-wise operations
applied to the diagonal elements, and note
that $\mathbf{D}$ is a diagonal matrix.
Therefore, we can transform \eqref{equ:phase_mat_form_1}
into 
\begin{equation}
\mathbf{B}_2\nabla^2\left(\nabla^2\bm{\Phi}_{2} \right)
- \frac{1}{\eta^2}\mathbf{S}_{2}\nabla^2 \bm{\Phi}_2
+ \bm{\Phi}_2
= \mathbf{D}^{-1}\mathbf{T}^T \mathbf{Q},
\label{equ:phase_mat_form_2}
\end{equation}
where
\begin{equation}
\mathbf{B}_2 = \mathbf{D}^{-1}\mathbf{B}_1\mathbf{D}^{-1},
\qquad
\mathbf{S}_2 = \mathbf{D}^{-1}\mathbf{S}_1\mathbf{D}^{-1},
\qquad
\bm{\Phi}_2 = \mathbf{D}\bm{\Phi}_1.
\label{equ:B2_expr}
\end{equation}
It is straightforward to verify that
$\mathbf{B}_2$ is SPD because $\mathbf{A}$ is SPD.

Because $\mathbf{B}_2$ is SPD, it can be
diagonalized as follows,
\begin{equation}
\mathbf{P}^T\mathbf{B}_2\mathbf{P} = \mathbf{K}
= \text{diag}(\hat{\lambda}_1, \hat{\lambda}_2, \dots,\hat{\lambda}_{N-1}),
\qquad
\mathbf{B}_2 = \mathbf{PKP}^T,
\label{equ:B2_diagonalization}
\end{equation}
where $\hat{\lambda}_i>0$ ($1\leqslant i\leqslant N-1$)
are the eigenvalues of $\mathbf{B}_2$,
$\mathbf{K}$ is the diagonal matrix of $\hat{\lambda}_i$,
$\mathbf{P}$ is the orthogonal matrix formed
by the eigenvectors of $\mathbf{B}_2$,
and note that $\mathbf{P}^{-1}=\mathbf{P}^T$.

Now we choose $S_{ij}$ ($1\leqslant i,j\leqslant N-1$)
in \eqref{equ:phase_1_reform}
such that
\begin{equation}
\mathbf{P}^T\mathbf{S}_2\mathbf{P} = \hat{\mathbf{S}}
=\text{diag}(\hat{s}_1, \hat{s}_2, \dots, \hat{s}_{N-1}),
\qquad
\mathbf{S}_2 = \mathbf{P}\hat{\mathbf{S}}\mathbf{P}^T,
\label{equ:S2_diagonalization}
\end{equation}
where $\hat{s}_i$ ($1\leqslant i\leqslant N-1$) are
($N-1$) chosen constants to be determined below. 

Using the expression for $\mathbf{B}_2$ in 
\eqref{equ:B2_diagonalization} and
the expression for $\mathbf{S}_2$ in 
\eqref{equ:S2_diagonalization},
we can transform \eqref{equ:phase_mat_form_2}
into
\begin{equation}
\begin{split}
\mathbf{K}\nabla^2\left(\nabla^2 \mathbf{X}  \right)
- \frac{1}{\eta^2}\hat{\mathbf{S}}\nabla^2\mathbf{X}
+ \mathbf{X} 
& = \mathbf{P}^T\mathbf{D}^{-1}\mathbf{T}^T\mathbf{Q} \\
& = \left(\mathbf{P}^T\mathbf{D}^{-1}\mathbf{T}^T\right)\mathbf{Q}^{(1)}
+ \left(\mathbf{P}^T\mathbf{D}^{-1}\mathbf{T}^T\right)\nabla^2\mathbf{Q}^{(2)}
\end{split}
\label{equ:phase_mat_form_3}
\end{equation}
where
\begin{equation}
\mathbf{X} = \mathbf{P}^T\bm{\Phi}_2,
\qquad
\bm{\Phi}_2 = \mathbf{PX}.
\label{equ:X_expr}
\end{equation}

Let
\begin{equation}
\left\{
\begin{split}
&
\mathbf{X} = \begin{bmatrix}
\vdots \\
\xi_i^{n+1} \\
\vdots
\end{bmatrix}_{(N-1)\times 1},
\ 
\mathbf{P}^T\mathbf{D}^{-1}\mathbf{T}^T\mathbf{Q}
= \begin{bmatrix}
\vdots \\
q_i \\
\vdots
\end{bmatrix}_{(N-1)\times 1}, \
\mathbf{P}^T\mathbf{D}^{-1}\mathbf{T}^T\mathbf{Q}^{(1)}
= \begin{bmatrix}
\vdots \\
q_i^{(1)} \\
\vdots
\end{bmatrix}_{(N-1)\times 1}, \
\\
&
\mathbf{P}^T\mathbf{D}^{-1}\mathbf{T}^T\mathbf{Q}^{(2)}
= \begin{bmatrix}
\vdots \\
q_i^{(2)} \\
\vdots
\end{bmatrix}_{(N-1)\times 1}.
\end{split}
\right.
\label{equ:X_vec_expr}
\end{equation}
Then \eqref{equ:phase_mat_form_3}
can be written into ($N-1$) de-coupled individual
equations in terms of the components,
\begin{equation}
\begin{split}
\nabla^2\left(\nabla^2\xi_i^{n+1}  \right)
- \frac{\hat{s}_i}{\hat{\lambda}_i\eta^2}\nabla^2\xi_i^{n+1}
+ \frac{1}{\hat{\lambda}_i}\xi_i^{n+1} 
& = \frac{1}{\hat{\lambda}_i}q_i \\
& = \frac{1}{\hat{\lambda}_i}q_i^{(1)}
  + \frac{1}{\hat{\lambda}_i}\nabla^2 q_i^{(2)},
\qquad
1\leqslant i\leqslant N-1.
\end{split}
\label{equ:phase_decoupled_1}
\end{equation}
We have now transformed the system of 
strongly-coupled fourth-order equations \eqref{equ:phase_1_reform}
into ($N-1$) de-coupled 
fourth-order scalar equations \eqref{equ:phase_decoupled_1}.


Each equation in \eqref{equ:phase_decoupled_1}
has the same form as that of \cite{Dong2014}, and 
therefore can be dealt with in a similar manner.
They each can be re-formulated into two
de-coupled Helmholtz-type equations using a technique 
originated from two-phase flows \cite{YueFLS2004,DongS2012,Dong2012}.
We provide below only the final re-formulated
equations; see \cite{DongS2012,Dong2014} for the process
of reformulations.

The final reformulated forms
for \eqref{equ:phase_decoupled_1} 
are,
\begin{subequations}
\begin{equation}
\nabla^2\psi_i^{n+1} 
- \frac{1}{\hat{\lambda}_i}\left(
    \alpha_i + \frac{\hat{s}_i}{\eta^2}
  \right) \psi_i^{n+1}
= \frac{1}{\hat{\lambda}_i} q_i,
\qquad
1\leqslant i\leqslant N-1,
\label{equ:psi_1}
\end{equation}
\begin{equation}
\nabla^2\xi_i^{n+1} + \frac{\alpha_i}{\hat{\lambda}_i}\xi_i^{n+1}
= \psi_i^{n+1},
\qquad
1\leqslant i\leqslant N-1,
\label{equ:phi_1}
\end{equation}
\end{subequations}
where $\psi_i^{n+1}$ ($1\leqslant i\leqslant N-1$)
are auxiliary variables defined by \eqref{equ:phi_1},
$\alpha_i$ ($1\leqslant i\leqslant N-1$) are
constants given by
\begin{equation}
\alpha_i = \frac{\hat{s}_i}{2\eta^2}\left(
  -1 - \sqrt{1 - 4\hat{\lambda}_i\frac{\eta^4}{\hat{s}_i^2} }
\right),
\qquad
\text{or} \ \
\alpha_i = \frac{\hat{s}_i}{2\eta^2}\left(
  -1 + \sqrt{1 - 4\hat{\lambda}_i\frac{\eta^4}{\hat{s}_i^2} }
\right),
\qquad
1\leqslant i\leqslant N-1,
\end{equation}
and $\hat{s}_i$ ($1\leqslant i \leqslant N-1$) are 
($N-1$) chosen constants that must satisfy 
the condition,
\begin{equation}
\hat{s}_i \geqslant 2\eta^2\sqrt{\hat{\lambda}_i},
\qquad
1\leqslant i\leqslant N-1.
\label{equ:s_hat_condition}
\end{equation}
Note that
the two equations \eqref{equ:psi_1} and \eqref{equ:phi_1}
are apparently de-coupled. One can first
solve \eqref{equ:psi_1} for $\psi_i^{n+1}$, and
then solve \eqref{equ:phi_1} for $\xi_i^{n+1}$.

With the above formulations, in order to compute
$\phi_i^{n+1}$ from the coupled system \eqref{equ:phase_1},
we only need to solve $2(N-1)$ {\em de-coupled individual}
Helmholtz-type equations given by \eqref{equ:psi_1}
and \eqref{equ:phi_1}.

Let us now consider the boundary conditions
\eqref{equ:phase_2} and \eqref{equ:phase_3}.
They can be written in matrix form,
\begin{equation}
\left.\mathbf{n}\cdot\nabla\left(\nabla^2 \bm{\Phi} \right)\right|_{\partial\Omega}
= \left(\mathbf{TD}^{-1}\mathbf{P}\right)\left.\mathbf{n}\cdot\nabla\left(\nabla^2\mathbf{X} \right)
  \right|_{\partial\Omega}
= 0,
\end{equation}
\begin{equation}
\left.
  \mathbf{n}\cdot\nabla\bm{\Phi}
\right|_{\partial\Omega}
= \left(\mathbf{TD}^{-1}\mathbf{P}\right) \left.
  \mathbf{n}\cdot\nabla\mathbf{X}
\right|_{\partial\Omega}
=0,
\end{equation}
where we have used the relations in
\eqref{equ:B1_expr}, \eqref{equ:B2_expr}
and \eqref{equ:X_expr}.
It follows from the above equations that
\begin{equation}
\left.\mathbf{n}\cdot\nabla\left(\nabla^2\xi_i^{n+1}  \right) \right|_{\partial\Omega}=0, \qquad 1\leqslant i\leqslant N-1,
\label{equ:xi_bc_1}
\end{equation}
\begin{equation}
\left.\mathbf{n}\cdot\nabla\xi_i^{n+1} \right|_{\partial\Omega}=0, \qquad 1\leqslant i\leqslant N-1.
\label{equ:xi_bc_2}
\end{equation}
By using equations \eqref{equ:phi_1}
and \eqref{equ:xi_bc_2},
we can transform \eqref{equ:xi_bc_1} into
\begin{equation}
\left. \mathbf{n}\cdot\nabla\psi_i^{n+1}  \right|_{\partial\Omega} = 0,
\qquad 1\leqslant i\leqslant N-1.
\label{equ:psi_bc_1}
\end{equation}


In order to facilitate the implementation with
$C^0$ spectral elements (or finite elements),
we next derive the weak forms for the
equations \eqref{equ:psi_1} and \eqref{equ:phi_1},
incorporating the boundary conditions 
\eqref{equ:xi_bc_2} and \eqref{equ:psi_bc_1}.
Let $\varpi \in H^1(\Omega)$
denote the test function. Taking the 
$L^2$ inner product between equation \eqref{equ:psi_1}
and $\varpi$ and integrating by part,
we get the weak form about $\psi_i^{n+1}$,
\begin{equation}
\int_{\Omega} \nabla\psi_i^{n+1}\cdot\nabla\varpi
+ \frac{1}{\hat{\lambda}_i}\left(
    \alpha_i + \frac{\hat{s}_i}{\eta^2}
  \right) \int_{\Omega} \psi_i^{n+1} \varpi
= -\frac{1}{\hat{\lambda}_i} \int_{\Omega} q_i^{(1)}\varpi
+ \frac{1}{\hat{\lambda}_i} 
    \int_{\Omega} \nabla q_i^{(2)}\cdot \nabla\varpi,
\qquad
\forall \varpi \in H^{1}(\Omega),
\label{equ:psi_weakform}
\end{equation}
where we have used the boundary conditions
\eqref{equ:phase_3} and \eqref{equ:psi_bc_1}.
Taking the $L^2$ inner product between
equation \eqref{equ:phi_1} and $\varpi$
and integrating by part,
we get the weak form about $\xi_i^{n+1}$,
\begin{equation}
\int_{\Omega} \nabla\xi_i^{n+1}\cdot\nabla\varpi
- \frac{\alpha_i}{\hat{\lambda}_i} \int_{\Omega} \xi_i^{n+1}\varpi
= -\int_{\Omega} \psi_i^{n+1} \varpi,
\qquad
\forall \varpi \in H^1(\Omega),
\label{equ:xi_weakform}
\end{equation}
where we have used the boundary condition
\eqref{equ:xi_bc_2}.
The two weak forms, \eqref{equ:psi_weakform}
and \eqref{equ:xi_weakform}, can be discretized
with $C^0$ spectral elements or 
 finite elements in a straightforward
fashion.
We employ $C^0$ spectral 
elements \cite{KarniadakisS2005,ZhengD2011} for spatial
discretizations in the current paper.


\vspace{0.2in}
Overall, employing general order parameters for
N-phase formulations involves several operations
during pre-processing:
\begin{enumerate}

\item
Choose a specific set of order parameters, by specifying
the matrix $\mathbf{A}_1$ and the vector $\mathbf{b}_1$
defined in \eqref{equ:order_param}.
Compute $\mathbf{A}_2$ from equation \eqref{equ:A2_expr}.
Compute $\left(\mathbf{A}_1^{-1}\right)^T$ in
equation \eqref{equ:A1_inverse_expr}.

\item
Compute $\lambda_{ij}$ ($1\leqslant i,j\leqslant N-1$)
from equations \eqref{equ:lambda_ij_gop},
\eqref{equ:lambda_kl_expr} and \eqref{equ:lambda_kk_expr}
based on the pairwise surface tensions
$\sigma_{kl}$ ($1\leqslant k<l\leqslant N$)
among the N fluids.
Form matrix $\mathbf{A}$ according to \eqref{equ:A_matrix_expr}.

\item
Solve the eigenvalue problem about matrix $\mathbf{A}_2$.
Form matrices $\mathbf{E}$ and $\mathbf{T}$
in \eqref{equ:A2_diagonalization}.
Compute matrix $\mathbf{D}$ in \eqref{equ:D_matrix_expr}.

\item
Compute matrix $\mathbf{B}_2$ in \eqref{equ:B2_expr}.
Solve the eigenvalue problem about $\mathbf{B}_2$.
Form matrices $\mathbf{K}$ and $\mathbf{P}$
in \eqref{equ:B2_diagonalization}.

\item
Choose ($N-1$) constants $\hat{s}_i$ ($1\leqslant i\leqslant N-1$)
that satisfy the conditions \eqref{equ:s_hat_condition}.
Form the diagonal matrix $\hat{\mathbf{S}}$ in
\eqref{equ:S2_diagonalization}.
Compute matrix $\mathbf{S}$ based on
\begin{equation}
\mathbf{S} = \left[ S_{ij} \right]_{(N-1)\times(N-1)} 
= \left(\mathbf{TDP}\right)\hat{\mathbf{S}}\left(\mathbf{TDP}\right)^T.
\label{equ:S_matrix_expr}
\end{equation}

\end{enumerate}

During each time step, given ($\phi_i^n$, $\mathbf{u}^n$),
we compute ($\phi_i^{n+1}$, $\nabla^2\phi_i^{n+1}$)
with the following procedure. We refer to this procedure
as {\bf Advance-Phase-GOP} (``GOP'' standing for general
order parameters) hereafter in this paper. It is comprised of
several steps: \\[0.1in]
\noindent\underline{{\bf Advance-Phase-GOP} procedure:}
\begin{enumerate}

\item
Compute $Q_i^{(1)}$ and $Q_i^{(2)}$ ($1\leqslant i\leqslant N-1$)
based on \eqref{equ:Q_expr}.
Form vectors $\mathbf{Q}^{(1)}$ and $\mathbf{Q}^{(2)}$
in \eqref{equ:Q_vec_expr}.

\item
Compute vectors
$
\mathbf{P}^T\mathbf{D}^{-1}\mathbf{T}^T\mathbf{Q}^{(1)}
$
and 
$
\mathbf{P}^T\mathbf{D}^{-1}\mathbf{T}^T\mathbf{Q}^{(2)}
$
in \eqref{equ:X_vec_expr}.
Then $q_i^{(1)}$ and $q_i^{(2)}$ are known.

\item
Solve equations \eqref{equ:psi_weakform}
for $\psi_i^{n+1}$ ($1\leqslant i\leqslant N-1$).

\item
Solve equations \eqref{equ:xi_weakform} for
$\xi_i^{n+1}$ ($1\leqslant i\leqslant N-1$).
Form vector $\mathbf{X}$ in \eqref{equ:X_vec_expr}.

\item
Compute $\bm{\Phi}$ based on
\begin{equation}
\bm{\Phi} = \mathbf{T}\mathbf{D}^{-1}\mathbf{P} \mathbf{X}.
\end{equation}
This provides $\phi_i^{n+1}$ ($1\leqslant i\leqslant N-1$).

\item
Compute $\nabla^2\phi_i^{n+1}$ ($1\leqslant i\leqslant N-1$)
based on 
\begin{equation}
\begin{bmatrix}
\vdots \\
\nabla^2\phi_i^{n+1} \\
\vdots
\end{bmatrix}
= \nabla^2\bm{\Phi}
= \mathbf{T}\mathbf{D}^{-1}\mathbf{P} \nabla^2\mathbf{X}
= \mathbf{T}\mathbf{D}^{-1}\mathbf{P} \begin{bmatrix}
\vdots \\
\psi_i^{n+1} - \frac{\alpha_i}{\hat{\lambda}_i}\xi_i^{n+1} \\
\vdots
\end{bmatrix}.
\label{equ:laplace_phi}
\end{equation}

\item
Compute 
\begin{equation}
\tilde{\mathbf{J}}^{n+1}
= \tilde{\mathbf{J}}(\vec{\phi}^{n+1},\nabla\vec{\phi}^{n+1})
\label{equ:J_tilde_discretized}
\end{equation}
based on equation \eqref{equ:J_tilde_expr},
where $\nabla^2\phi_i^{n+1}$ are obtained from
the previous step.

\item
Compute 
\begin{equation}
\rho^{n+1} = \rho(\vec{\phi}^{n+1}), \qquad
\mu^{n+1} = \mu (\vec{\phi}^{n+1}),
\label{equ:rho_mu_discretized}
\end{equation}
based on equations \eqref{equ:rho_expr}
and \eqref{equ:mu_expr}.
When the maximum density ratio among the
N fluids is large (typically beyond about $10^2$),
we further clamp the values of $\rho^{n+1}$
and $\mu^{n+1}$ as follows,
\begin{equation}
\rho^{n+1} = \left\{
\begin{array}{ll}
\rho^{n+1}, & \text{if} \ \rho^{n+1}\in\left[\tilde{\rho}_{\min},\tilde{\rho}_{\max} \right] \\
\tilde{\rho}_{\max}, & \text{if} \ \rho^{n+1} > \tilde{\rho}_{\max} \\
\tilde{\rho}_{\min}, & \text{if} \ \rho^{n+1} < \tilde{\rho}_{\min},
\end{array}
\right.
\qquad
\mu^{n+1} = \left\{
\begin{array}{ll}
\mu^{n+1}, & \text{if} \ \mu^{n+1}\in\left[\tilde{\mu}_{\min},\tilde{\mu}_{\max} \right] \\
\tilde{\mu}_{\max}, & \text{if} \ \mu^{n+1} > \tilde{\mu}_{\max} \\
\tilde{\mu}_{\min}, & \text{if} \ \mu^{n+1} < \tilde{\mu}_{\min},
\end{array}
\right.
\label{equ:rho_mu_clamp}
\end{equation}
where 
$\tilde{\rho}_{\max} = \max\left\{\tilde{\rho}_i  \right\}_{1\leqslant i\leqslant N}$,
$\tilde{\rho}_{\min} = \min\left\{\tilde{\rho}_i  \right\}_{1\leqslant i\leqslant N}$,
$\tilde{\mu}_{\max} = \max\left\{\tilde{\mu}_i  \right\}_{1\leqslant i\leqslant N}$,
and
$\tilde{\mu}_{\min} = \min\left\{\tilde{\mu}_i  \right\}_{1\leqslant i\leqslant N}$.

\end{enumerate}


The need for clamping the $\rho^{n+1}$ and $\mu^{n+1}$
values in \eqref{equ:rho_mu_clamp} when 
the maximum density ratio
among the N fluids becomes large
has been pointed out in \cite{Dong2014};
see also the same situation
in two-phase flows for large density ratios \cite{DongS2012,Dong2012}.
Practical simulations have shown that 
the numerical values for the phase field variables $\phi_i$
may slightly go out of range at certain spatial points.
When the maximum density ratio among the N fluids
is large, this may produce un-physical (negative) 
values for $\rho^{n+1}$ and $\mu^{n+1}$ computed
from \eqref{equ:rho_mu_discretized} at
certain points in the domain, and cause
numerical difficulties.
Therefore, when the maximum density ratio among
the N fluids becomes large, we employ
the operations in equation \eqref{equ:rho_mu_clamp} 
to avoid this issue.



Let us briefly contrast the above algorithm for
general order parameters with that of \cite{Dong2014}
for the set of special order parameters defined by
\eqref{equ:special_order_param}.
While the phase field equations \eqref{equ:CH}  for the
general order parameters are considerably more
complicated and more strongly coupled than those 
in \cite{Dong2014},  with the algorithm presented above
 these equations do not pose essential computational difficulties.
The computational complexity per time step of 
the above algorithm for general order parameters
is comparable to that of \cite{Dong2014} 
for the simpler phase field equations.


\subsection{Overall Method}

%

Let us now discuss the overall method for simulating
the coupled system of governing equations, \eqref{equ:nse}--\eqref{equ:CH},
for general order parameters.


The introduction of the general order parameters considered here, 
when compared 
with the special set of order parameters of \cite{Dong2014},
does not alter
the overall form of 
the N-phase momentum equations, although the 
 $\tilde{\mathbf{J}}(\vec{\phi},\nabla\vec{\phi})$ here
is very different than in \cite{Dong2014}.
Therefore, the N-phase momentum equations, 
\eqref{equ:nse}--\eqref{equ:continuity},
can in principle be 
solved numerically using the  algorithm
 discussed in \cite{Dong2014}.
However, in the current paper
we will employ an alternative scheme for
the N-phase momentum equations.
This alternative algorithm has been presented in Appendix B. 
It is also a velocity correction-type scheme,
but the algorithmic formulation
is  different from that of \cite{Dong2014}.
It is observed that the current algorithm results in
comparable velocity errors to, but significantly smaller
pressure errors than, the scheme of \cite{Dong2014}
in numerical tests with analytic solutions.
The current algorithm similarly only results in
constant and time-independent coefficient matrices
for the pressure and velocity linear algebraic
systems after discretization, despite
the variable nature of the mixture density
and dynamic viscosity in the N-phase momentum equations.


We combine the algorithm for  
the system of ($N-1$) phase field equations
discussed in Section \ref{sec:algorithm}
and the algorithm in Appendix B
for the N-phase momentum equations 
into an overall method for solving
the coupled system \eqref{equ:nse}--\eqref{equ:CH}
with general order parameters.
The overall procedure can be summarized 
as follows.

Given ($\mathbf{u}^n$, $P^n$, $\phi_i^n$), 
where $P$ is an auxiliary pressure defined by \eqref{equ:effective_P}
in Appendix B,
we compute $\phi_i^{n+1}$, $P^{n+1}$ and
$\mathbf{u}^{n+1}$ successively in
a de-coupled fashion using these steps:
\begin{enumerate}

\item
Use the {\bf Advance-Phase-GOP} procedure
from Section \ref{sec:algorithm} 
to compute $\phi_i^{n+1}$ and
$\nabla^2\phi_i^{n+1}$ ($1\leqslant i\leqslant N-1$),
$\tilde{\mathbf{J}}^{n+1}$,
$\rho^{n+1}$ and $\mu^{n+1}$;

\item
Solve equation \eqref{equ:p_weakform} for $P^{n+1}$
(see Appendix B);

\item
Solve equation \eqref{equ:vel_weakform}, together
with the Dirichlet condition \eqref{equ:velocity_2},
for $\mathbf{u}^{n+1}$ (see Appendix B).

\end{enumerate}


This method has the following characteristics:
\begin{itemize}

\item
It can employ arbitrary order parameters 
of the form \eqref{equ:order_param}
for N-phase flows.

\item
The mixing energy density coefficients $\lambda_{ij}$
($1\leqslant i,j\leqslant N-1$) are explicitly 
given in terms of the pairwise surface
tensions among the N fluids.

\item
The computations for different flow variables
are completely de-coupled. The computations
for the ($N-1$) phase field variables are completely
de-coupled. The computations for the three velocity
components (see \eqref{equ:vel_weakform} in Appendix B) are
also completely de-coupled.

\item
The linear algebraic systems 
for all flow variables involve only
{\em constant} and {\em time-independent} coefficient
matrices after discretization, 
which can be pre-computed during pre-processing,
despite the variable nature of the density and dynamic
viscosity of the N-phase mixture.

\item
Within each time step, the method only requires 
the solution of individual Helmholtz-type (including
Poisson) equations.

\item
It can deal with large density ratios and large viscosity
ratios among the N fluids.

\end{itemize}


\section{Representative Numerical Tests}
\label{sec:tests}

In this section we use several multiphase flow problems
in two dimensions
to demonstrate the 
accuracy and capabilities
of the N-phase physical formulation and numerical
algorithm presented in Section \ref{sec:method}.
These problems involve large density contrasts,
large viscosity contrasts, and pair-wise surface
tensions.
We compare our simulations with the Langmuir-de Gennes theory
of floating liquid lenses
for a three-phase problem to show that our method
 produces physically accurate results.
The majority of simulation results in
this section are obtained using 
the volume fractions $c_i$ ($1\leqslant i\leqslant N-1$)
as the order parameters. 
For several cases, the results obtained
using other order parameters
(e.g. $c_i-c_N$, $1\leqslant i\leqslant N-1$)
are also presented.

\begin{table}
\begin{center}
\begin{tabular*}{1.0\textwidth}{@{\extracolsep{\fill}}
l c | l c}
\hline
variables/parameters & normalization constants 
& variables/parameters & normalization constants \\
$\mathbf{x}$, $\eta$ & $L$ 
& $m_i$ & $L/(\tilde{\rho}_1U_0)$ \\
$\mathbf{u}$, $\mathbf{w}$ & $U_0$
& $\tilde{\gamma}_i$, $\Gamma$ & $1/\tilde{\rho}_1$ \\
$t$, $\Delta t$ & $L/U_0$
& $\sigma_{ij}$ & $\tilde{\rho}_1U_0^2L$ \\
$\mathbf{g}_r$ (gravity) & $U_0^2/L$
& $\lambda_{ij}$, $S_{ij}$ & $\tilde{\rho}_1U_0^2L^2$ \\
$p$, $P$, $h(\vec{\phi})$, $W(\vec{\phi},\nabla\vec{\phi})$ & $\tilde{\rho}_1U_0^2$
& $d_{ij}$  & $\tilde{\rho}_1U_0/L$  \\
$\phi_i$, $\vec{\phi}$, $c_i$ & $1$
& $\mathbf{f}$ & $\tilde{\rho}_1U_0^2/L$ \\
$\beta$ & $\sqrt{\tilde{\rho}_1}U_0 L$
& $g_i$ & $\tilde{\rho}_1U_0^2/L^2$ \\
$\tilde{\rho}_i$, $\rho_i$, $\rho$, $a_{ij}$, $b_i$, $\varphi_i(\vec{\phi})$, $\rho_0$  
& $\tilde{\rho}_1$
& $\tilde{\mathbf{J}}$ & $\tilde{\rho}_1U_0$ \\
$\tilde{\mu}_i$, $\mu$ & $\tilde{\rho}_1 U_0 L$
& $\nu_0$ & $U_0L$ \\
\hline
\end{tabular*}
\caption{
Normalization of flow variables and physical/numerical parameters.
}
\label{tab:normalization}
\end{center}
\end{table}


Let us now briefly mention  the normalizations 
of flow variables, governing equations, and the
boundary/initial conditions.
One can show that,
when the flow variables and physical parameters 
are normalized in a proper fashion,
the forms of the N-phase governing equations 
and the boundary/initial conditions
will remain un-changed upon non-dimensionalization.
Specifically, 
the normalization constants for various flow variables
and parameters are summarized in 
Table \ref{tab:normalization}, where
$L$ and $U_0$ are respectively the 
characteristic length and velocity scales,
and $\tilde{\rho}_1$ is the density of the
first fluid.
For example, the non-dimensional pairwise surface tensions 
are given by $\frac{\sigma_{ij}}{\tilde{\rho}_1U_0^2L}$
(inverse of Weber numbers)
based on this table.
In the following discussions, all variables
are non-dimensional unless otherwise specified, and have been
normalized according to Table \ref{tab:normalization}.

\subsection{Convergence Rates}

The goal of this section is to demonstrate
 the spatial and temporal convergence rates of
the algorithm developed in Section \ref{sec:method}
by using a contrived analytic solution
to the N-phase  governing equations.

Consider the flow domain defined by
$
\Omega = \left\{ \ (x,y) \ : \
0\leqslant x\leqslant 2, \
-1\leqslant y\leqslant 1 \ 
\right\}
$, and a four-phase fluid mixture  contained in $\Omega$.
We assume the following analytic expressions for 
the flow and phase field variables
\begin{equation}
\left\{
\begin{split}
&
u = A_u \cos \pi y \sin a x \sin \omega t \\
&
v = -\frac{A_ua}{\pi}\sin \pi y \cos ax \sin \omega t \\
&
P = A_u\sin\pi y\sin ax \cos \omega t \\
&
\phi_1 = \frac{1}{6}\left( 
      1 + A_1\cos a_1x \cos b_1y \sin \omega_1 t
    \right) \\
&
\phi_2 = \frac{1}{6}\left( 
      1 + A_2\cos a_2x \cos b_2y \sin \omega_2 t
    \right) \\
& 
\phi_3 = \frac{1}{6}\left( 
      1 + A_3\cos a_3x \cos b_3y \sin \omega_3 t
    \right)
\end{split}
\right.
\label{equ:anal_soln}
\end{equation}
where $(u,v)$ are the velocity components of $\mathbf{u}$,
and 
$A_u$, $a$, $\omega$, $A_i$, $a_i$, $b_i$, $\omega_i$
($1\leqslant i\leqslant 3$)
are prescribed constants to be given below.
It is straightforward to verify that
$(u,v)$ in \eqref{equ:anal_soln} satisfy the
equation \eqref{equ:continuity},
and that the expressions for $\phi_1$, $\phi_2$ and $\phi_3$
satisfy the boundary conditions
\eqref{equ:phi_bc_1} and \eqref{equ:phi_bc_2}
with the parameter values given below.
We choose the body force $\mathbf{f}(\mathbf{x},t)$ 
in \eqref{equ:nse_reform} (in Appendix B)
and the source terms $g_i(\mathbf{x},t)$
($1\leqslant i\leqslant N-1$) in
\eqref{equ:CH} such that 
the analytic expressions in \eqref{equ:anal_soln}
satisfy the governing 
equations \eqref{equ:nse_reform} and \eqref{equ:CH}.
In addition, we choose the boundary velocity 
$\mathbf{w}(\mathbf{x},t)$ in \eqref{equ:vel_bc}
according to the velocity analytic 
expressions in \eqref{equ:anal_soln},
and choose the initial velocity and the initial
phase field functions by setting $t=0$
to the analytic expressions of \eqref{equ:anal_soln}.


\begin{table}
\begin{center}
\begin{tabular*}{0.9\textwidth}{@{\extracolsep{\fill}}
l c | l c}
\hline
parameters & values 
& parameters & values \\
$A_u$ & $2.0$ 
 & $\tilde{\rho}_3$   & $2.0$ \\
$A_1$, $A_2$, $A_3$ & $1.0$ 
 & $\tilde{\rho}_4$  & $4.0$ \\
$a$, $a_1$, $a_2$, $a_3$ & $\pi$
 & $\tilde{\mu}_1$  & $0.01$  \\
$b_1$, $b_2$, $b_3$ & $\pi$
 & $\tilde{\mu}_2$  & $0.02$  \\
$\omega$, $\omega_1$ & $1.0$
  & $\tilde{\mu}_3$  & $0.03$ \\
$\omega_2$ & $1.2$
  & $\tilde{\mu}_4$  & $0.04$ \\
$\omega_3$ & $0.8$
  & $m_1$  & $10^{-3}$  \\
$\eta$ & $0.1$
 & $m_2$  & $2\times 10^{-3}$ \\
$\beta$ & $0.05$
 & $m_3$  & $3\times 10^{-3}$ \\
$\tilde{\rho}_1$ & $1.0$
 & $\rho_0$  & $\min (\tilde{\rho}_1,\dots,\tilde{\rho}_4)$ \\
$\tilde{\rho}_2$ & $3.0$
 & $\nu_0$ & $\max\left(\frac{\tilde{\mu}_1}{\tilde{\rho}_1},\dots, \frac{\tilde{\mu}_4}{\tilde{\rho}_4} \right)$ \\
$\sigma_{12}$ & $6.236\times 10^{-3}$ 
 & $\sigma_{23}$  & $8.165\times 10^{-3}$ \\
$\sigma_{13}$ & $7.265 \times 10^{-3}$
 & $\sigma_{24}$   & $5.270\times 10^{-3}$ \\
$\sigma_{14}$ & $3.727\times 10^{-3}$
 & $\sigma_{34}$  & $6.455\times 10^{-3}$  \\
$J$ (integration order) & $2$
  & $\lambda_{ij}$  & computed from \eqref{equ:matrix_lambda_ij_gop}  \\
\hline
\end{tabular*}
\caption{
Parameter values for the convergence-rate tests.
}
\label{tab:anal_param}
\end{center}
\end{table}


We partition the domain along the $x$ direction
 into two quadrilateral
elements of equal size. The element order
is varied systematically in the
tests. We employ the algorithm developed
in Section \ref{sec:method} to integrate 
the governing equations in time from
$t=0$ to $t=t_f$ ($t_f$ to be specified later),
and then compute and monitor the errors of the numerical
solution at $t=t_f$ against the analytic solution from
\eqref{equ:anal_soln}. 
The parameter values for this problem 
are listed in Table \ref{tab:anal_param}.

We have tested two sets of order parameters.
The first set employs the volume
fractions $c_i$ ($1\leqslant i\leqslant N-1$)
as the order parameters, which is defined 
in equation \eqref{equ:order_param_volfrac}.
The other set employs the 
re-scaled density differences
$\rho_i-\rho_{i+1}$ ($1\leqslant i\leqslant N-1$)
as the order parameters; see the definition
in \eqref{equ:id_gop_4}.

\begin{figure}
\centerline{
\psfig{file=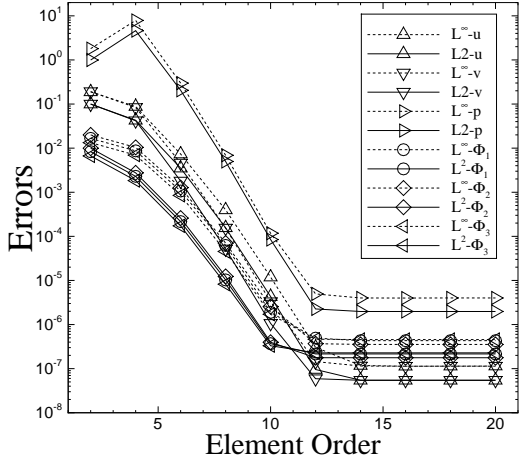,height=2.8in}(a)
\psfig{file=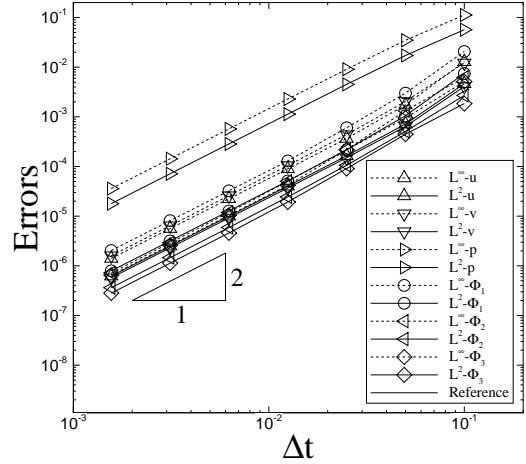,height=2.8in}(b)
}
\centerline{
\psfig{file=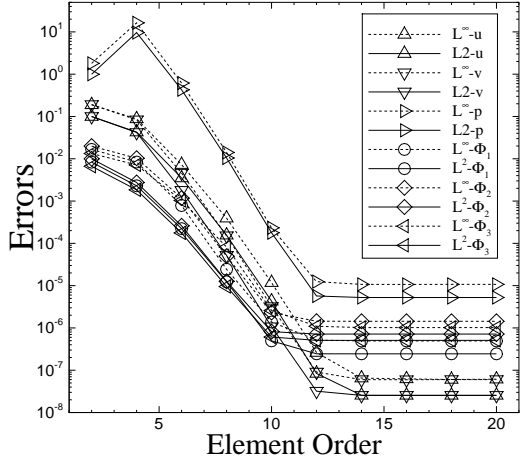,height=2.8in}(c)
\psfig{file=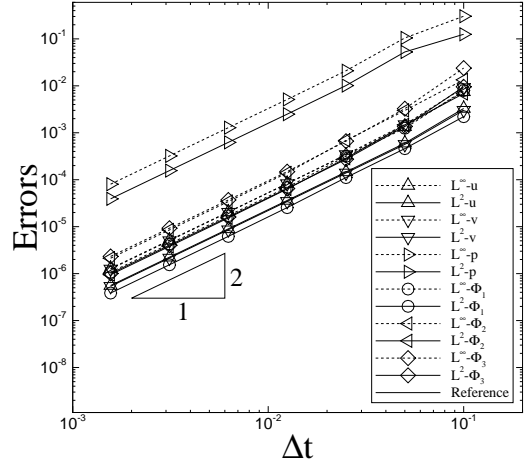,height=2.8in}(d)
}
\caption{
Convergence rates:
(a) and (c), numerical errors as a function of element order
(fixed $\Delta t=0.001$) showing spatial exponential convergence
and error saturation at large element orders due to temporal
truncation error. (b) and (d), numerical errors as a function
of $\Delta t$ (fixed element order $18$) showing temporal
second-order convergence rate. 
Results in (a) and (b) are obtained using formulation
with volume fractions $c_i$ ($1\leqslant i\leqslant N-1$) as order parameters. 
Those in (c) and (d)
are obtained using formulation with re-scaled
density differences, 
$\rho_i-\rho_{i+1}=-\tilde{\rho}_{i+1}
+(\tilde{\rho}_i+\tilde{\rho}_{i+1})\phi_i$ ($1\leqslant i\leqslant N-1$),
as order parameters. 
}
\label{fig:conv}
\end{figure}


In the first group of tests we fix the time step size
at $\Delta t=0.001$ and final integration time at
$t_f = 0.1$, and vary the element order
systematically between $2$ and $20$.
Figure \ref{fig:conv}(a) shows 
the errors of the numerical solution 
in $L^{\infty}$ and $L^2$ norms
for the velocity, pressure and the 
phase-field variables ($\phi_1$, $\phi_2$, $\phi_3$) at $t=t_f$
as a function of the element order,
obtained with the volume fractions as
the order parameters (see equation \eqref{equ:order_param_volfrac}).
It is evident that the numerical errors
decrease exponentially as the element order
increases and is below about $12$. 
As the element order increases beyond $12$,
the error curves level off owing to the saturation
with the temporal truncation errors.
Figure \ref{fig:conv}(c) shows the corresponding
error results obtained using the re-scaled
density differences as the order parameters
(see \eqref{equ:id_gop_4}).
One can similarly observe an exponential
 spatial convergence rate.

In the second group of convergence tests 
we fix the element order at a large value $18$
and the final integration time at $t_f=1.0$.
Then we vary the time step size systematically
between $\Delta t=0.0015625$ and $\Delta t=0.1$.
In Figure \ref{fig:conv}(b) we
plot the $L^{\infty}$ and $L^2$ errors of the numerical
solution at $t=t_f$ against the analytic solution
for different variables as a function of
$\Delta t$, obtained with the volume
fractions as the order parameters.
The error curves exhibit a second-order
temporal convergence rate.
Figure \ref{fig:conv}(d) contains the corresponding
results obtained with the re-scaled
density differences as the order parameters,
showing similarly a second-order 
temporal convergence rate.


\subsection{Floating Oil Lens on Water Surface}


\begin{figure}
\centerline{
\includegraphics[height=1.6in]{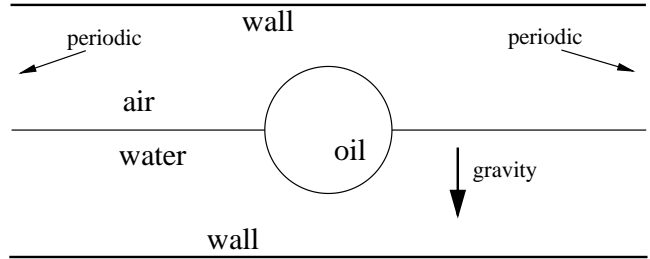}
}
\caption{
Initial configuration for the three-phase problem of floating oil lens 
on water surface.
}
\label{fig:3phase_config}
\end{figure}

In this section we look into an air-water-oil 
three-phase problem, where the oil forms a liquid lens floating
on the water surface at equilibrium. 
We  quantitatively compare  the simulation results 
with the theory by Langmuir and
de Gennes \cite{Langmuir1933,deGennesBQ2003}
to show that our method produces physically accurate
results. A similar liquid-lens problem was considered
in \cite{Dong2014} using the special set of order parameters 
defined by \eqref{equ:special_order_param}. 
Here we simulate the problem with general order parameters,
and investigate the effects of several sets of physical parameters
on the air-water-oil configurations.


Consider the flow domain depicted in Figure \ref{fig:3phase_config}, 
$-\frac{L}{2}\leqslant x\leqslant \frac{L}{2}$
and $0\leqslant y\leqslant \frac{4}{5}L$,
where $L=4cm$.
The top and the bottom of the domain
($y=0$ and $\frac{4}{5}L$)
are two solid walls.
In the horizontal direction the domain
is periodic at $x=\frac{L}{2}$ and
$x=-\frac{L}{2}$.
At $t=0$, the domain is filled with air
in its top half and filled with water
in its bottom half, and
a circular oil drop of radius $R_0=\frac{L}{5}$
is held at rest on
the water surface with its center located
at $\mathbf{x}_c = (x_c,y_c) = (0, \frac{2}{5}L)$.
It is assumed that the gravity is in the
$-y$ direction, and that
there is no initial flow.
The system is then released, and evolves to
equilibrium
due to the interactions among the three fluid components and
the effects of the gravity and the surface
tensions.
Our goal is investigate the effects of several
physical parameters on the equilibrium configurations
of the oil.


\begin{table}
\begin{center}
\begin{tabular*}{1.0\textwidth}{@{\extracolsep{\fill}}
l| l l| l l|  l l}
\hline
density [$kg/m^3$]: & air & $1.2041$ & water & $998.207$ & oil & $577$ (or varied) \\
dynamic viscosity [$kg/(m\cdot s)$]:
  & air & $1.78E-5$ & water & $1.002E-3$ & oil & $9.15E-2$ \\
surface tension [$kg/s^2$]:
  & air/water & $0.0728$  & oil/water & $0.04$ & air/oil & $0.065$ (or varied) \\
gravity [$m/s^2$]: & $9.8$ &  &  &  &  &   \\
\hline
\end{tabular*}
\caption{
Physical parameter values for the air-water-oil three phase problem.
}
\label{tab:3phase_param}
\end{center}
\end{table}

Table \ref{tab:3phase_param} lists the values of
the physical parameters, 
including the densities, dynamic viscosities,
and the pairwise surface tensions of the three
fluids involved in this problem.
We use $L$ as the characteristic
length scale, and choose
a characteristic velocity scale
$U_0 = \sqrt{g_{r0}L}$,
where $g_{r0} = 1m/s^2$.
We assign the air, water and oil
as the first, second and third fluid, respectively.
All the flow variables and physical
parameters are then non-dimensionalized
based on the normalization
constants specified in Table \ref{tab:normalization}.
We set $g_i=0$ ($1\leqslant i\leqslant N-1$)
in \eqref{equ:CH} for the simulations.


In order to simulate the problem,
we discretize the domain using $160$ equal-sized
quadrilateral elements, with $20$ elements
along the $x$ direction and $8$
along the $y$ direction.
The element order is $16$ in
the simulations.
The algorithm developed in Section
\ref{sec:method} has been used to solve
 the coupled system of governing
equations. 
For the boundary conditions,
at the top/bottom walls, 
we impose the condition \eqref{equ:vel_bc}
with $\mathbf{w}=0$ for the velocity
and the conditions \eqref{equ:phi_bc_1}
and \eqref{equ:phi_bc_2}
for the phase field functions.
In the horizontal direction, all
the flow variables (velocity, pressure,
phase field functions) are set to be
periodic.
Table \ref{tab:3phase_simu_param} summarizes
the numerical parameter values 
in the simulations.


\begin{table}
\begin{center}
\begin{tabular*}{0.8\textwidth}{@{\extracolsep{\fill}}
l l }
\hline
parameters & values  \\
$\lambda_{ij}$ & computed based on \eqref{equ:matrix_lambda_ij_gop}
 \\
$\eta/L$ & $0.0075$ 
 \\
$\beta$ & computed based on \eqref{equ:beta_expr}
\\
$m_i\tilde{\rho}_1U_0/L$, $1\leqslant i\leqslant N-1$ &
$10^{-7}/\lambda_{max}$, 
where $\lambda_{max} = \frac{ \max\{\lambda_{ij} \} }{\tilde{\rho}_1U_0^2L^2}$
\\ 
$\hat{s}_i$ & $2\eta^2\sqrt{\hat{\lambda}_i}$ 
\\
$\rho_0$ & $\min\left\{\tilde{\rho}_i  \right\}_{1\leqslant i\leqslant N}$
\\
$\nu_0$ & $2\max\left\{\frac{\tilde{\mu}_i}{\tilde{\rho}_i} \right\}_{1\leqslant i\leqslant N} $
\\
$U_0\Delta t/L$ & $2.5\times 10^{-6}$
\\
$J$ (temporal order) & $2$ \\
\hline
\end{tabular*}
\caption{
Simulation parameter values for the air-water-oil three phase problem.
}
\label{tab:3phase_simu_param}
\end{center}
\end{table}


We employ the formulation with
the volume fractions as the order parameters,
defined in \eqref{equ:order_param_volfrac},
for simulations of all the cases in this section.
We have also simulated several selected cases 
using the formulation with
the re-scaled volume fraction differences as
the order parameters, as defined in \eqref{equ:id_gop_1}.
The initial velocity is set to zero.
The initial phase field functions are
set to 
\begin{equation}
\phi_i(\mathbf{x}, t=0) = c_{i0},
\quad 1\leqslant i\leqslant N-1,
\label{equ:ic_gop_0}
\end{equation}
for the order parameters defined by \eqref{equ:order_param_volfrac},
and 
\begin{equation}
\phi_i(\mathbf{x}, t=0) = \frac{1}{2} + \frac{1}{2}\left(c_{i0}-c_{N0} \right),
\quad 1\leqslant i\leqslant N-1,
\label{equ:ic_gop_1}
\end{equation}
for the order parameters defined by \eqref{equ:id_gop_1}.
In these equations, $c_{i0}$ ($1\leqslant i\leqslant N$)
are the initial volume fractions given by ($N=3$)
\begin{multline*}
c_{10} = \left[1 - \Theta(x,x_{c}-R_0) + \Theta(x,x_c+R_0)  \right]
    \frac{1}{2}\left(
      1 + \tanh\frac{y - y_c}{\sqrt{2}\eta}
    \right) \\
    + \left[\Theta(x,x_{c}-R_0) - \Theta(x,x_c+R_0) \right]
    \frac{1}{2}\left(
      1 + \tanh\frac{\left|\mathbf{x} -\mathbf{x}_c  \right| - R_0}{\sqrt{2}\eta}
    \right) \Theta(y,y_c),
\end{multline*}
\begin{multline*}
c_{20} = \left[1 - \Theta(x,x_{c}-R_0) + \Theta(x,x_c+R_0)  \right]
    \frac{1}{2}\left(
      1 - \tanh\frac{y - y_c}{\sqrt{2}\eta}
    \right)  \\
    + \left[\Theta(x,x_{c}-R_0) - \Theta(x,x_c+R_0) \right]
    \frac{1}{2}\left(
      1 + \tanh\frac{\left|\mathbf{x} -\mathbf{x}_c  \right| - R_0}{\sqrt{2}\eta}
    \right) \left[1 - \Theta(y,y_c)  \right],
\end{multline*}
\begin{equation*}
c_{30} = \frac{1}{2}\left(
        1 - \tanh\frac{\left|\mathbf{x} -\mathbf{x}_c  \right| - R_0}{\sqrt{2}\eta}
    \right),
\end{equation*}
where $\Theta(x,a)$ is the unit step function,
assuming unit value if $x\geqslant a$ and zero otherwise.


\begin{figure}
\centering
\includegraphics[height=1.5in]{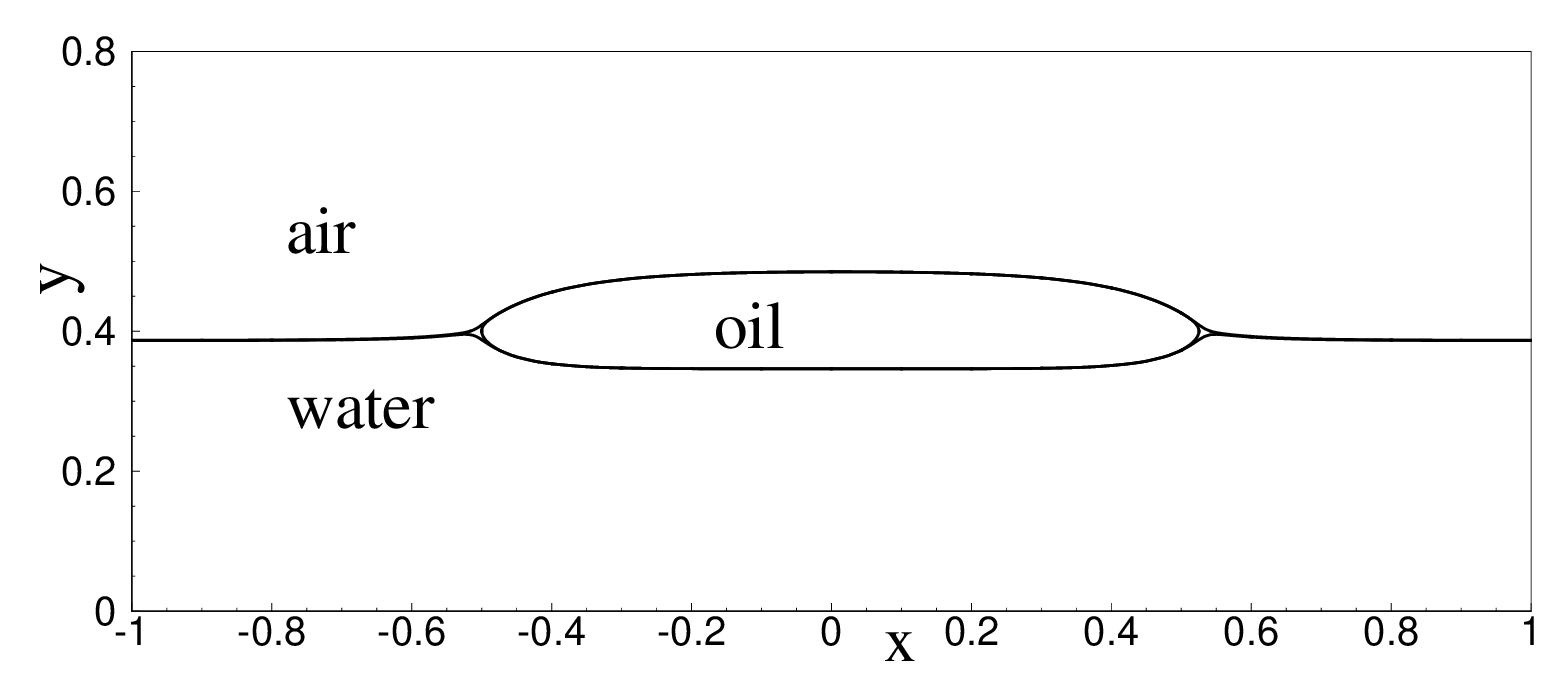}(a)
\includegraphics[height=1.5in]{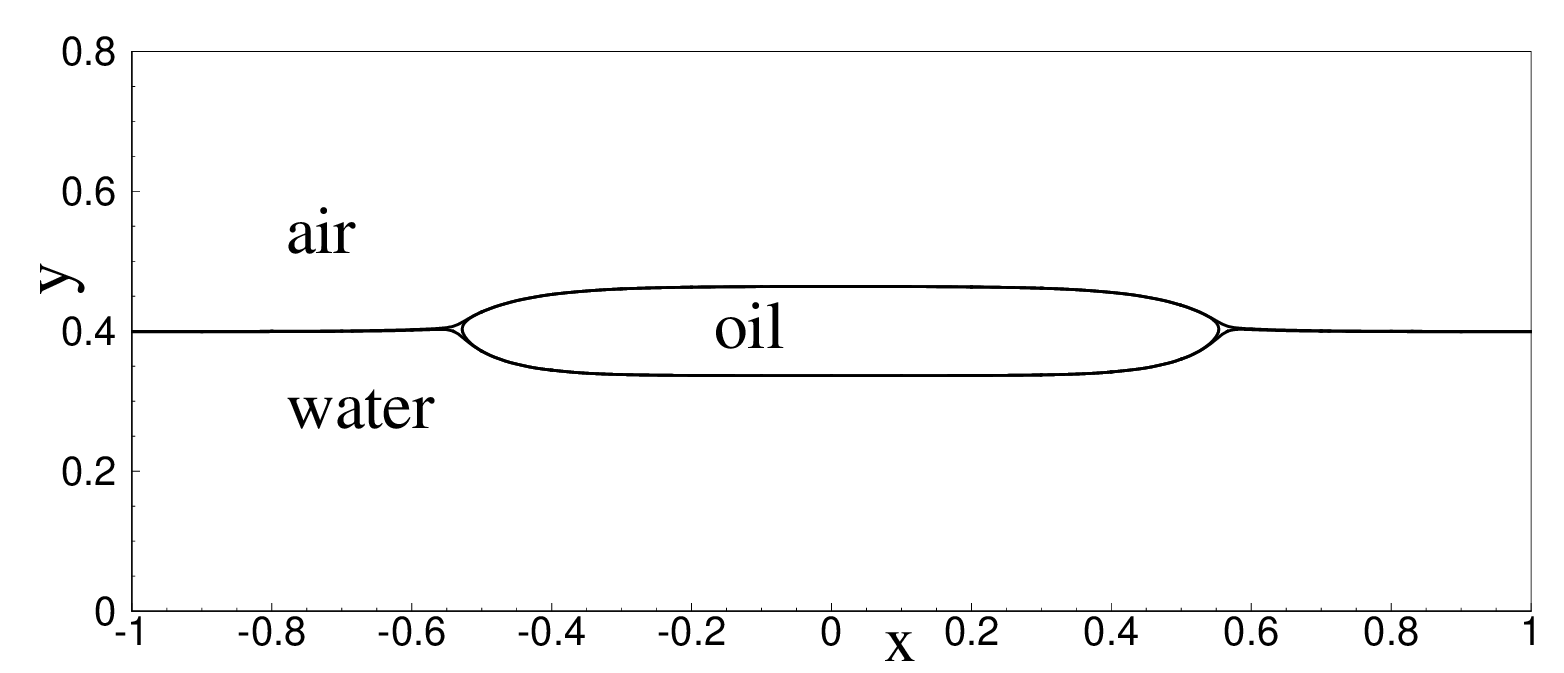}(b)
\includegraphics[height=1.5in]{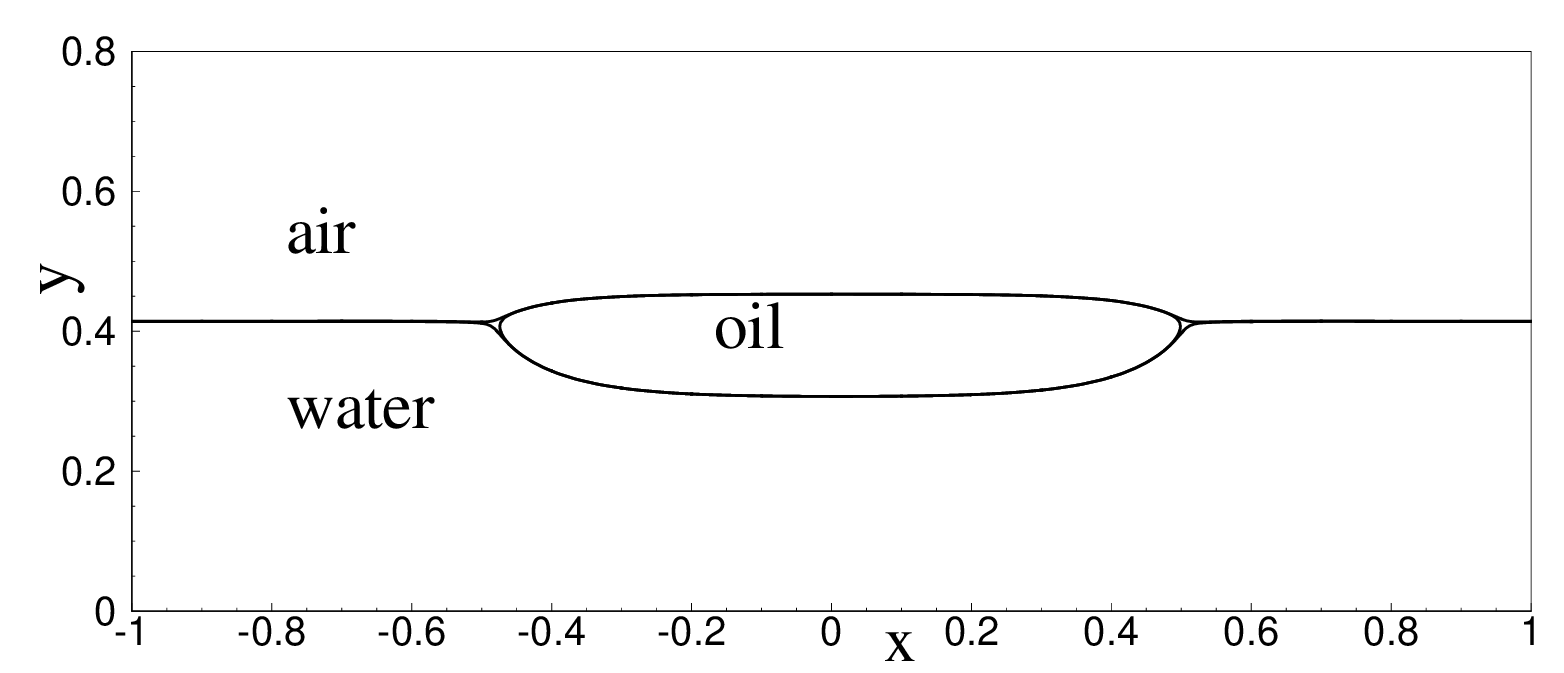}(c)
\caption{Effect of oil density on equilibrium oil puddle 
configurations (3 fluid phases): (a) $\rho_o = 300 kg/m^3$,
(b) $\rho_o = 500 kg/m^3$,
(c) $\rho_o = 750 kg/m^3$.
Results obtained using volume fractions $c_i$ ($1\leqslant i\leqslant N-1$)
as order parameters.
}
\label{fig:3phase_density_config}
\end{figure}


The physics of floating liquid lenses for three
phases was discussed in Langmuir \cite{Langmuir1933} and
 de Gennes et al \cite{deGennesBQ2003} (pages 54--56).
The equilibrium oil-drop shape is determined by
the interplay of the gravity
and the three pair-wise surface tensions,
and also influenced by the three densities.
When the gravity effect dominates, the oil
will form a puddle on the water surface.
If the surface tension effects dominate,
the oil-drop shape will consist of  two
circular caps in two dimensions
or two spherical caps
in three dimensions.
One can approximately 
determine which effect is dominant
by comparing the characteristic drop size
with the three capillary lengths associated
with the three types of interfaces; 
see \cite{deGennesBQ2003}.

We will investigate the effects on 
the equilibrium oil configurations
of two physical parameters:
the oil density and the air-oil surface 
tension.

Let us first consider the effect of
the oil density.
In this group of tests, we vary 
the density of the oil systematically
ranging from $300kg/m^3$ to $750kg/m^3$,
and fix all the other physical parameters
at values given in Table
\ref{tab:3phase_param}
(air-oil surface tension fixed at
$0.065kg/s^2$).
In Figure \ref{fig:3phase_density_config}
we show the equilibrium oil configurations
corresponding to three
oil-density values
$\rho_o=300kg/m^3$, $500kg/m^3$
and $750kg/m^3$.
They are obtained using
the volume fractions
as the order parameters.
%
Plotted are the contours of
volume fractions $c_i = \frac{1}{2}$
($i=1, 2, 3$) for the three fluids.
One can note the  ``star''-shaped
regions around the three-phase contact
lines in these figures. 
This is because in these regions
none of the three fluids has a volume fraction
larger than $\frac{1}{2}$.
It is evident that the oil forms
puddles floating on the water surface.
Subtle differences can be noticed in their shapes,
for example, in the curvature of 
the oil profiles not far from
the three-phase contact lines. 
In addition, the immersion depths of
the oil in the water are notably different
as the oil density changes.


\begin{figure}
\centering
\includegraphics[height=1.5in]{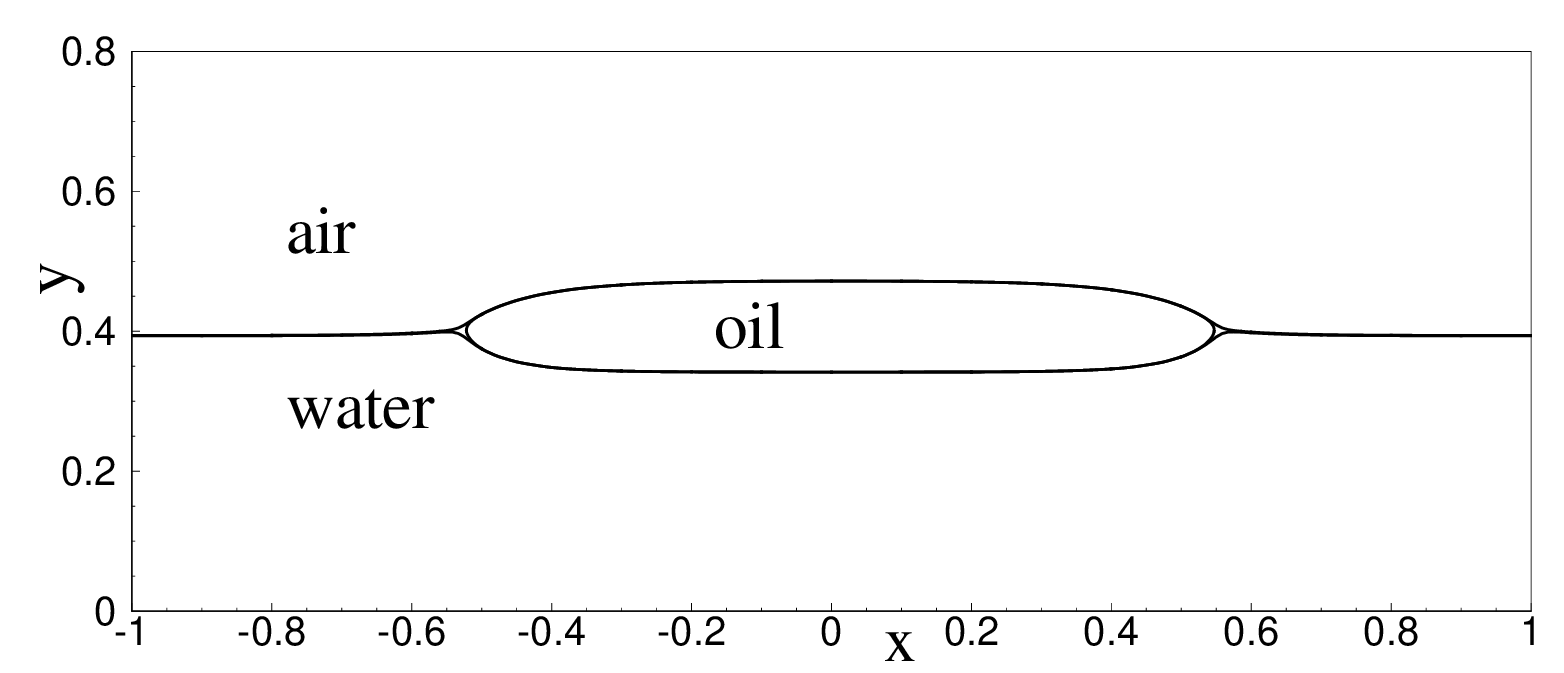}(a)
\includegraphics[height=1.5in]{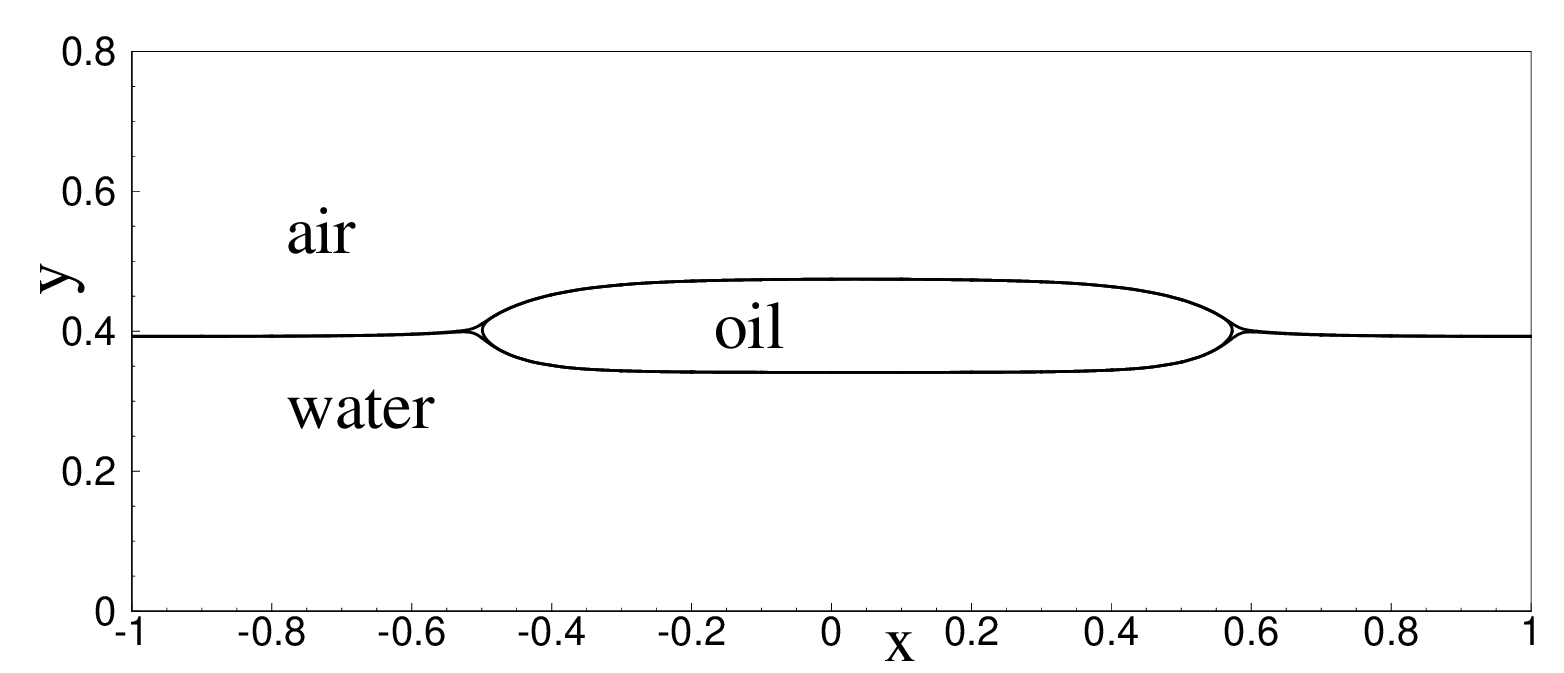}(b)
\caption{
Equilibrium oil-puddle configurations for oil
density $\rho_o=400 kg/m^3$ obtained using
(a) volume fractions $c_i$ ($1\leqslant i\leqslant N-1$) as order parameters,
and (b) re-scaled volume fraction 
differences  
as order parameters (equation \eqref{equ:id_gop_1}).
}
\label{fig:3phase_density_compare_gop}
\end{figure}

The simulation results obtained using  different
sets of order parameters are similar,
as is expected.
This is shown by Figure \ref{fig:3phase_density_compare_gop}.
Here we compare the equilibrium oil configurations,
as shown by the contour levels $c_i=\frac{1}{2}$,
corresponding to the oil density $\rho_0 = 400kg/m^3$
obtained using the volume fractions as
the order parameters (equation \eqref{equ:order_param_volfrac}),
see Figure \ref{fig:3phase_density_compare_gop}(a),
and using the re-scaled volume fraction differences
as the order parameters (equation \eqref{equ:id_gop_1}),
see Figure \ref{fig:3phase_density_compare_gop}(b).
The results are qualitatively similar.


\begin{figure}
\centerline{
\includegraphics[width=3in]{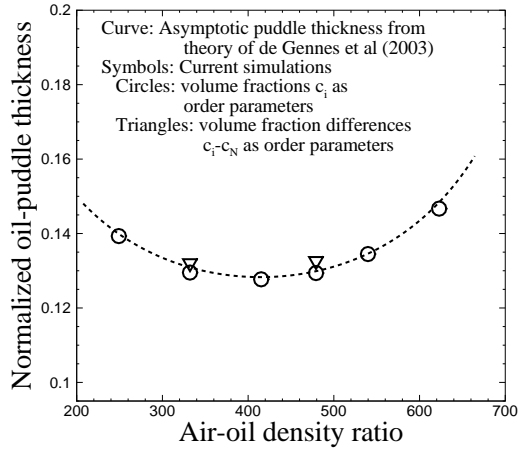}
}
\caption{
Comparison of the oil-puddle thickness as a function of the oil density
between current simulations and the Langmuir-de Gennes theory 
\cite{deGennesBQ2003}
for three fluid phases.
}
\label{fig:3phase_thickness_density}
\end{figure}

We have computed the thickness of the oil puddles
at equilibrium, defined as the distance along the vertical direction
between the upper and lower puddle surfaces, 
 corresponding to 
different oil densities.
The symbols in
Figure \ref{fig:3phase_thickness_density}
show the normalized oil-puddle thickness as
a function of the normalized oil density obtained
from the current simulations.
The results from both sets of order parameters
have been included in this figure, differentiated
using different symbols.
It is shown in \cite{Langmuir1933,deGennesBQ2003}
that for three fluid phases, 
when the gravity is dominant (i.e. oil forming puddles),
the puddle thickness can be explicitly expressed
in terms of the known physical parameters as
follows,
\begin{equation}
e_c = \sqrt{
  \frac{2\left(\sigma_{ao}+\sigma_{ow}-\sigma_{ow} \right)\rho_w}
       {(\rho_w - \rho_o)\rho_o g_r}
},
\label{equ:puddle_thickness}
\end{equation}
where $e_c$ is the asymptotic puddle
thickness when the gravity is dominant,
$g_r$ is the gravitational acceleration,
$\rho_w$ and $\rho_o$ are respectively
the water and oil densities,
and $\sigma_{ao}$, $\sigma_{aw}$
and $\sigma_{ow}$ are respectively the
air-oil, air-water, and oil-water
surface tensions.
For comparison, the dashed curve in
Figure \ref{fig:3phase_thickness_density} shows
the relation between $e_c$ and $\rho_o$
given by \eqref{equ:puddle_thickness}.
The results from the current simulations
agree with the results based
on the theory of \cite{Langmuir1933,deGennesBQ2003}
quite well.


\begin{figure}
\centering
\includegraphics[height=1.5in]{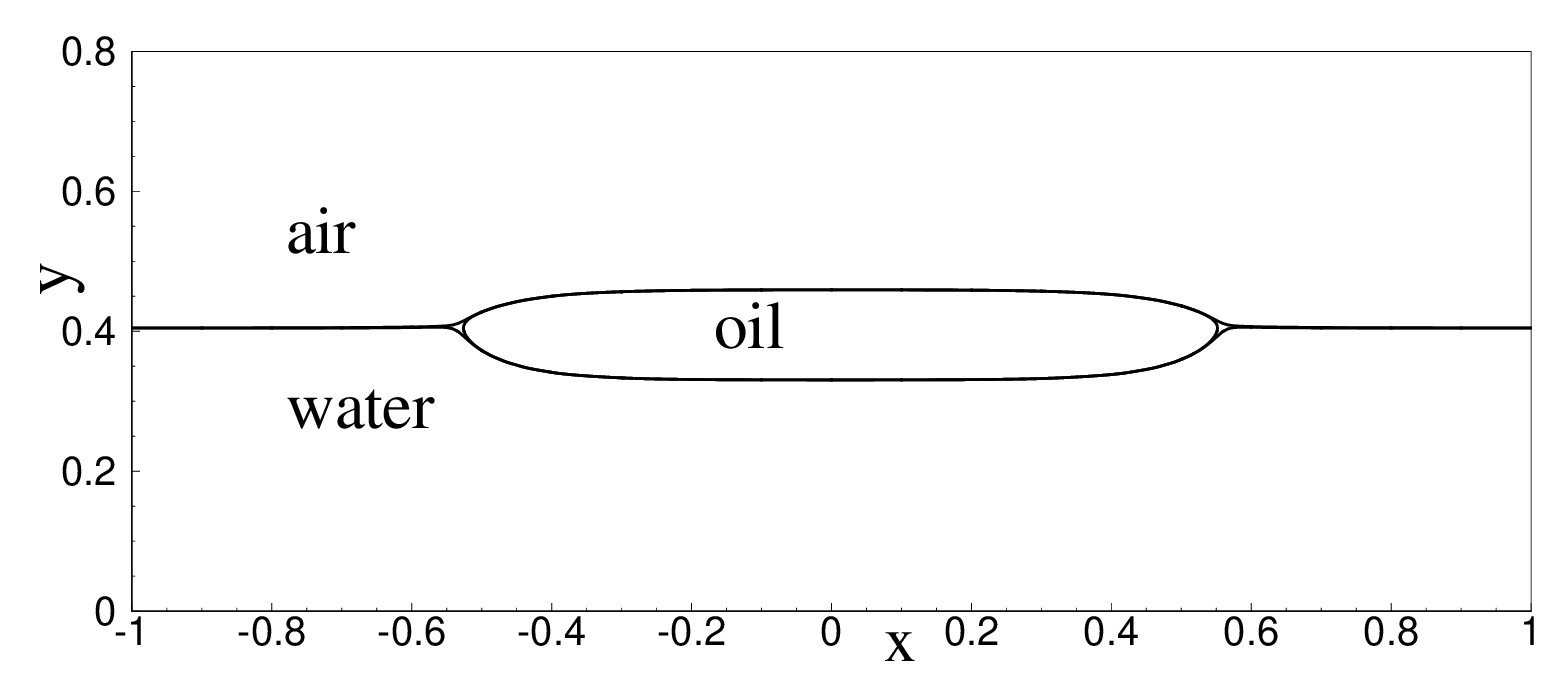}(a)
\includegraphics[height=1.5in]{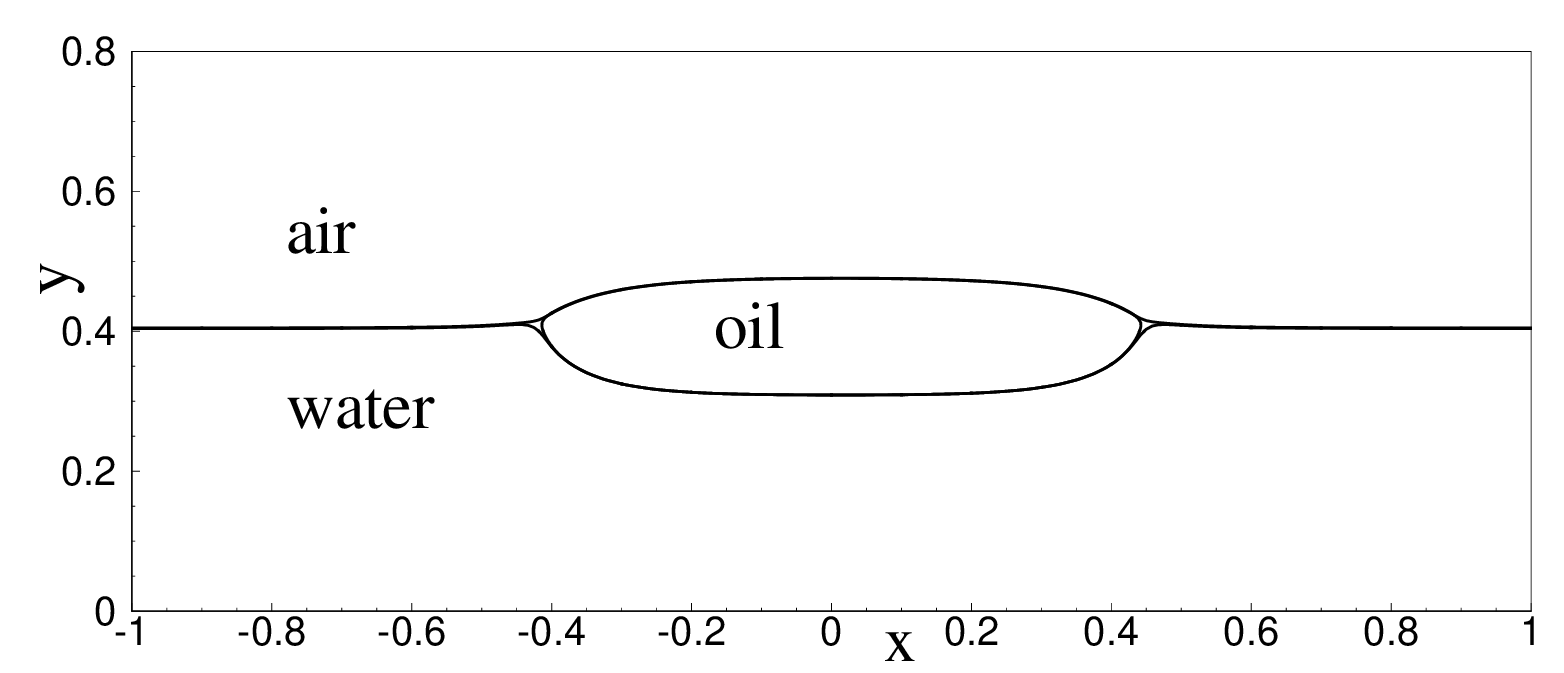}(b)
\caption{
Equilibrium configurations of an oil puddle on the air-water interface 
corresponding to air-oil surface tensions: 
(a) $0.065 kg/s^2$, (b) $0.085 kg/s^2$.
All other physical parameters are fixed.
Results are obtained with the volume 
fractions $c_i$ ($1\leqslant i\leqslant N-1$) 
as the order parameters.
}
\label{fig:3phase_ao_surften}
\end{figure}

We next look into the effect of the air-oil
surface tension on the equilibrium configuration
of this three-phase system.
In this group of tests we vary the air-oil
surface tension systematically between
$0.055kg/s^2$ and $0.095kg/s^2$,
while fixing all the other parameters
at those values in Table \ref{tab:3phase_param}
(oil density is $577kg/m^3$).
Figure \ref{fig:3phase_ao_surften}
shows the equilibrium configurations of the system
(contour lines $c_i=\frac{1}{2}$, $1\leqslant i\leqslant 3$)
corresponding to the air-oil surface tensions
$0.065kg/s^2$ and $0.085kg/s^2$.
They are obtained with the volume fractions
as the order parameters.
The oil forms a puddle, and the puddle size and
thickness have a clear dependence on
the value of the air-oil surface tension.
A larger air-oil surface tension 
leads to a smaller but thicker oil puddle at equilibrium.


\begin{figure}
\centerline{
\includegraphics[width=3in]{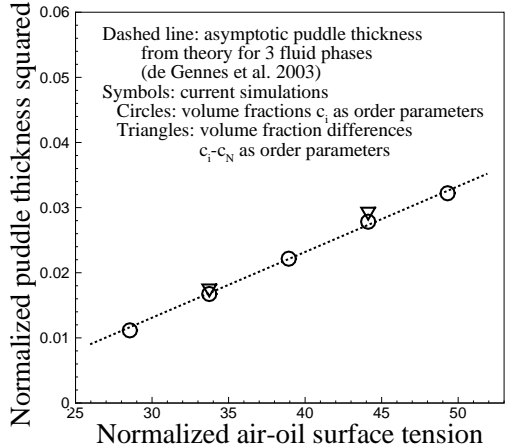}
}
\caption{
Comparison between current simulations and the
Langmuir-de Gennes theory \cite{deGennesBQ2003} for 
air-water-oil three fluid phases:
Oil-puddle thickness squared as a function of the
air-oil surface tension.
Circles denote results obtained with the volume fractions
as order parameters.
Triangles denote results obtained with
the re-scaled volume-fraction differences as
order parameters.
}
\label{fig:3phase_thickness_ao_surften}
\end{figure}

The quantitative relationship between
the oil-puddle thickness and the air-oil
surface tension is demonstrated 
by Figure \ref{fig:3phase_thickness_ao_surften}.
Here we plot the puddle-thickness {\em squared}
as a function of the air-oil surface tension.
The symbols represent results from
the simulations. The circles denote the results
 obtained using
the volume fractions as the order parameters
(equation \eqref{equ:order_param_volfrac}),
and the triangles denote those obtained
using the re-scaled volume-fraction differences
(equation \eqref{equ:id_gop_1}) as
the order parameters.
For comparison, the theoretical relationship
between these two quantities,
see equation \eqref{equ:puddle_thickness},
due to \cite{Langmuir1933,deGennesBQ2003}
is also shown in this figure, 
by the dashed line.
It can be observed that the data from current simulations
are in good agreement with the theory.


\vspace{0.15in}
To summarize, the floating liquid lens problem studied
in this section involves multiple fluid phases,
multiple pairwise surface tensions,
gravity, large density ratios and viscosity ratios.
The quantitative comparisons between current simulations
and the theory of Langmuir and de Gennes \cite{Langmuir1933,deGennesBQ2003}
show that, the physical formulations and the numerical
algorithm with general order
parameters developed in the current work produce 
physically accurate results.
The results of this section also 
demonstrate the significant effects
of the densities and the pairwise surface tensions
on the configurations of this three-phase
system.

\subsection{Dynamics of a Four-Phase Fluid Mixture}

In this section we look into the dynamics of a mixture
of four immiscible incompressible fluids. 
The goal is to demonstrate the capability 
of our method from Section \ref{sec:method}
for simulating 
dynamical problems.


\begin{figure}
\centerline{
\includegraphics[width=1.35in]{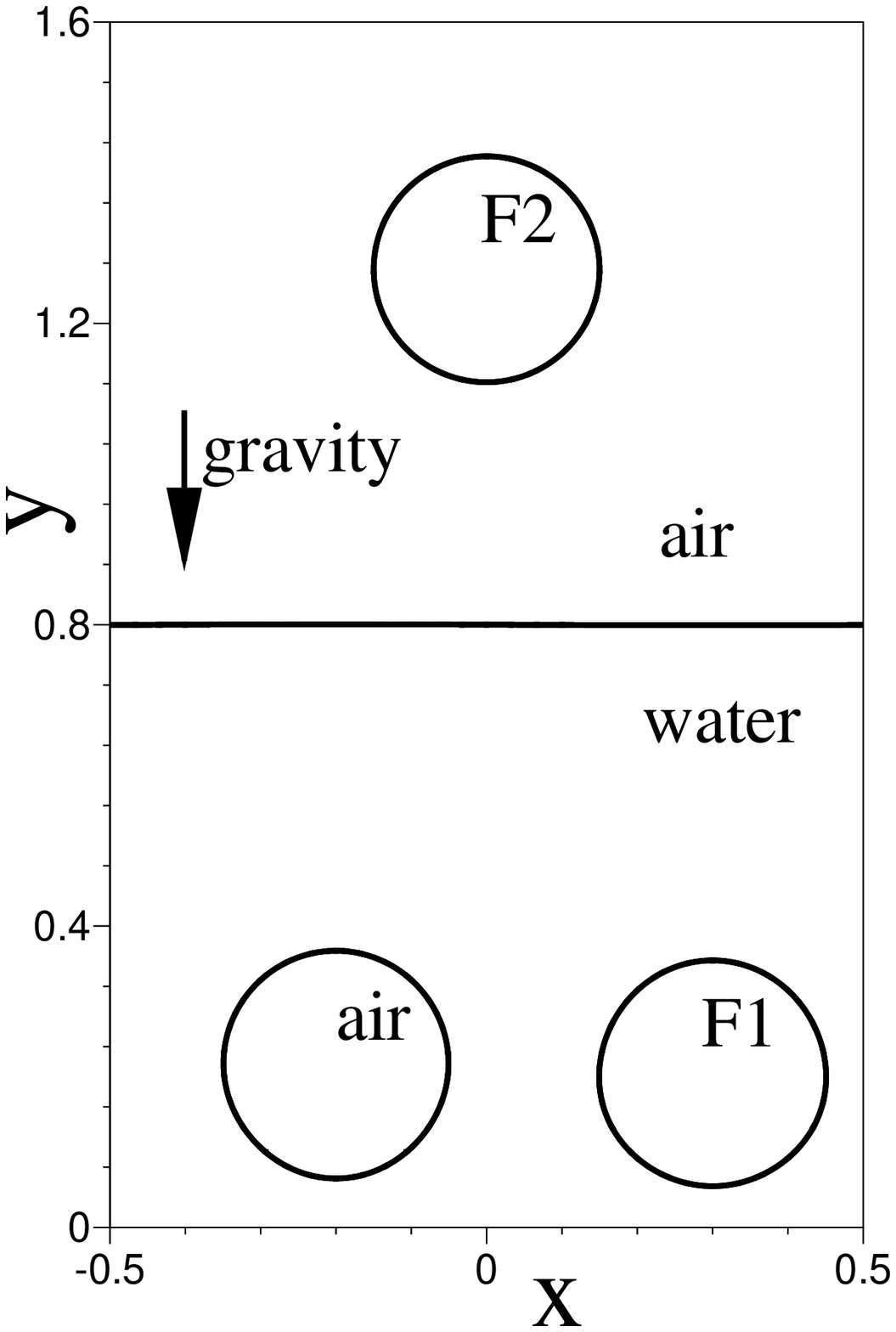}(a)
\includegraphics[width=1.35in]{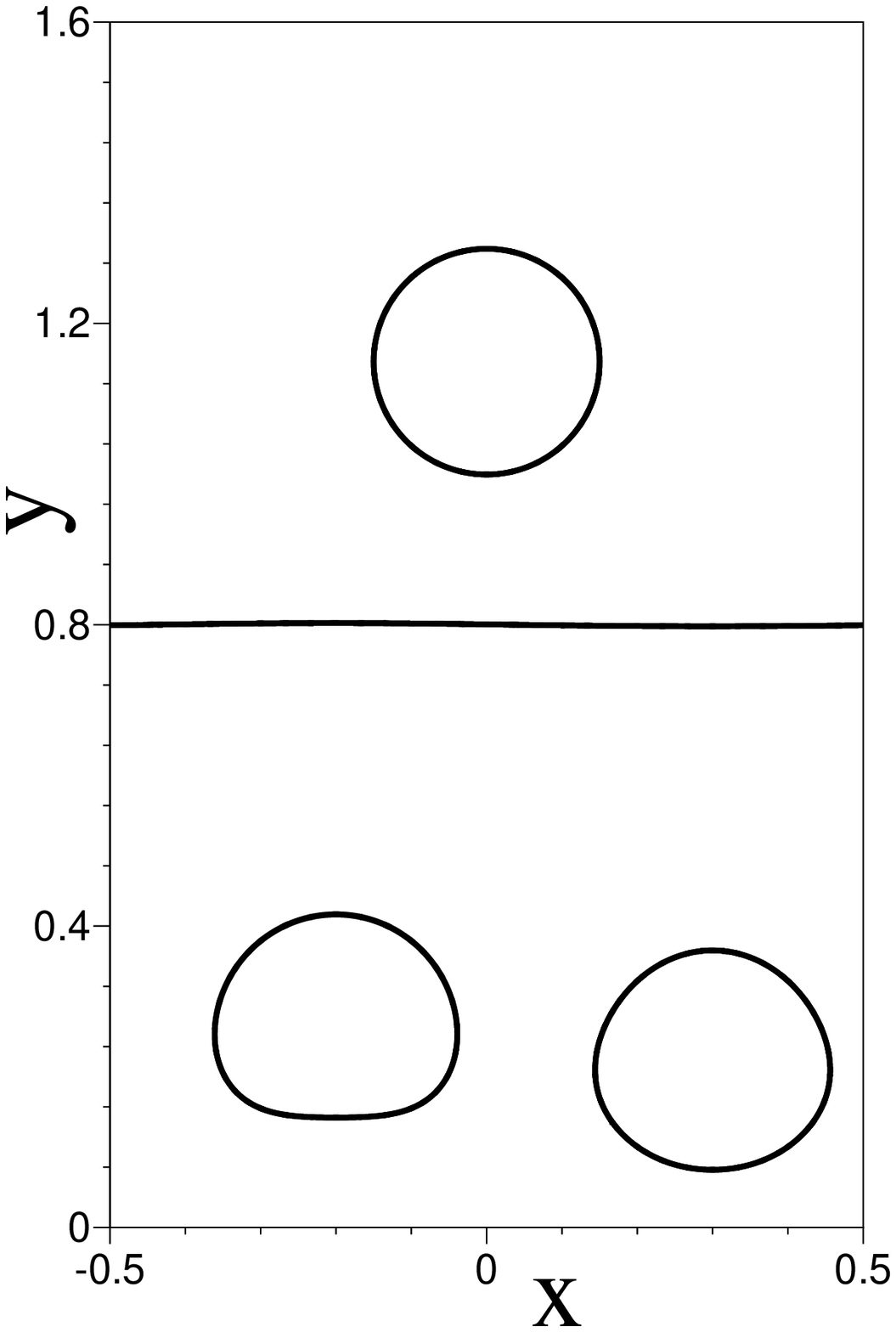}(b)
\includegraphics[width=1.35in]{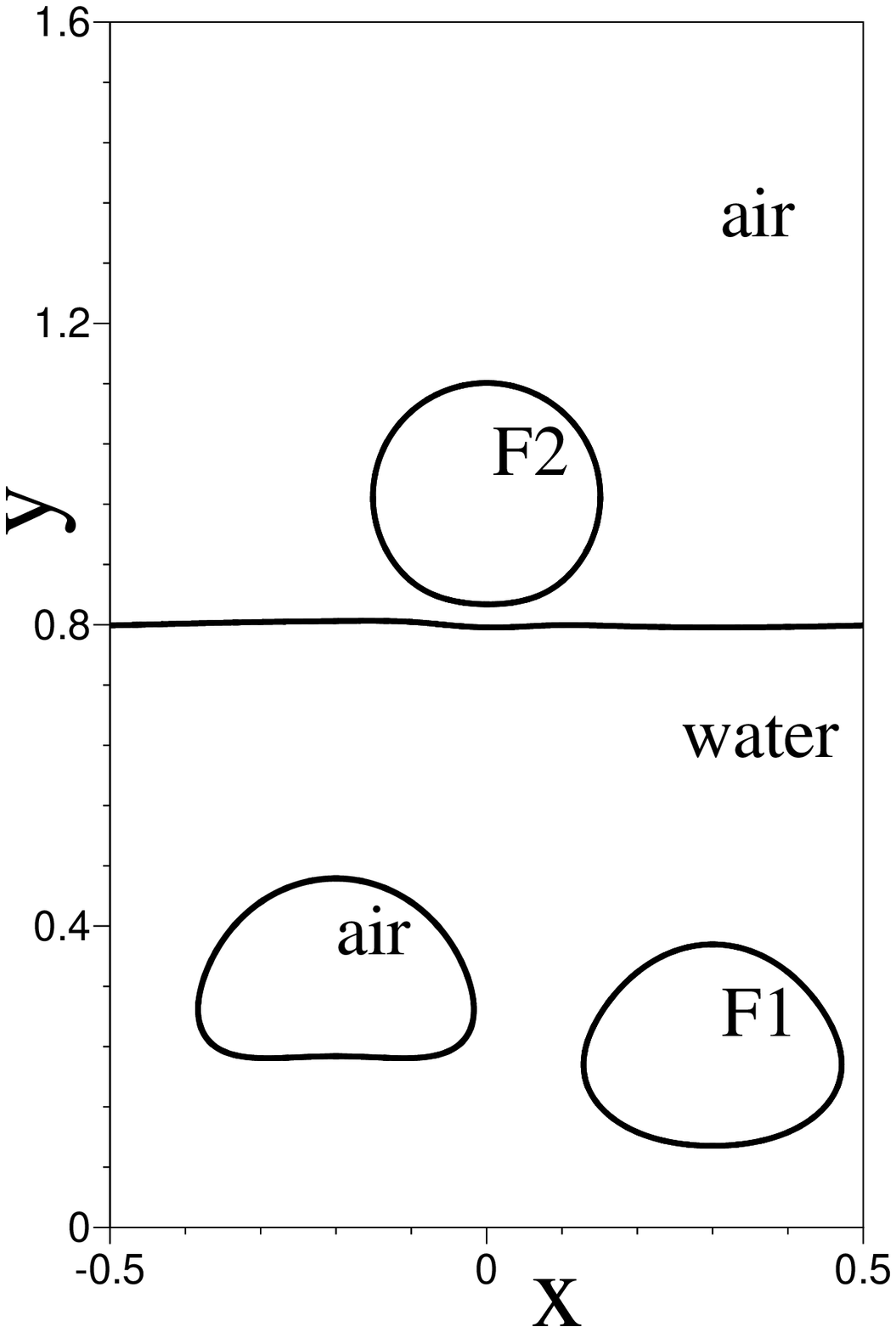}(c)
\includegraphics[width=1.35in]{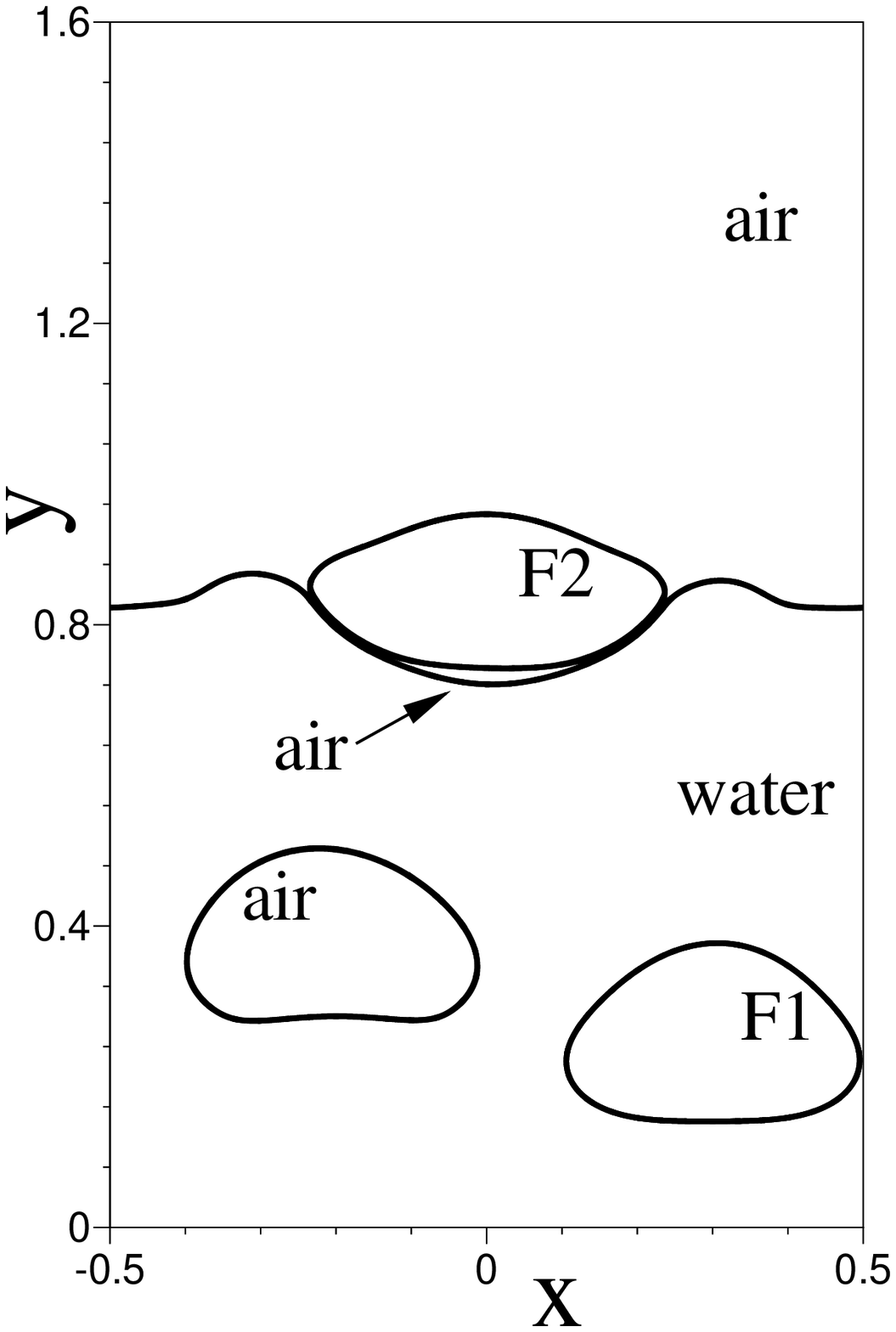}(d)
}
\centerline{
\includegraphics[width=1.35in]{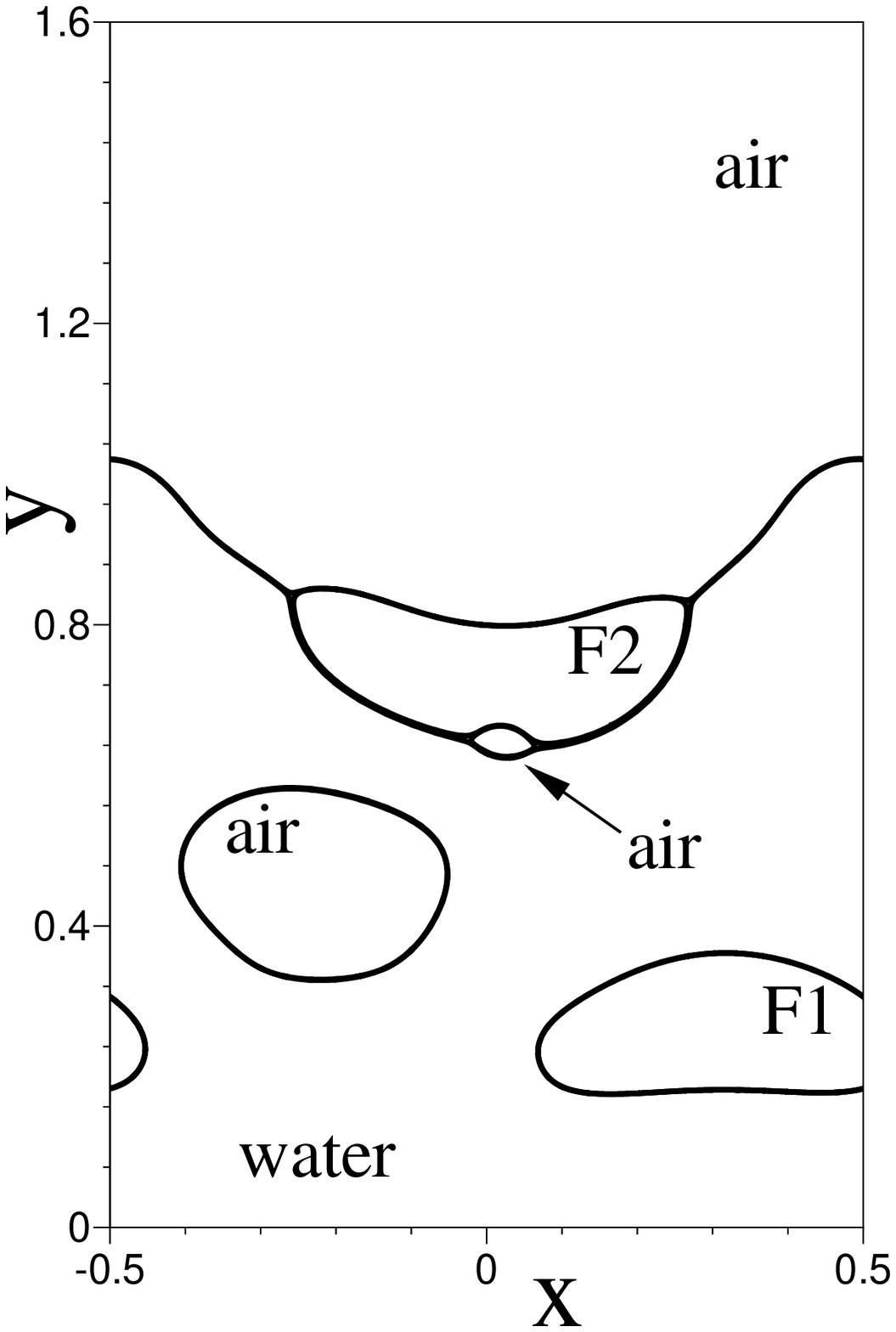}(e)
\includegraphics[width=1.35in]{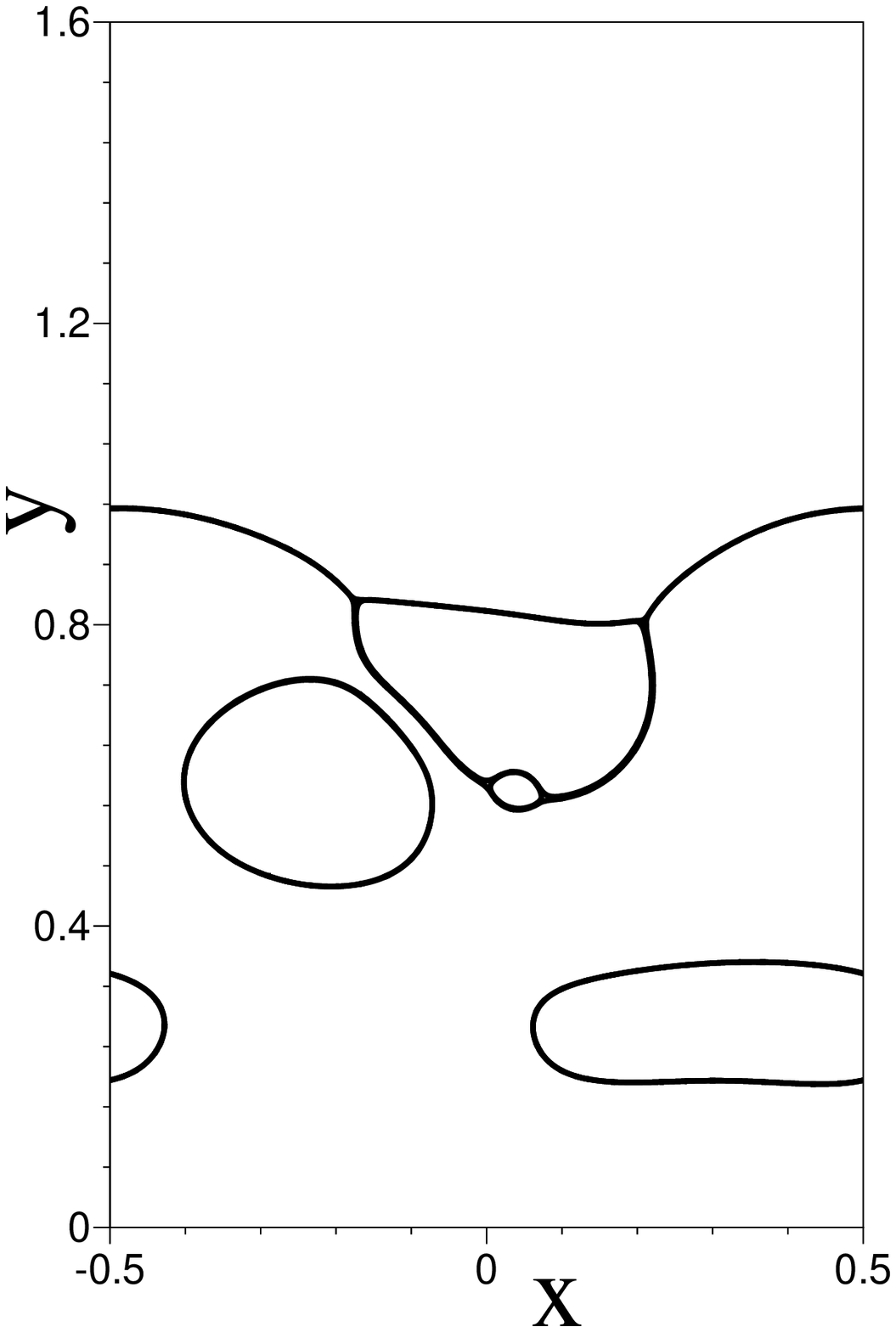}(f)
\includegraphics[width=1.35in]{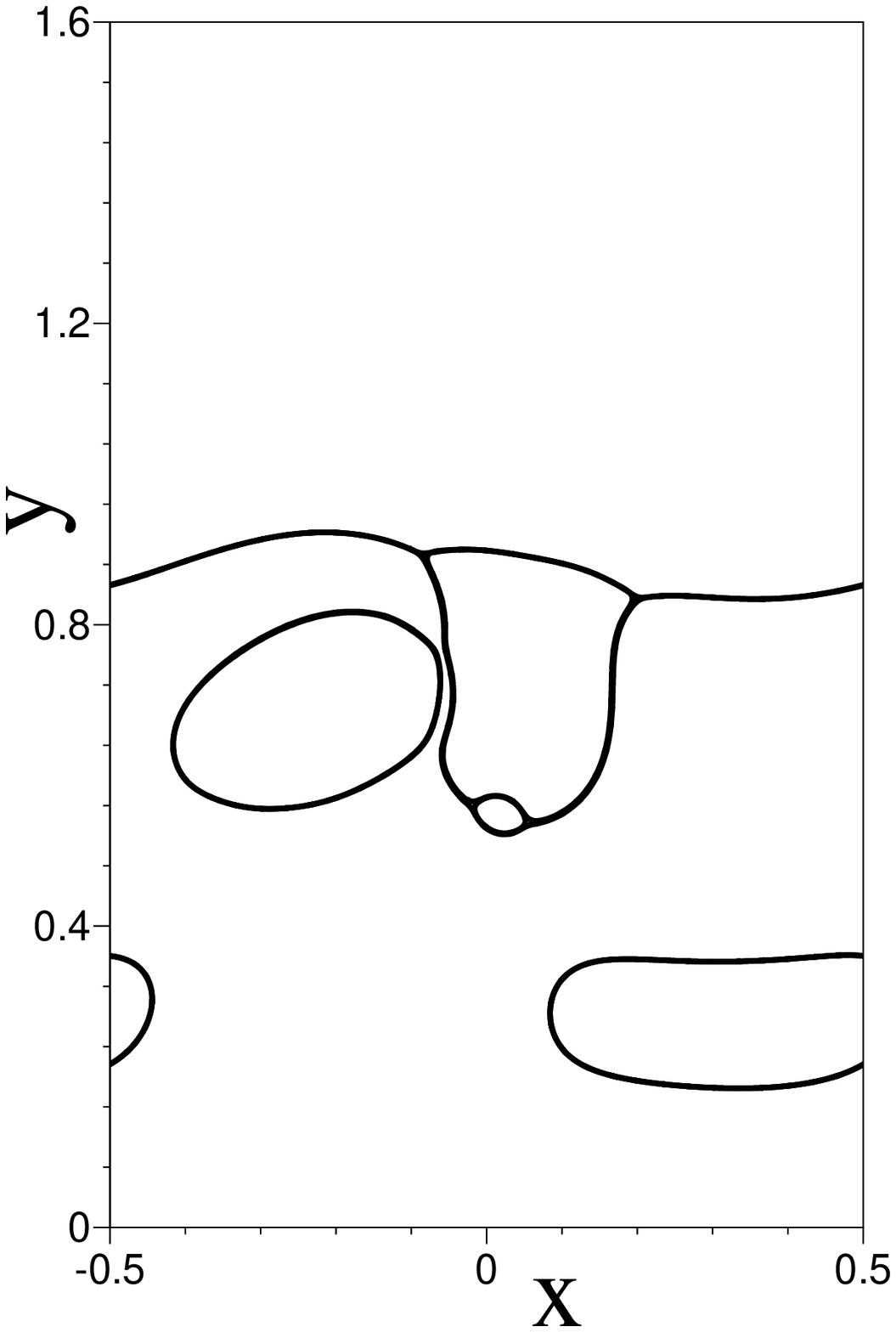}(g)
\includegraphics[width=1.35in]{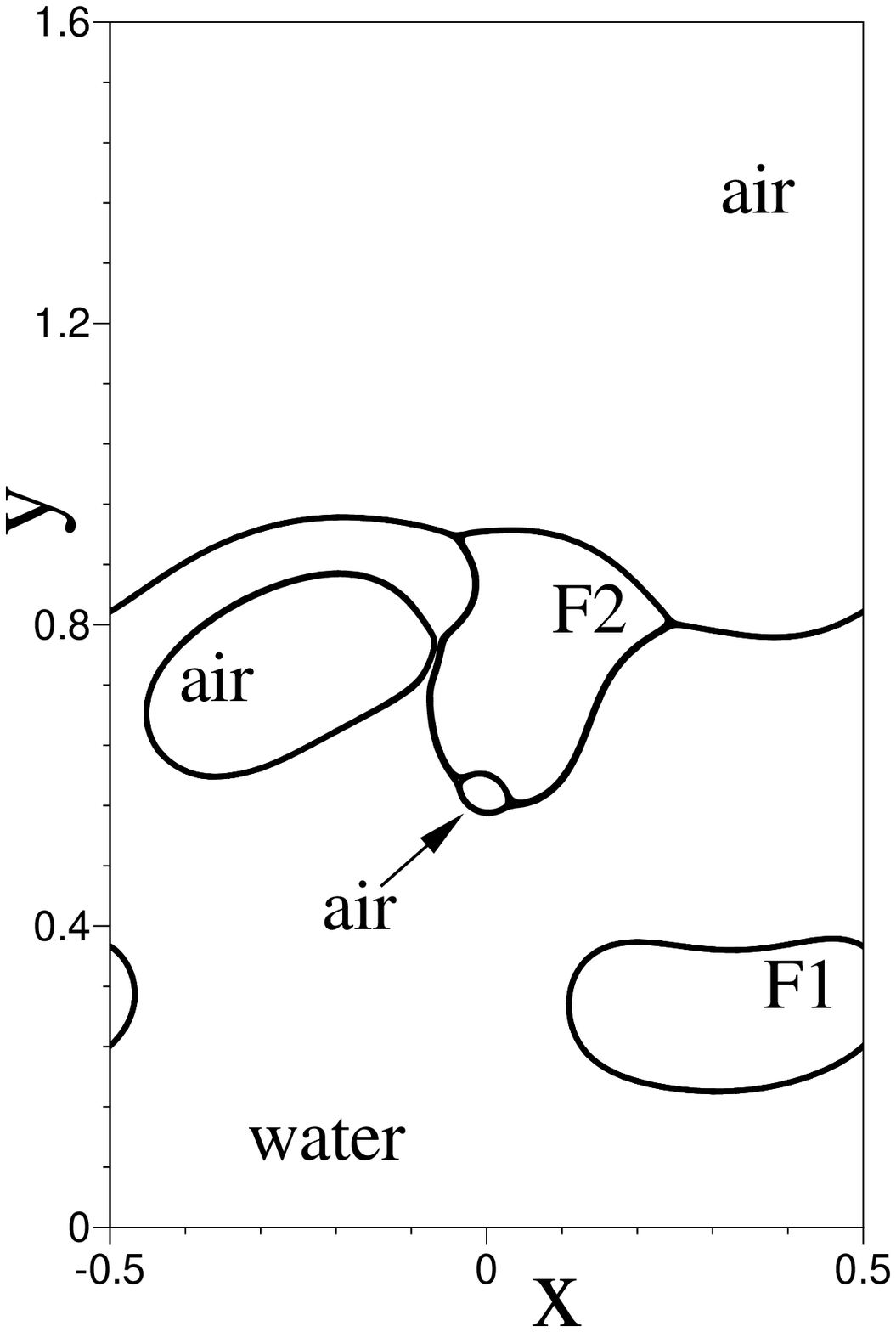}(h)
}
\centerline{
\includegraphics[width=1.35in]{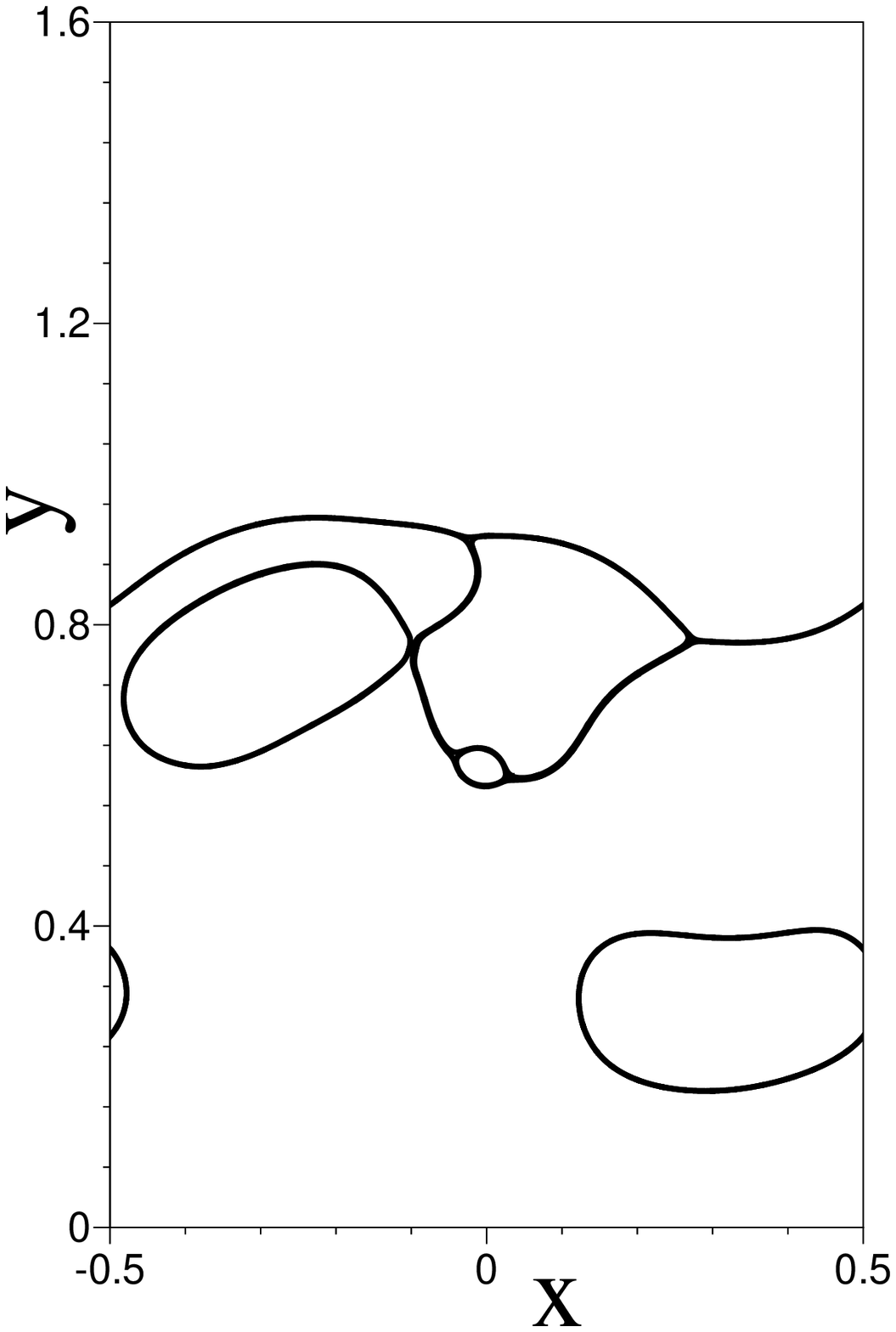}(i)
\includegraphics[width=1.35in]{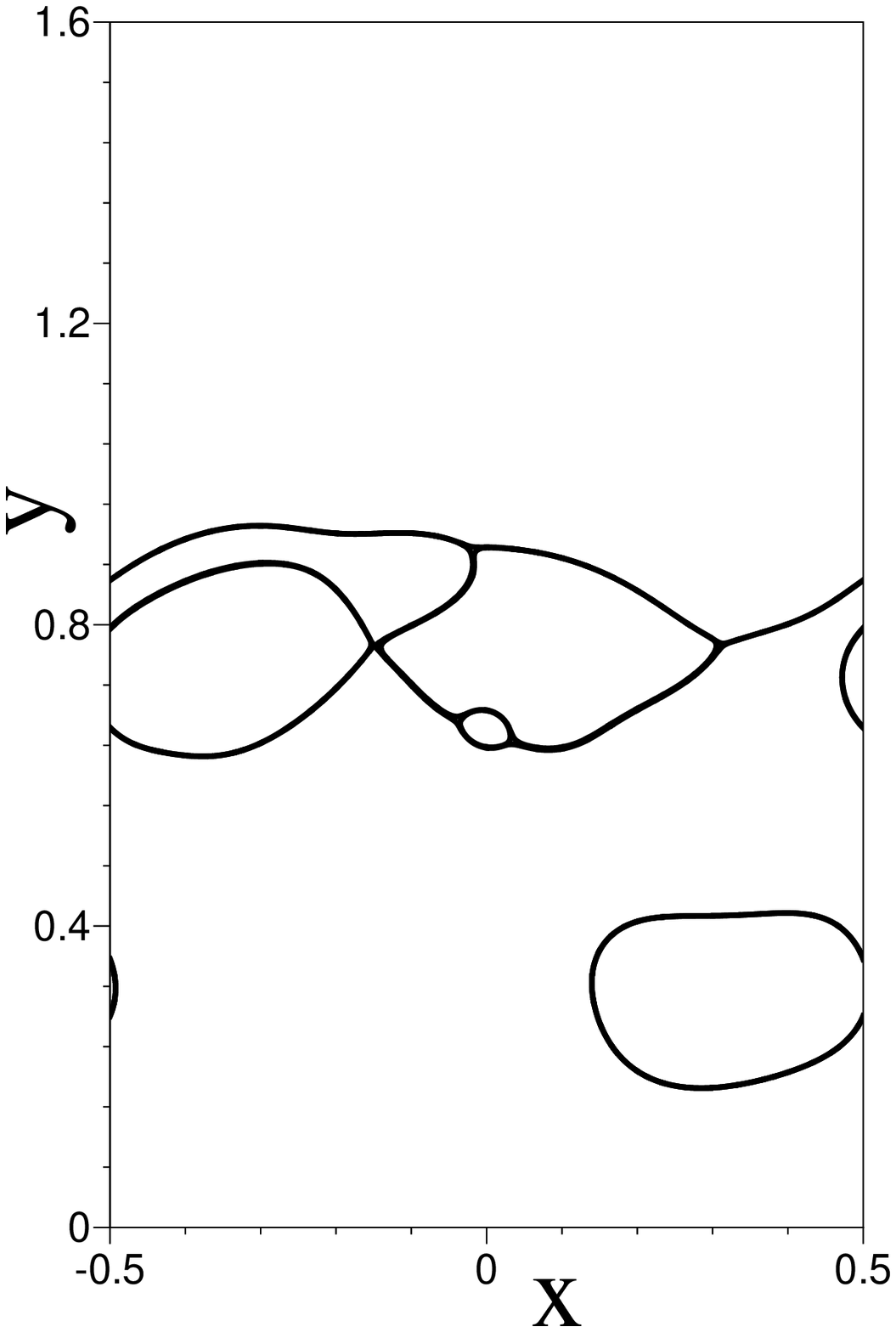}(j)
\includegraphics[width=1.35in]{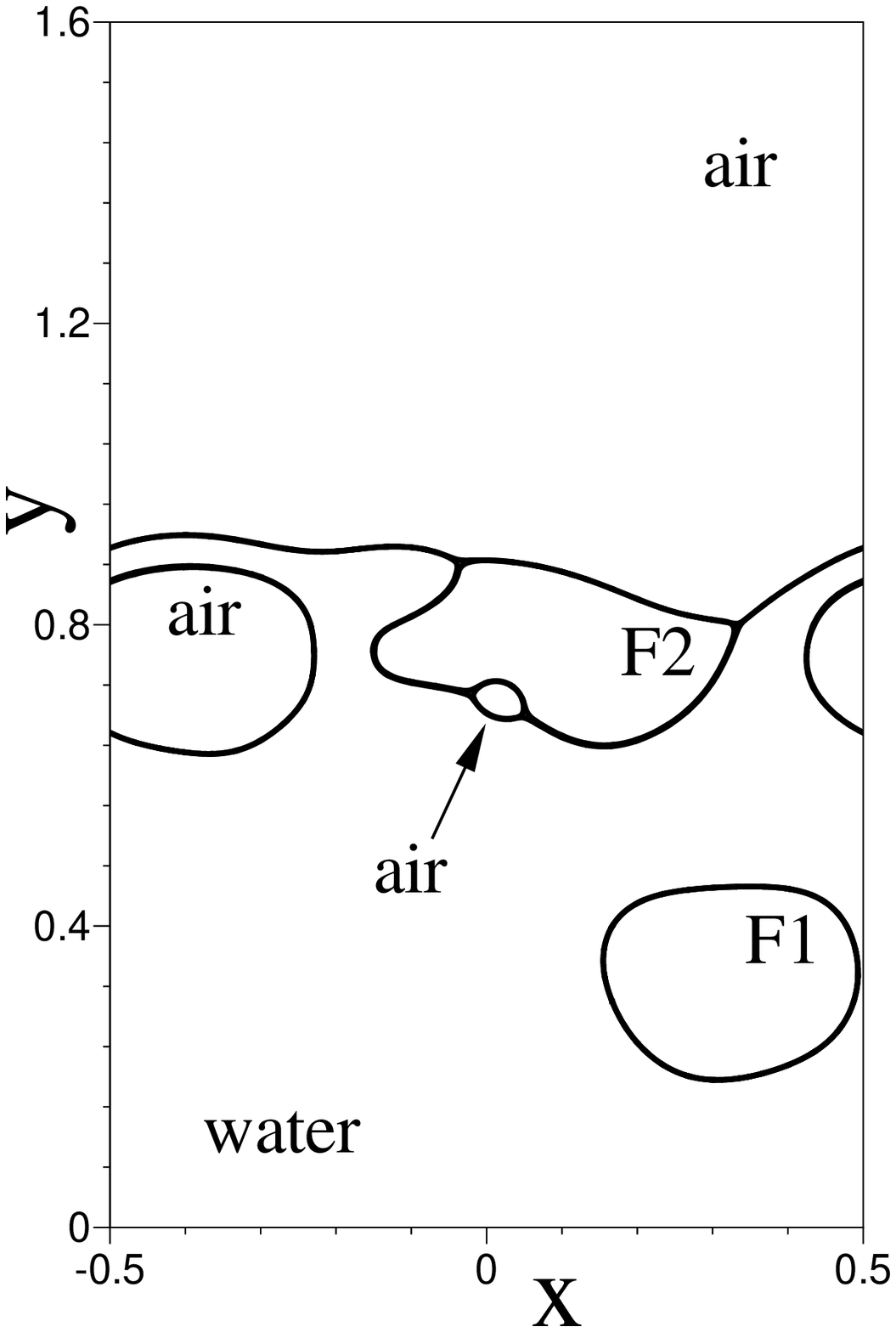}(k)
\includegraphics[width=1.35in]{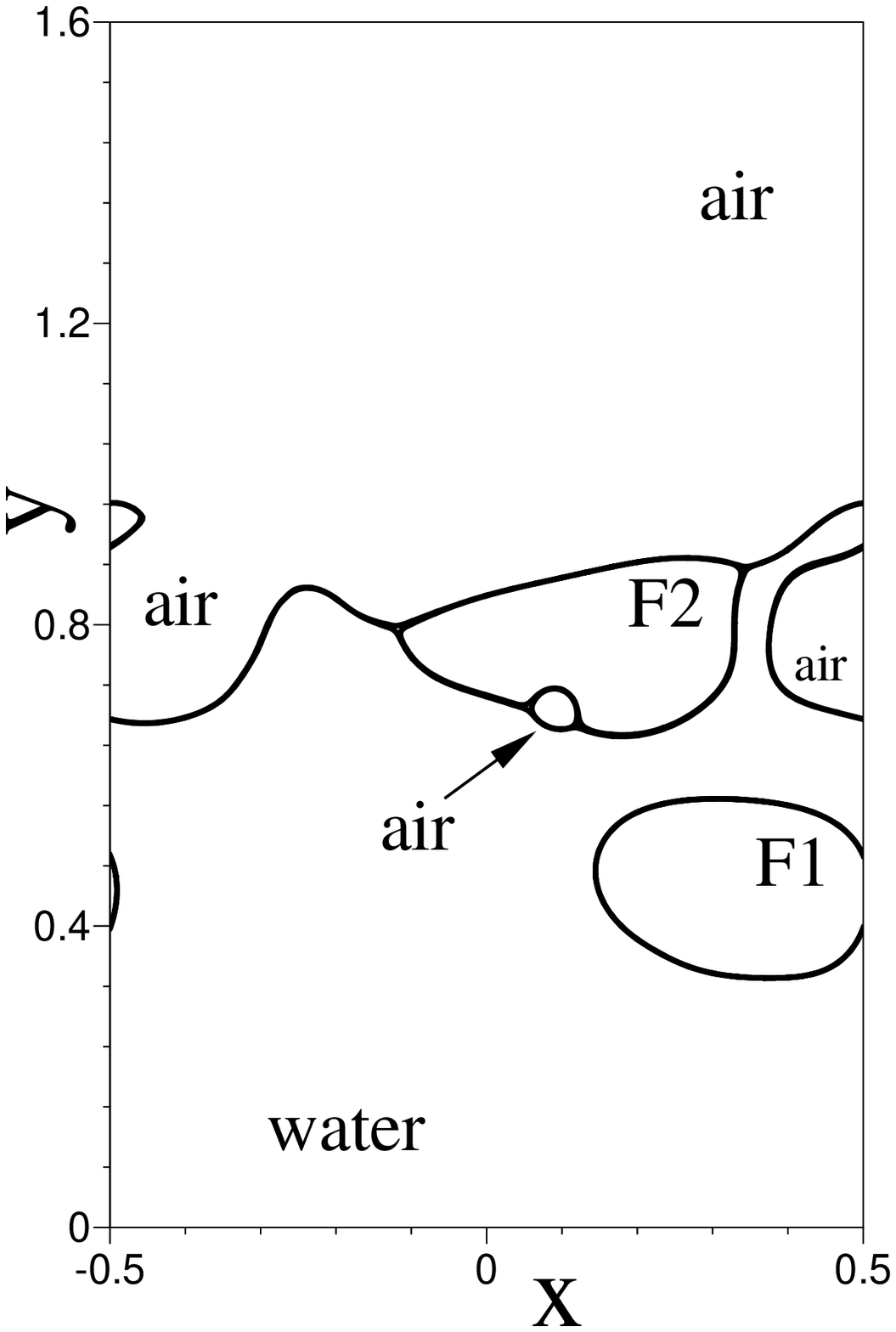}(l)
}
\centerline{
\includegraphics[width=1.35in]{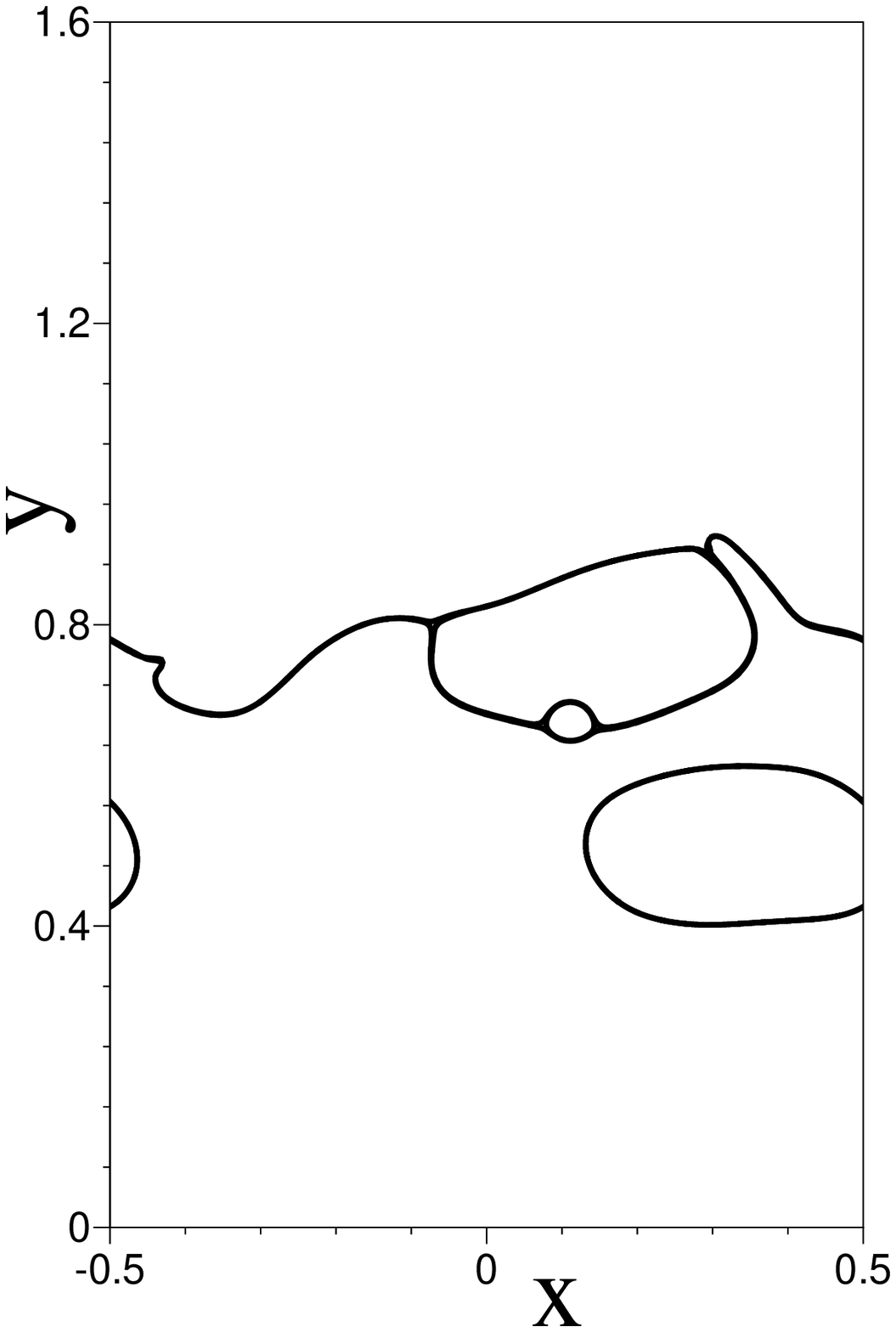}(m)
\includegraphics[width=1.35in]{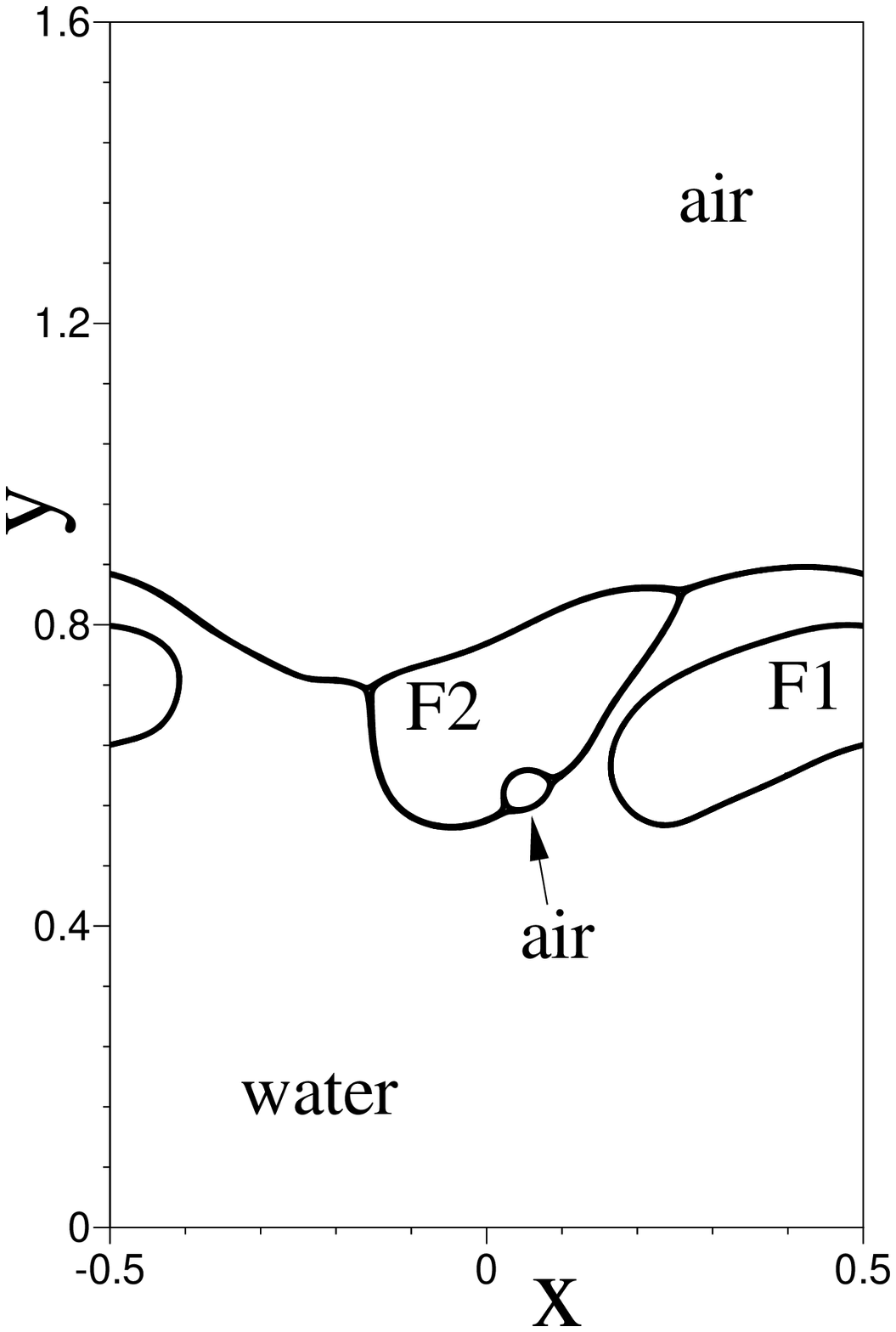}(n)
\includegraphics[width=1.35in]{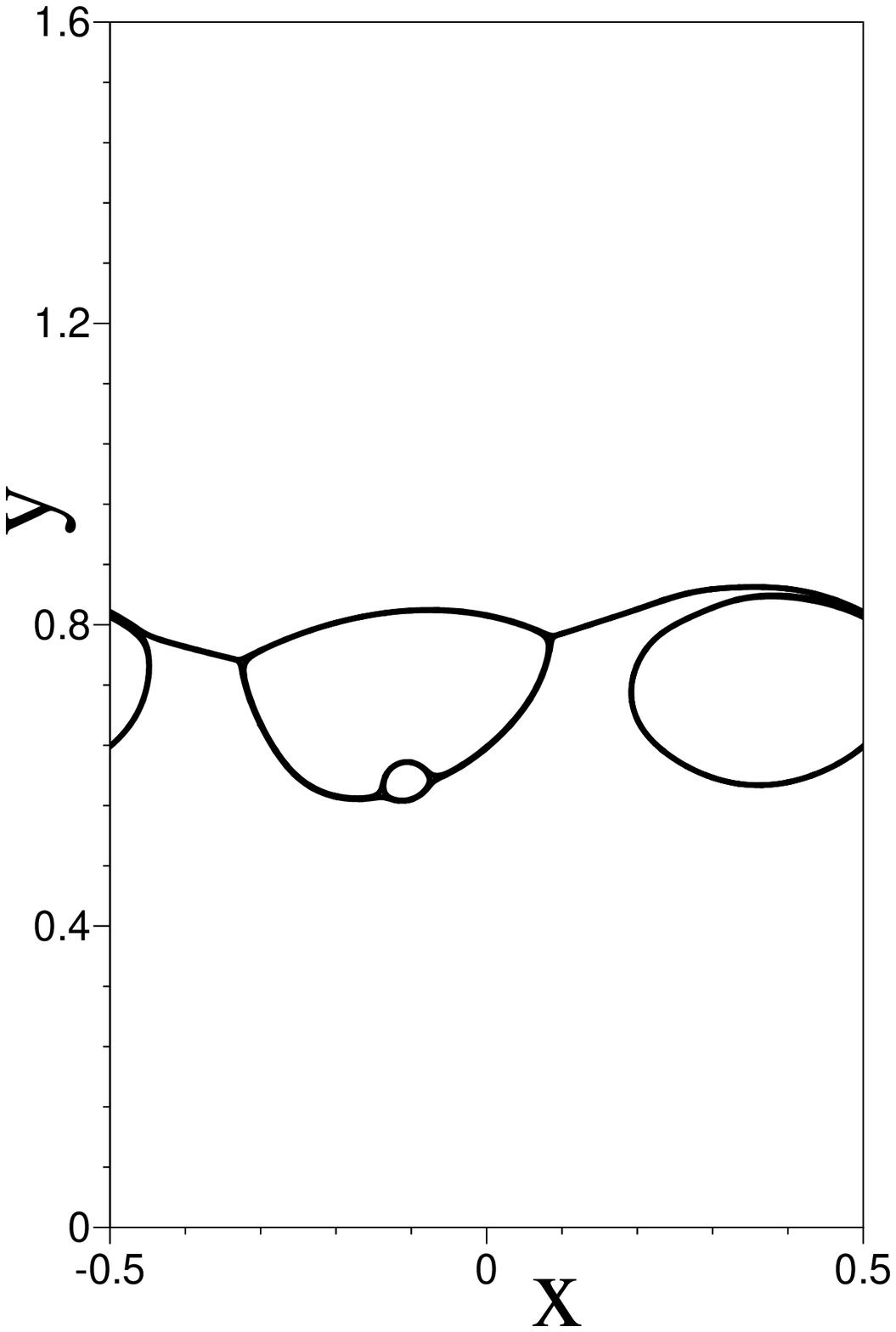}(o)
\includegraphics[width=1.35in]{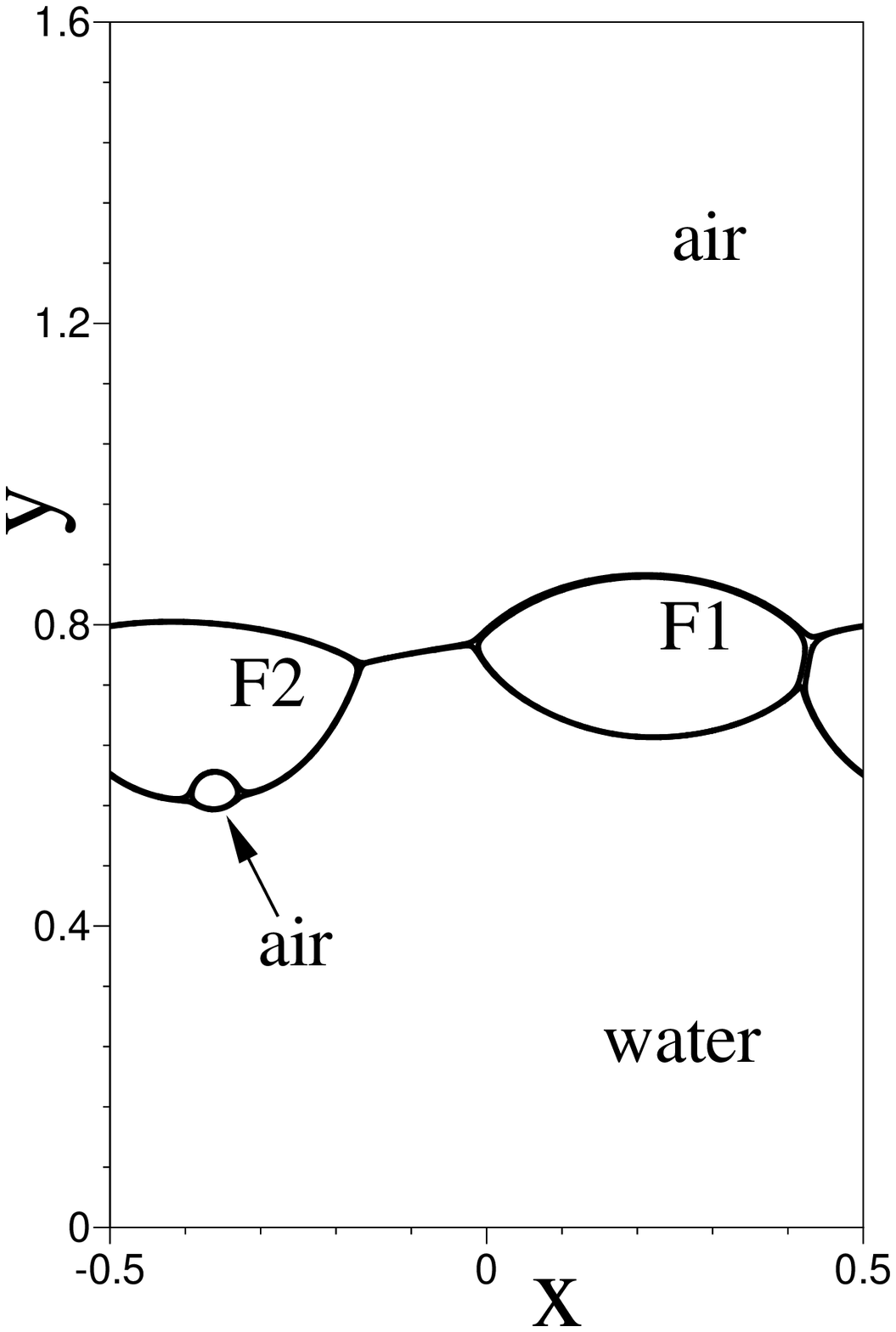}(p)
}
\caption{
Volume fraction contours ($c_i=\frac{1}{2}$) showing
fluid drops impacting water surface (4 fluid phases):
(a) $t=0.076$,
(b) $t=0.176$,
(c) $t=0.26$,
(d) $t=0.332$,
(e) $t=0.46$,
(f) $t=0.608$,
(g) $t=0.712$,
(h) $t=0.796$,
(i) $t=0.836$,
(j) $t=0.888$,
(k) $t=0.944$,
(l) $t=1.128$,
(m) $t=1.196$,
(n) $t=1.504$,
(o) $t=2.208$,
(p) $t=3.384$.
}
\label{fig:4p_phase}
\end{figure}


The setting of this problem is as follows.
Consider the rectangular flow domain
as shown
in Figure \ref{fig:4p_phase}(a),
$-\frac{L}{2}\leqslant x\leqslant \frac{L}{2}$
and $0\leqslant y\leqslant \frac{8}{5}L$, 
where $L=2cm$.
The top and bottom sides of the domain
are solid walls.  If any of the 
 fluid interfaces involved in this 
problem intersects with the walls,
 the contact angle
is assumed to be $90^0$.
In the horizontal direction
the domain is assumed to be periodic
at $x=\pm \frac{L}{2}$.
The gravity is assumed be along the $-y$ direction.
At $t=0$, the top half of the domain
is filled with air, and
the bottom half is filled with water.
A circular drop (diameter $0.3L$) of a fluid, denoted by ``F2'',
is suspended in the air at rest.
Simultaneously, an air bubble
and a drop of another fluid, denoted by ``F1'',
both  circular initially with diameter $0.3L$,
are trapped in the water.
%
The centers of the air bubble and the fluid drops
have the following coordinates
\begin{equation}
\left\{
\begin{split}
& 
\mathbf{x}_{F1} = \left(x_{F1}, y_{F1}  \right) = (0.3L, 0.2L),
\quad \text{(F1 drop)} \\
&
\mathbf{x}_{F2} = \left(x_{F2}, y_{F2}  \right) = (0, 1.3L),
\quad \text{(F2 drop)} \\
&
\mathbf{x}_{a} = \left(x_{a}, y_{a}  \right) = (-0.2L, 0.2L),
\quad \text{(air bubble)}. 
\end{split}
\right.
\end{equation}
The four types of fluids (air, water, F1, and F2)
are assumed to be incompressible and all immiscible
with one another, and it
 is assumed that there is no initial velocity.
The system is then released.
The liquid drops and the air bubble fall through the air
or rise through the water, and then impact the water surface.
The objective is to simulate the dynamics of this
process.


\begin{table}
\begin{center}
\begin{tabular*}{1.0\textwidth}{@{\extracolsep{\fill}}
l| l l| l l|  l l | l l}
\hline 
density[kg/m$^3$] & air & $1.2041$ & water & $998.207$ & F1 & $400$ & F2 & $870$ \\
\hline
dynamic viscosity[kg/(m$\cdot$s)] & air & $1.78E-5$ & water & $1.002E-3$ &
F1 & $0.02$ & F2 & $0.0915$ \\ \hline
surface tension[kg/s$^2$] & air/water & $0.0728$ & air/F1 & $0.06$ & air/F2 & $0.055$ \\ \cline{2-7}
 & water/F1 & $0.045$ & water/F2 & $0.044$ & F1/F2 & $0.048$  \\ \hline
gravity [m/s$^2$] & 9.8 \\
\hline
\end{tabular*}
\caption{
Physical parameter values for the air-water-F1-F2 four phase problem.
}
\label{tab:4p_param}
\end{center}
\end{table}


In Table \ref{tab:4p_param} we list the values of the physical
parameters involved in this problem, including 
the densities and dynamic viscosities of the
four fluids, and the six pairwise surface tensions
among them.
We assign the air, water, F2 and F1 fluids
respectively as the first, second, third and fourth
fluid.
%
We choose $L$ as the characteristic length scale
and $U_0=\sqrt{g_{r0}L}$ as the characteristic
velocity scale, where $g_{r0}=1m/s^2$.
The non-dimensionalization of the problem then
follows in a straightforward fashion
based on the constants given in Table \ref{tab:normalization}.
We set $g_i=0$ ($1\leqslant i\leqslant N-1$)
in \eqref{equ:CH} in the simulations.


In order to simulate the problem, we discretize
the domain using $1440$ quadrilateral elements of
equal sizes, with $30$ elements in the $x$ direction
and $48$ elements in the $y$ direction.
An element order of $12$ (with over-integration) 
has been used in the simulations
for all elements.
The algorithm developed in Section \ref{sec:method},
and the  formulation
with the volume fractions $c_i$ ($1\leqslant i\leqslant N-1$, where
$N=4$) as the order parameters as defined by
\eqref{equ:order_param_volfrac}, have been employed
to integrate this four-phase system in time.
%
For the boundary conditions
on the top and bottom walls, the  
condition \eqref{equ:vel_bc} with $\mathbf{w}=0$
has been imposed for the velocity,
and the conditions \eqref{equ:phi_bc_1} and \eqref{equ:phi_bc_2}
have been imposed for the phase field
functions $\phi_i$ ($1\leqslant i\leqslant 3$).
In the horizontal direction,
periodic conditions have been imposed
at $x=\pm \frac{L}{2}$ for all flow variables.
%
The initial velocity is set to zero.
The initial phase field functions are set according to
equation \eqref{equ:ic_gop_0} with 
$N=4$, where the initial volume fractions are
\begin{equation*}
\begin{split}
c_{10} = & \frac{1}{2}\left(
      1 + \tanh\frac{y - y_{w}}{\sqrt{2}\eta}
    \right)
     \left[
      1 - \frac{1}{2}\left(
            1 - \tanh\frac{\left|\mathbf{x} -\mathbf{x}_{F2}  \right| - R_0}{\sqrt{2}\eta}
          \right)    
    \right] 
    + \frac{1}{2}\left(
            1 - \tanh\frac{\left|\mathbf{x} -\mathbf{x}_{a}  \right| - R_0}{\sqrt{2}\eta}
          \right),
\end{split}
\end{equation*}
\begin{equation*}
c_{20} = \frac{1}{2}\left(
      1 - \tanh\frac{y - y_{w}}{\sqrt{2}\eta}
    \right)
     \left[
      1 - \frac{1}{2}\left(
            1 - \tanh\frac{\left|\mathbf{x} -\mathbf{x}_{a}  \right| - R_0}{\sqrt{2}\eta}
          \right)
    - \frac{1}{2}\left(
            1 - \tanh\frac{\left|\mathbf{x} -\mathbf{x}_{F1}  \right| - R_0}{\sqrt{2}\eta}
          \right)
    \right],
\end{equation*}
\begin{equation*}
c_{30} = \frac{1}{2}\left(
        1 - \tanh\frac{\left|\mathbf{x} -\mathbf{x}_{F2}  \right| - R_0}{\sqrt{2}\eta}
    \right),
\end{equation*}
\begin{equation*}
c_{40} = \frac{1}{2}\left(
        1 - \tanh\frac{\left|\mathbf{x} -\mathbf{x}_{F1}  \right| - R_0}{\sqrt{2}\eta}
    \right),
\end{equation*}
where $y_w = \frac{4}{5}L$ is the initial position of
the water surface in the $y$ direction,
and $R_0=0.15L$ is the initial radii of the air bubble
and fluid drops.


\begin{table}
\begin{center}
\begin{tabular*}{0.8\textwidth}{@{\extracolsep{\fill}}
l l }
\hline
parameters & values  \\
$\lambda_{ij}$ & computed based on \eqref{equ:matrix_lambda_ij_gop}
 \\
$\eta/L$ & $0.005$ 
 \\
$\beta$ & computed based on \eqref{equ:beta_expr}
\\
$m_i\tilde{\rho}_1U_0/L$, $1\leqslant i\leqslant N-1$ &
$2\times 10^{-8}/\lambda_{max}$, 
where $\lambda_{max} = \frac{ \max\{\lambda_{ij} \} }{\tilde{\rho}_1U_0^2L^2}$
\\ 
$\hat{s}_i$ & $2\eta^2\sqrt{\hat{\lambda}_i}$ 
\\
$\rho_0$ & $\min\left\{\tilde{\rho}_i  \right\}_{1\leqslant i\leqslant N}$
\\
$\nu_0$ & $5\max\left\{\frac{\tilde{\mu}_i}{\tilde{\rho}_i} \right\}_{1\leqslant i\leqslant N} $
\\
$U_0\Delta t/L$ & $2\times 10^{-6}$
\\
$J$ (temporal order) & $2$ \\
\hline
\end{tabular*}
\caption{
Simulation parameter values for the air-water-F1-F2 four phase problem.
$N=4$ in this table.
}
\label{tab:4p_simu_param}
\end{center}
\end{table}

Table \ref{tab:4p_simu_param} summarizes the values of all
the numerical parameters involved in the 
algorithm and the simulations.



We now look into the dynamics of this four-phase flow.
In Figure \ref{fig:4p_phase} we show a temporal
sequence of snapshots of the fluid interfaces
in the flow by plotting the contour lines
of the volume fractions $c_i=\frac{1}{2}$ 
($1\leqslant i\leqslant 4$).
From Figures \ref{fig:4p_phase}(a)--(c),
one can observe that upon release 
the F2 drop falls rapidly through the air due
to the gravity,
and is about to impact the water surface (Figure \ref{fig:4p_phase}c).
Its profile maintains essentially the original shape
during the falling process. 
Simultaneously, the air bubble and the F1 drop
rise through the water due to buoyancy, albeit much more slowly
compared to the falling F2 drop.
They experience significant deformations
in their shapes. The air bubble
has the  shape of a ``cap'',
with a flat underside (Figure \ref{fig:4p_phase}c).
%
Figure \ref{fig:4p_phase}(d) shows that
the F2 drop impacts the water surface, 
and generates a ripple that spreads outward.
The F2 drop traps a pocket of air at its
underside.
Later on, the F2 fluid forms a pool floating
on the water surface and the trapped air pocket forms
a small bubble (Figure \ref{fig:4p_phase}(e)).
Figures \ref{fig:4p_phase}(f) through
\ref{fig:4p_phase}(k) show the interaction
between the rising air bubble and 
the floating F2 pool on the water surface.
Notice that the F2 fluid is mostly
immersed in the water owing to the 
small density contrast between F2 and water
(see e.g. Figure \ref{fig:4p_phase}(g)).
As the air bubble rises, it approaches and ``kisses''
the F2 fluid immersed in the water 
(Figures \ref{fig:4p_phase}(g)--(j)),
and then pulls apart (Figure \ref{fig:4p_phase}(k)).
This interaction and the motion
of the air bubble has caused a dramatic
deformation in the profile of the F2
fluid (Figures \ref{fig:4p_phase}(i)--(k)).
One also observes that, during this period of time,
the upward motion of the F1 drop appears to
 have stalled (Figures \ref{fig:4p_phase}(f)--(j)),
but the drop exhibits significant deformations
in shape.
Figures \ref{fig:4p_phase}(k)--(m)
show that the air bubble touches 
the water surface and merges with the bulk of air
above the water.
As time goes on, the F1 drop rises
slowly through the water and
 forms a floating F1 drop on the
water surface (Figures \ref{fig:4p_phase}(n)--(p)).
Eventually, the water surface becomes mostly covered
by the floating F1 and F2 drops,
and the F2 drop has a small air bubble
trapped at its underside (Figure \ref{fig:4p_phase}(p)).


\begin{figure}
\centerline{
\includegraphics[width=1.4in]{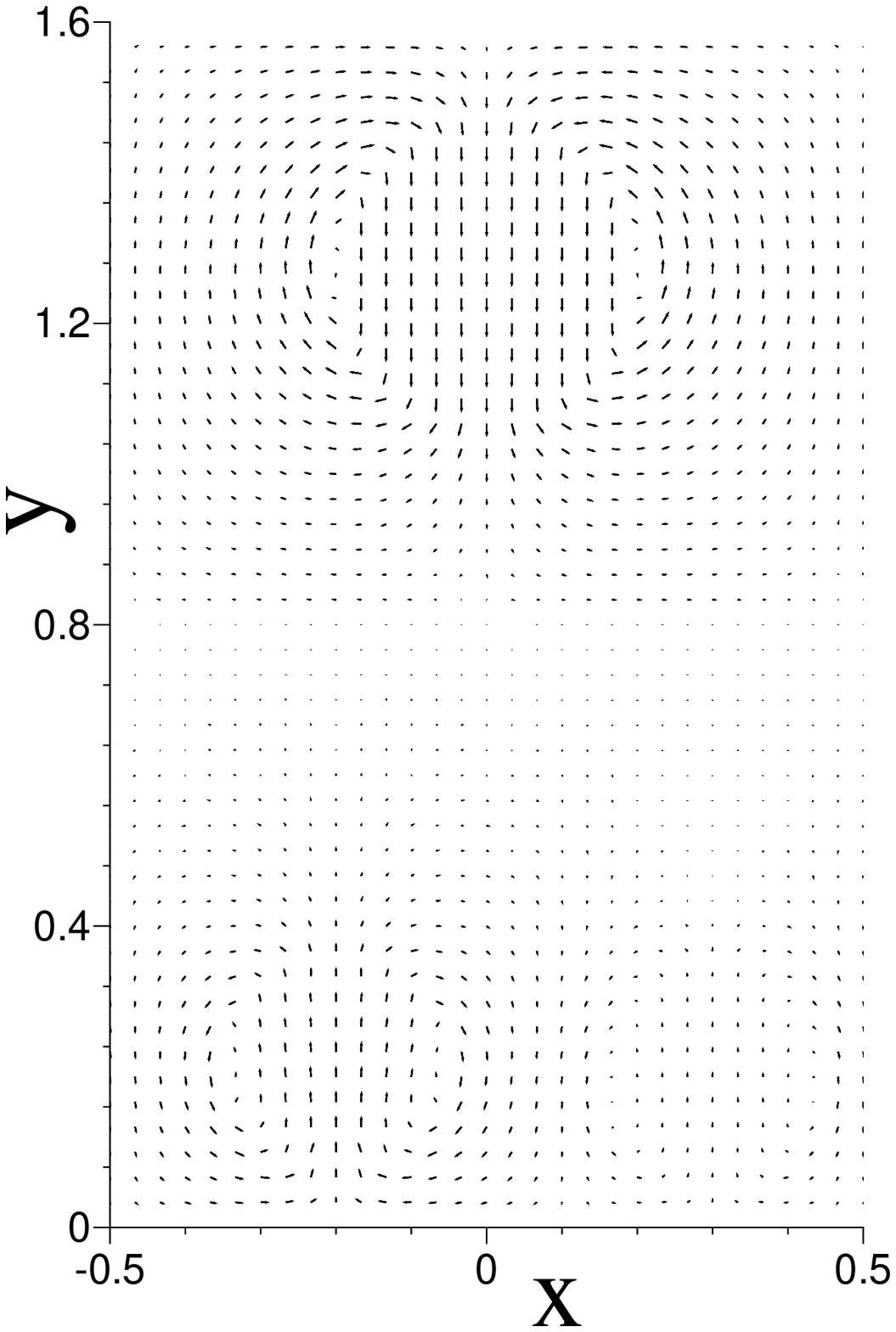}(a)
\includegraphics[width=1.4in]{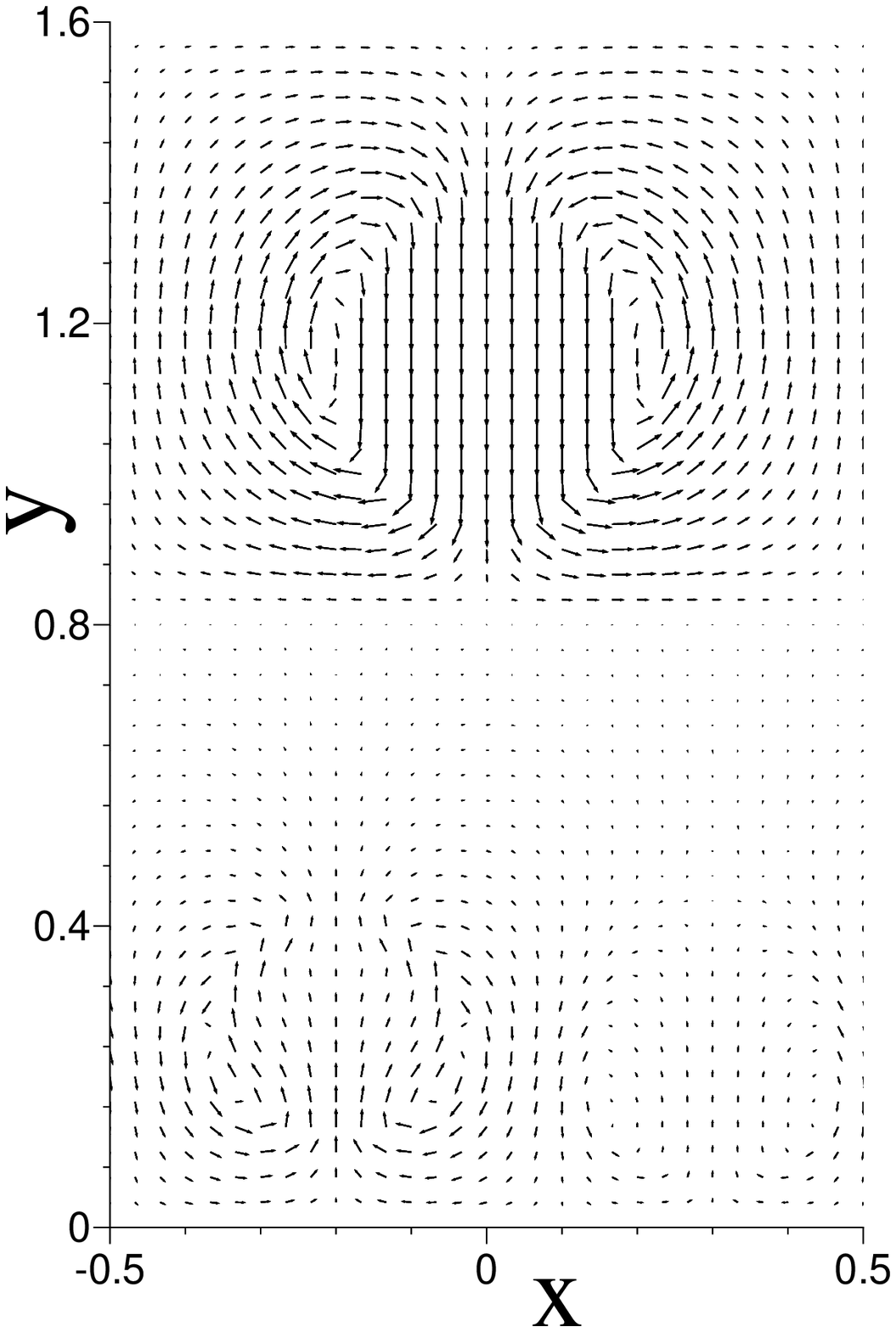}(b)
\includegraphics[width=1.4in]{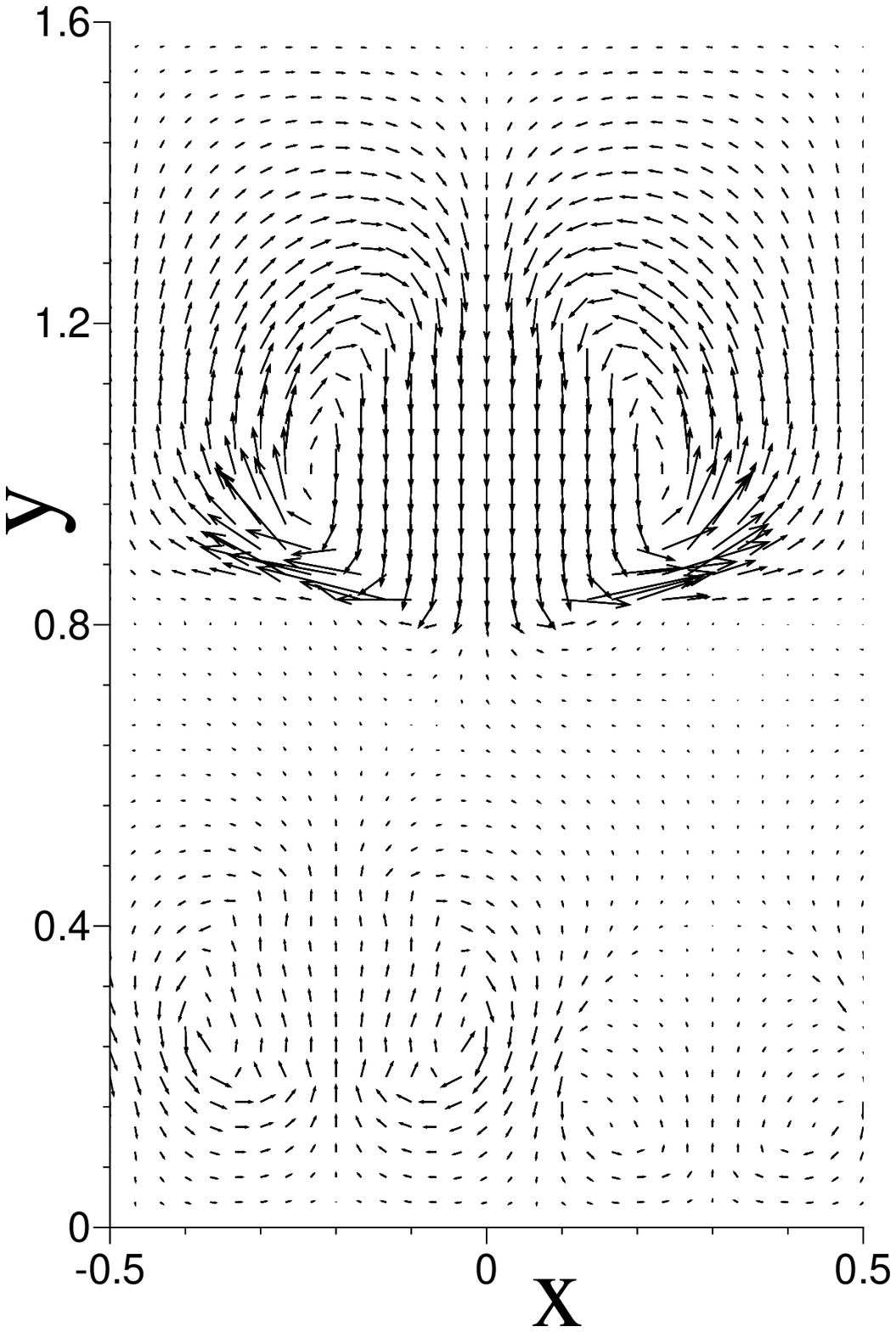}(c)
\includegraphics[width=1.4in]{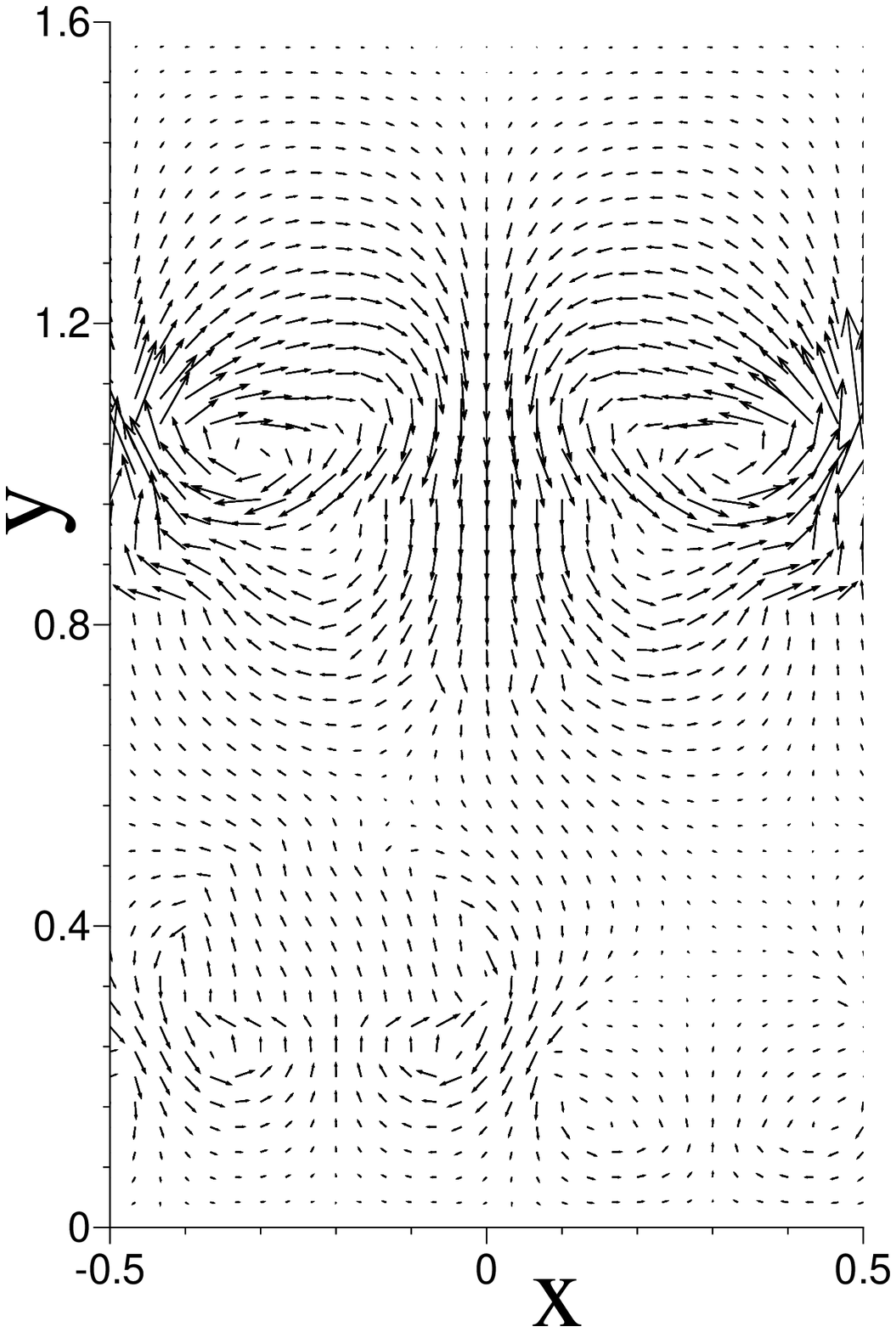}(d)
}
\centerline{
\includegraphics[width=1.4in]{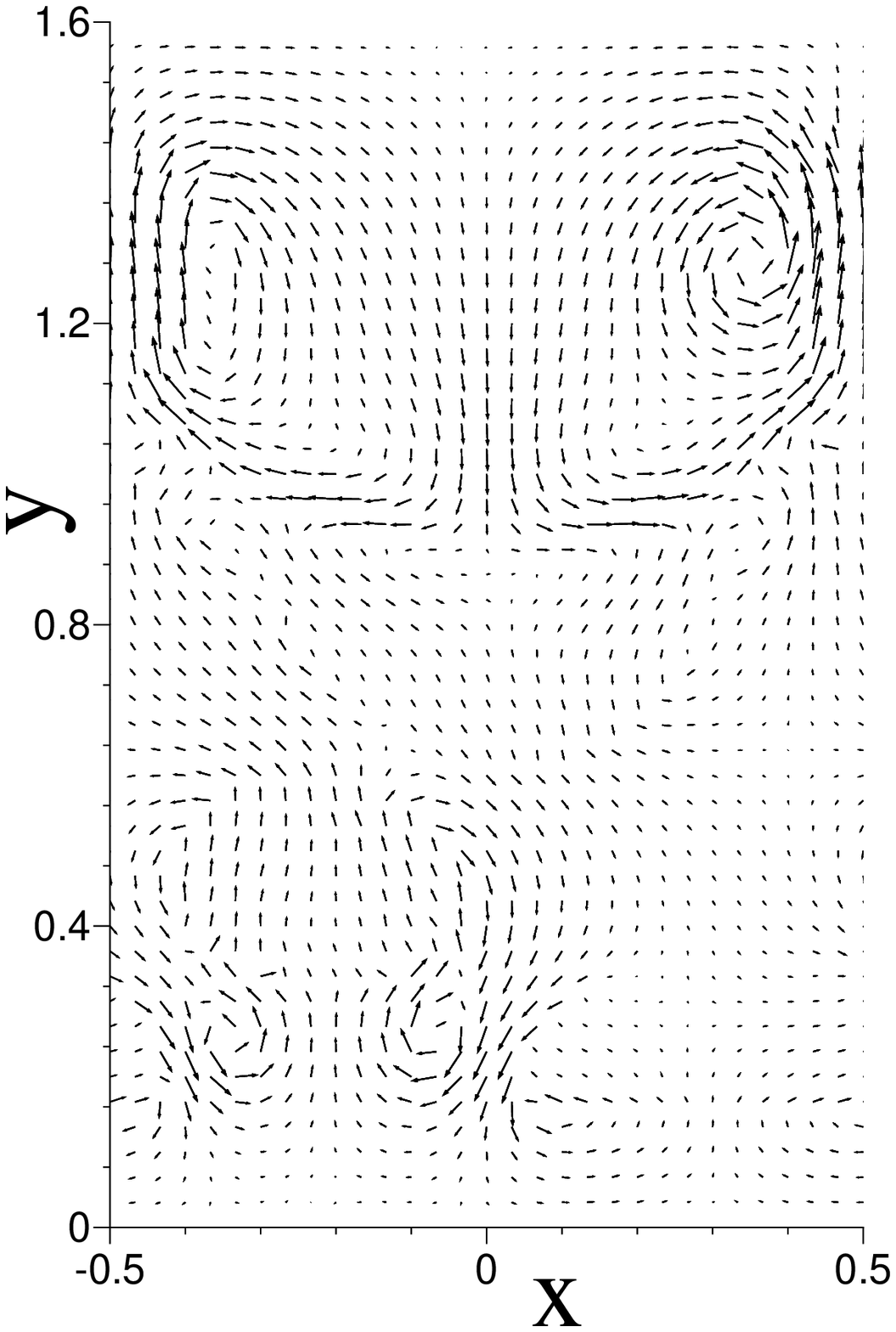}(e)
\includegraphics[width=1.4in]{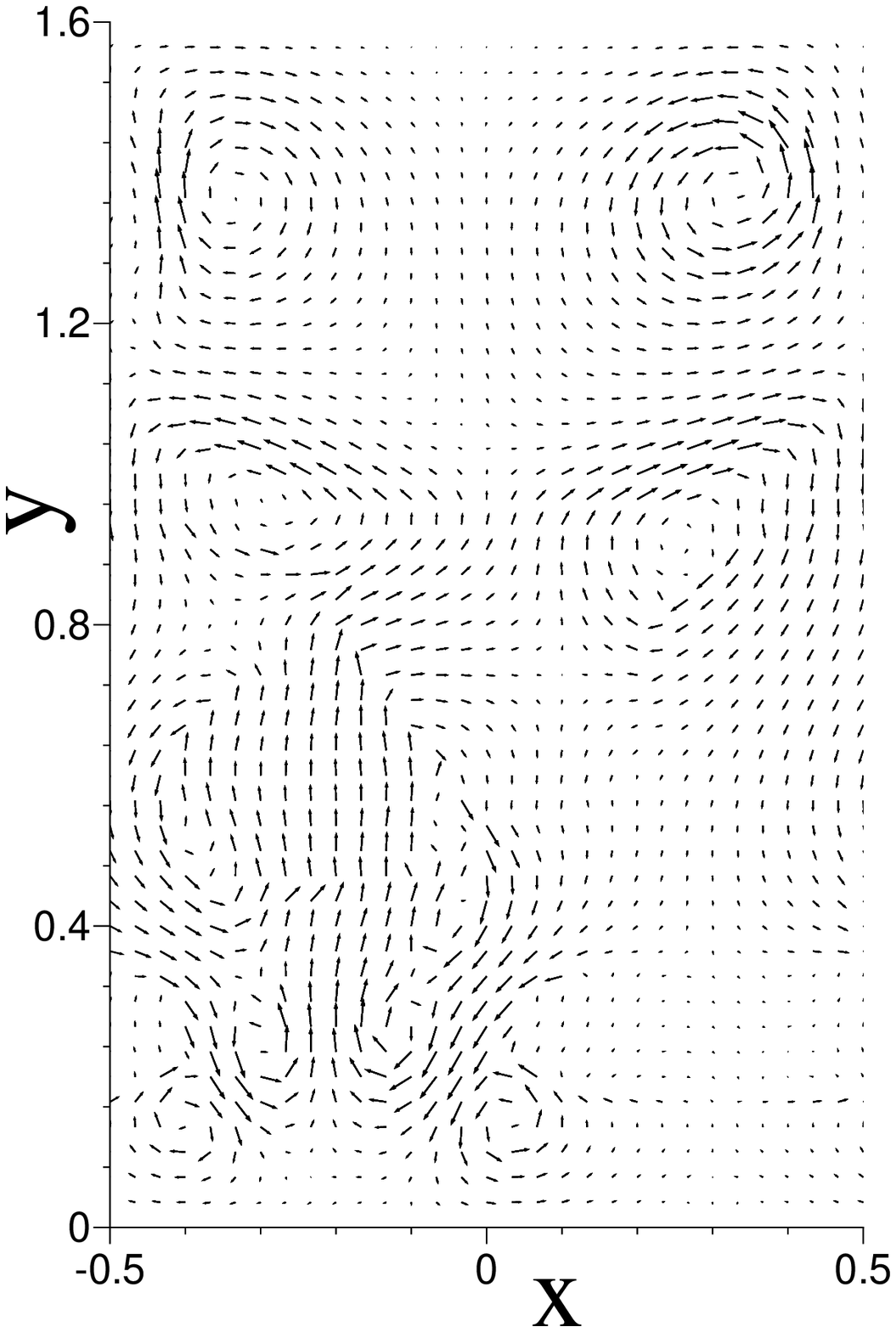}(f)
\includegraphics[width=1.4in]{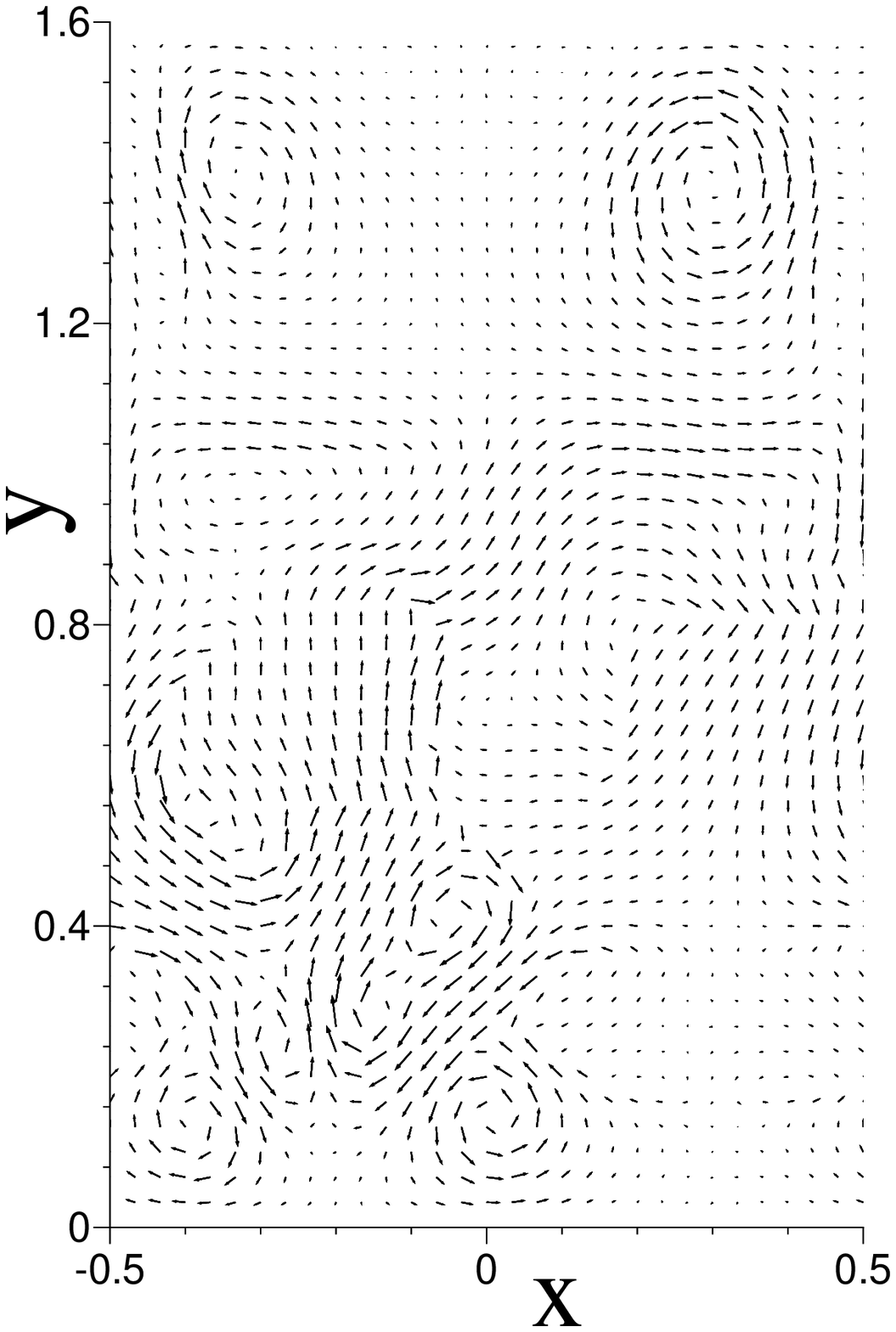}(g)
\includegraphics[width=1.4in]{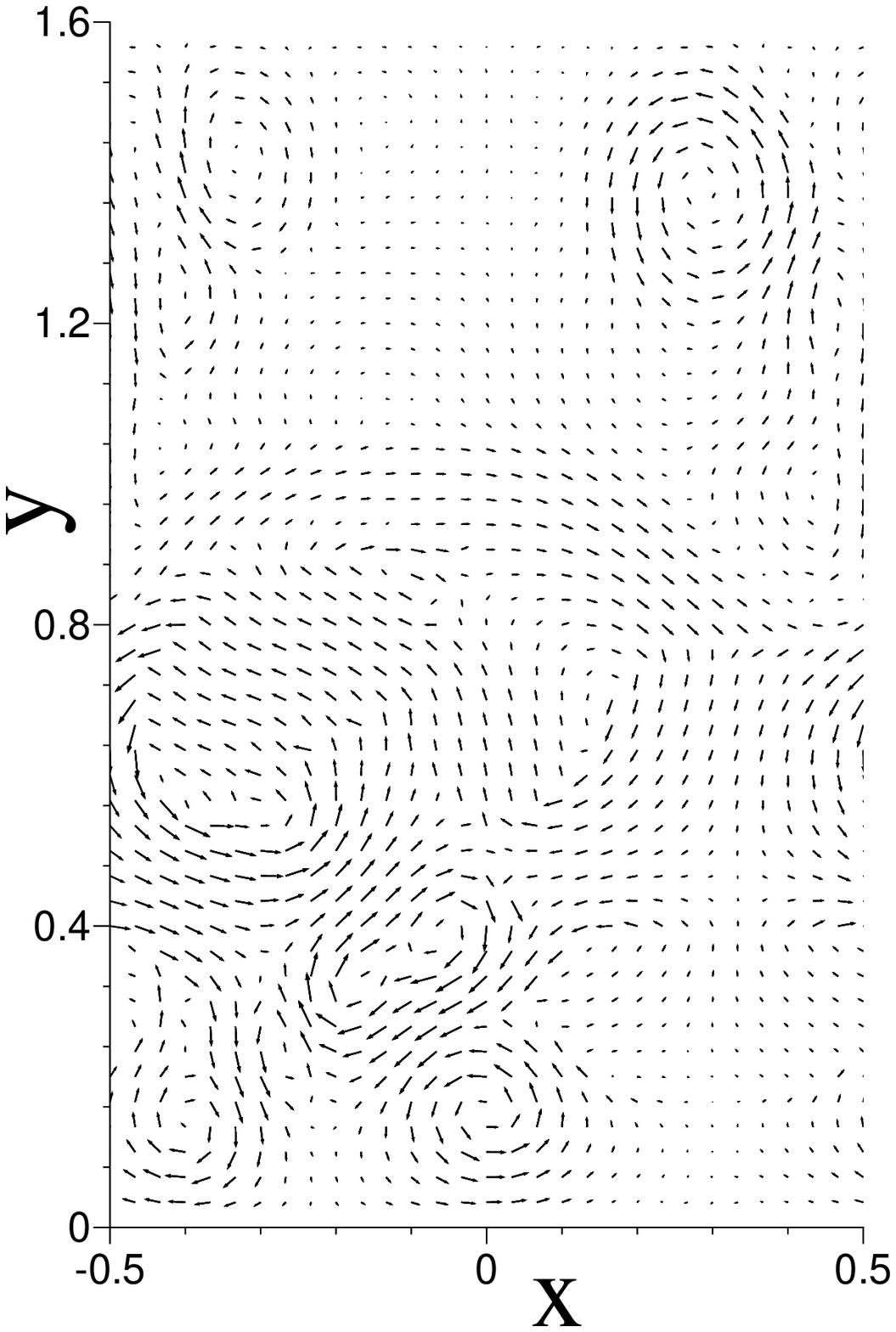}(h)
}
\centerline{
\includegraphics[width=1.4in]{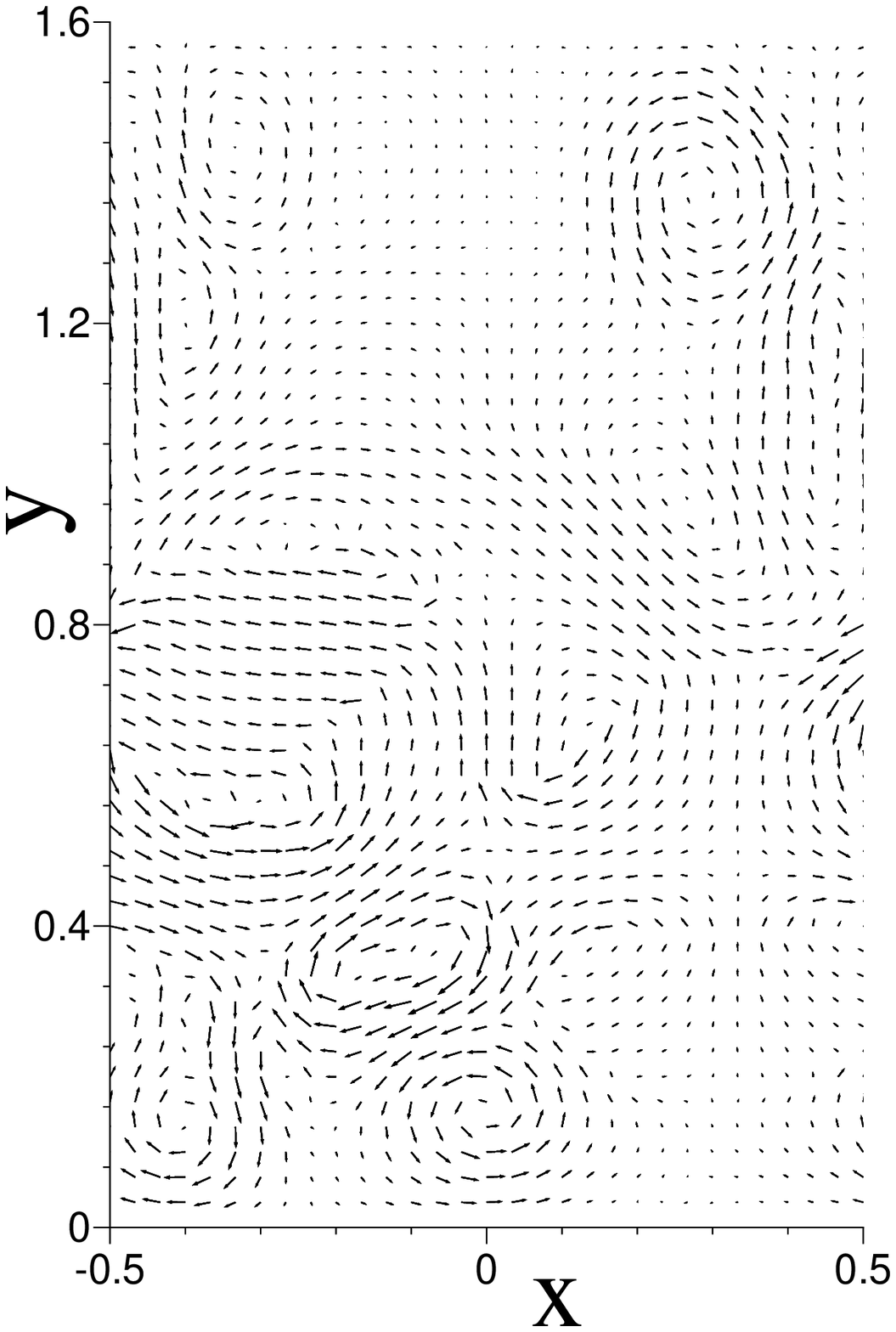}(i)
\includegraphics[width=1.4in]{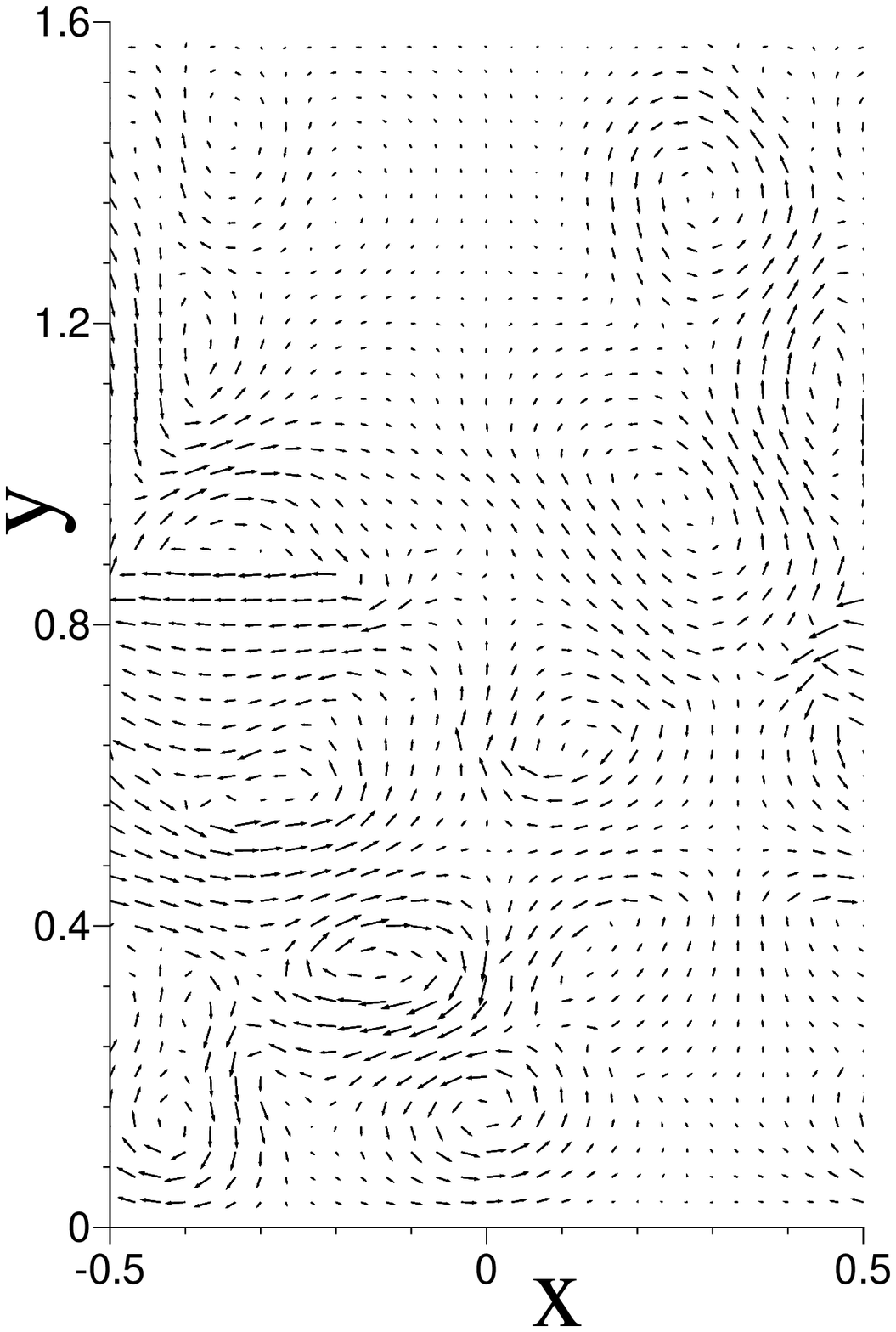}(j)
\includegraphics[width=1.4in]{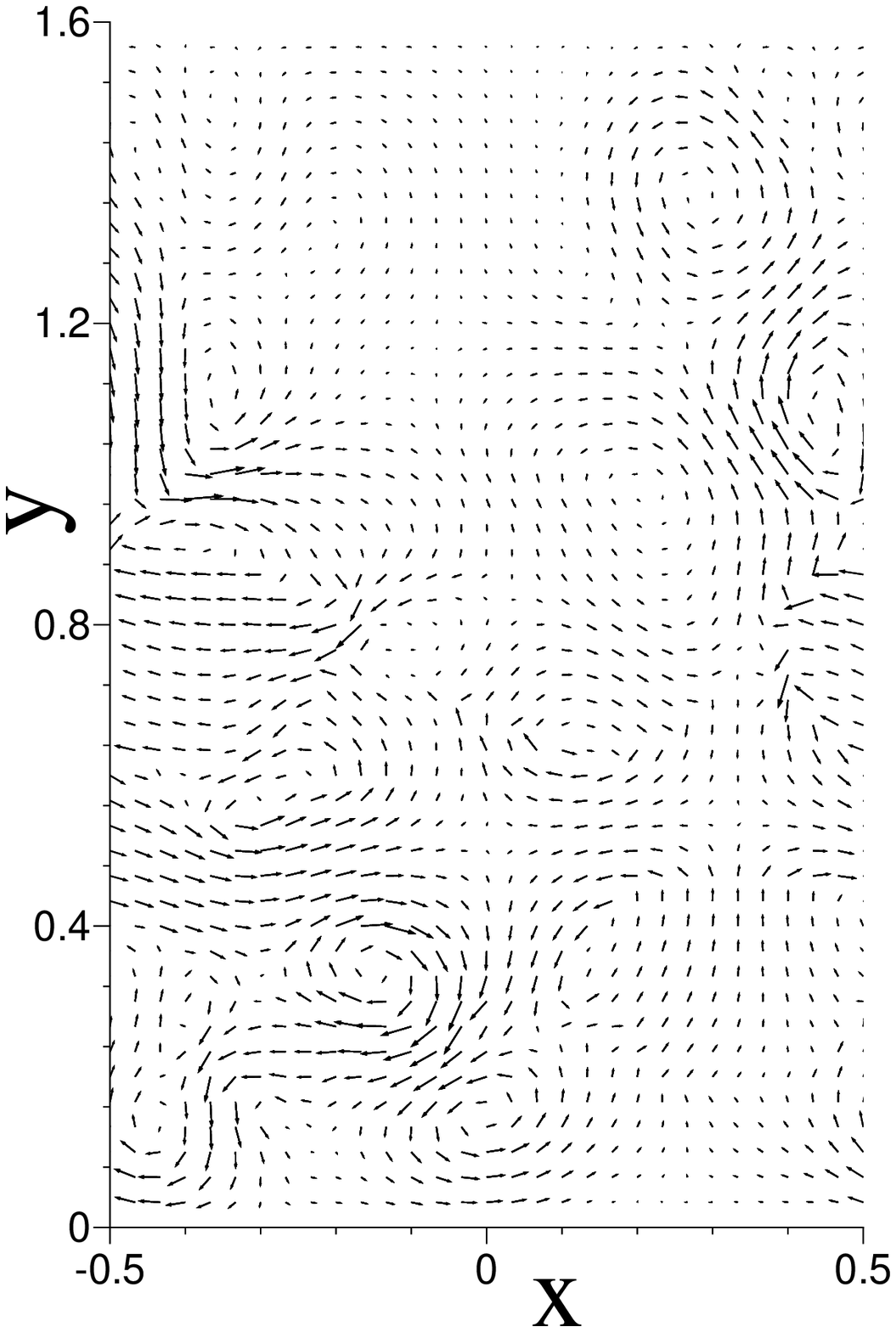}(k)
\includegraphics[width=1.4in]{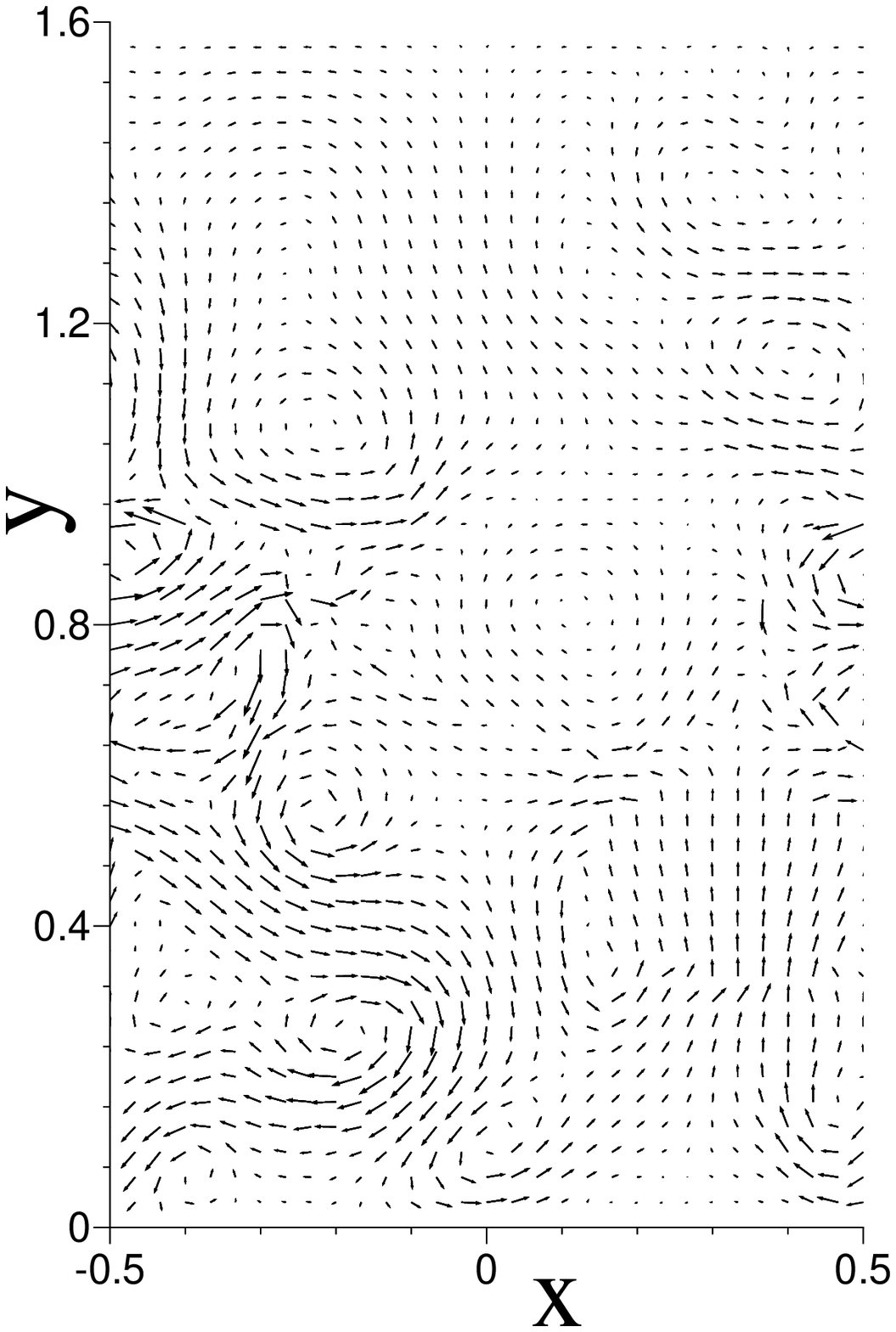}(l)
}
\centerline{
\includegraphics[width=1.4in]{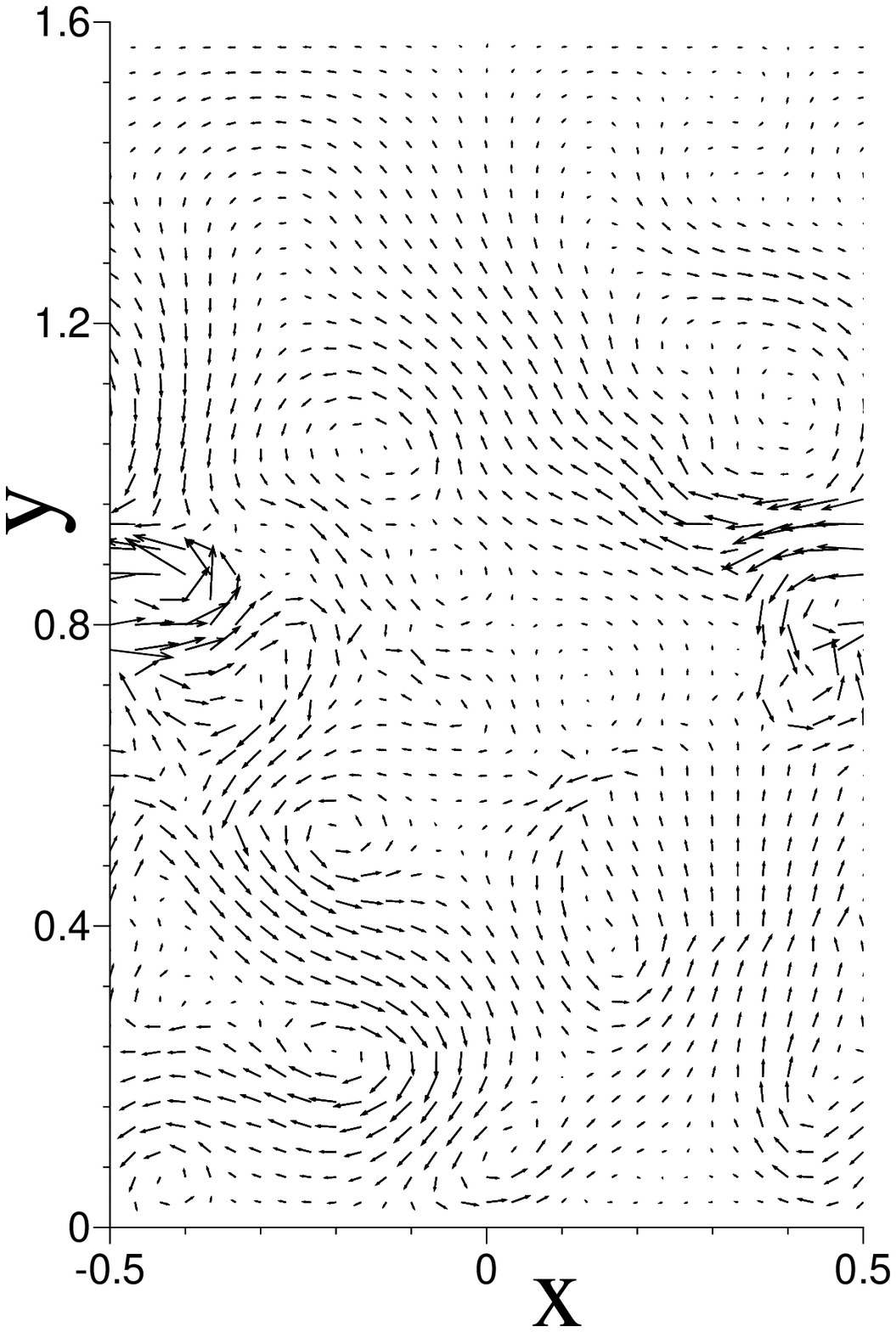}(m)
\includegraphics[width=1.4in]{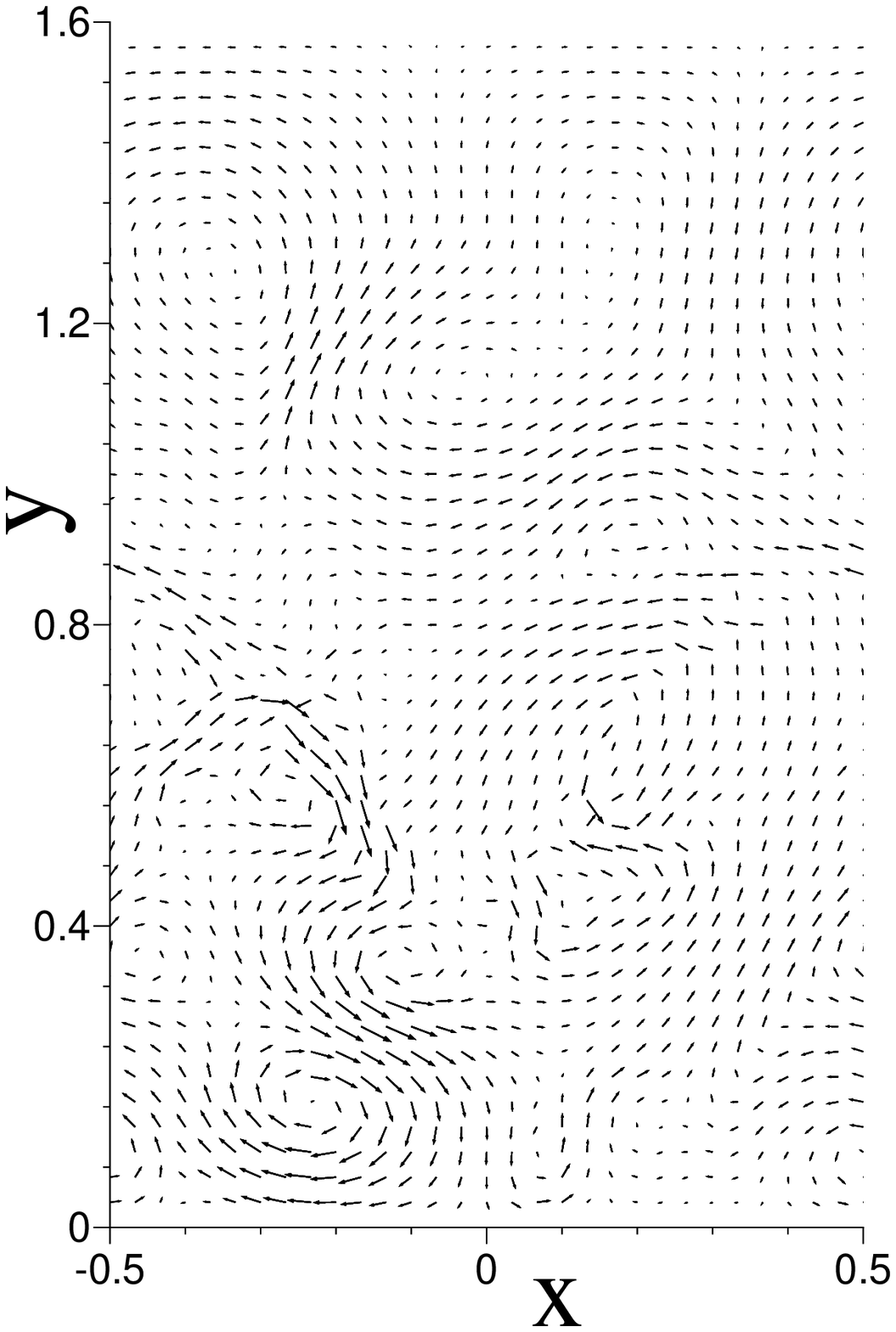}(n)
\includegraphics[width=1.4in]{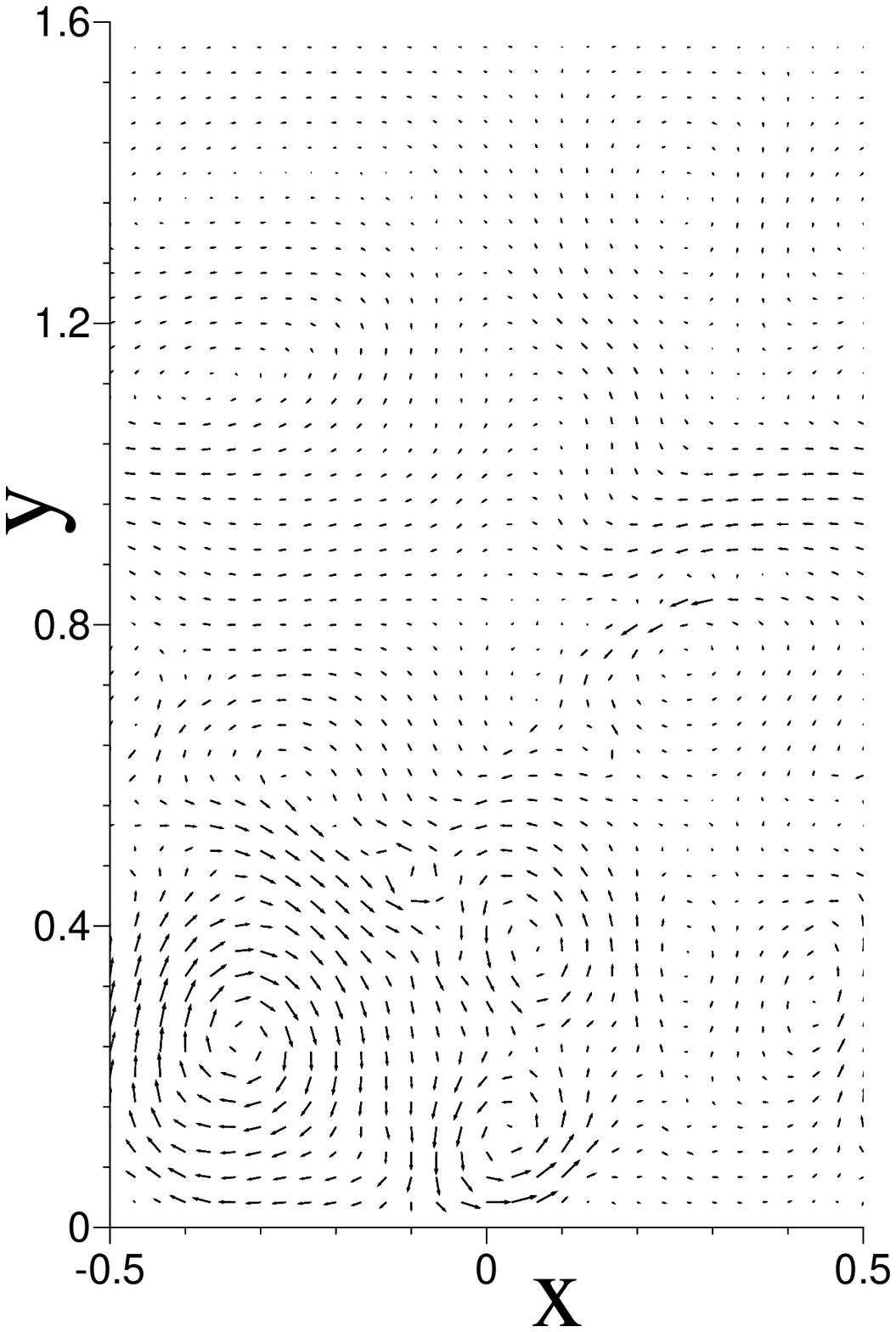}(o)
\includegraphics[width=1.4in]{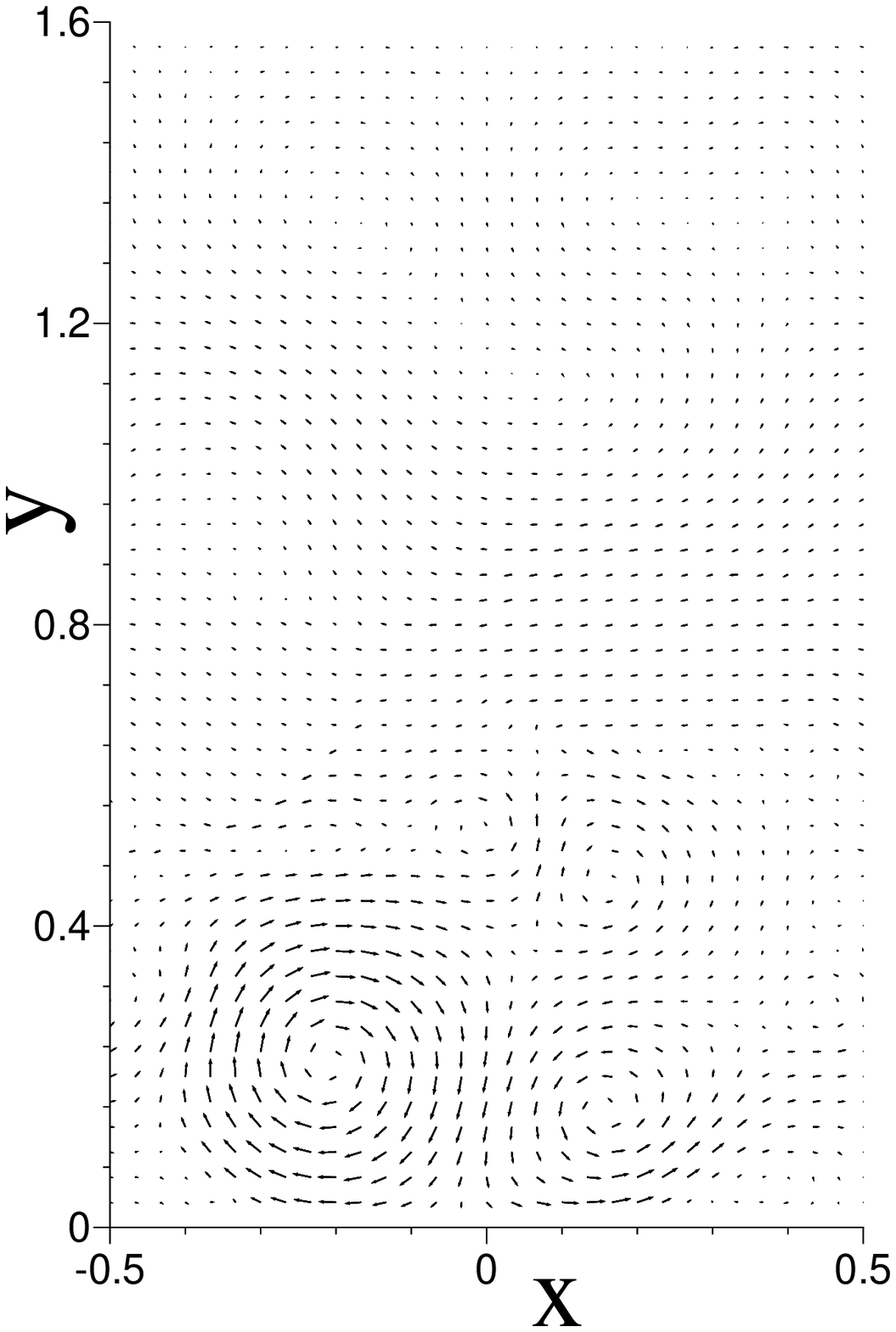}(p)
}
\caption{
Velocity fields of fluid drops impacting water surface (4 fluid phases):
(a) $t=0.076$,
(b) $t=0.176$,
(c) $t=0.26$,
(d) $t=0.332$,
(e) $t=0.46$,
(f) $t=0.608$,
(g) $t=0.712$,
(h) $t=0.796$,
(i) $t=0.836$,
(j) $t=0.888$,
(k) $t=0.944$,
(l) $t=1.128$,
(m) $t=1.196$,
(n) $t=1.504$,
(o) $t=2.208$,
(p) $t=3.384$.
Velocity vectors are plotted on every 15-th quadrature points in
each direction on each element.
}
\label{fig:4p_velocity}
\end{figure}


We further illustrate the dynamical features of this flow
by looking into the velocity  distributions.
Figure \ref{fig:4p_velocity} is a temporal sequence
of snapshots of the velocity fields of this four-phase
flow at the same time instants as those in
Figure \ref{fig:4p_phase}.
In order to make the figures clearer,
the velocity vectors have been plotted on every
$15$-th quadrature point in each direction within each element.
Figures \ref{fig:4p_velocity}(a)--(c) indicate
that the falling F2 drop induces 
a velocity field inside the air, forming a pair of
vortices near the shoulders of the F2 
drop (see e.g. Figures \ref{fig:4p_velocity}(b)--(c)).
On the other hand, the velocity inside the F2 drop is
largely uniform.
Prior to the impact on the water surface (Figure \ref{fig:4p_velocity}c),
the air is squeezed out from between 
the F2 drop and the water surface,
resulting in a strong lateral air flow just above
the water surface.
Upon impact, the strong air flow produces a pair
of vortices behind (i.e. above) either side of  
the F2 drop (Figure \ref{fig:4p_velocity}d),
noting the periodicity in the horizontal
direction.
The pair of vortices subsequently travel 
upward in the air and dies down gradually over
time (Figures \ref{fig:4p_velocity}(e)--(k)).
Simultaneously, the rise of the air bubble
through the water induces a pair of vortices
behind (see e.g. Figure \ref{fig:4p_velocity}d).
A pair of vortices can also be recognized behind
the F1 drop; see e.g. Figure \ref{fig:4p_velocity}(f).
An interaction of these two pairs of vortices
in the water can be observed 
(Figures \ref{fig:4p_velocity}(f)--(k)).
Subsequently, 
the merger of the air bubble
and the bulk of air generates an energetic air flow
near the water surface (Figures \ref{fig:4p_velocity}(l)--(m)).
Figures \ref{fig:4p_velocity}(n)--(p)
show that the velocity field inside the water,
and also in the air, dies down as the F1 drop rises to
the water surface and the system
approaches an equilibrium state.


\section{Concluding Remarks}
\label{sec:summary}


The contributions of the current work can be
summarized in terms of the following three aspects:
\begin{itemize}

\item
We have presented a set of N-phase physical formulations 
for a class of general order parameters.
They generalize the N-phase formulation presented
in \cite{Dong2014}, which is based on a 
special set of order parameters.
This  generalization has three implications: 
(1) The set of ($N-1$) phase field equations becomes  more
strongly coupled with one another, in particular,
the inertia terms $\frac{\partial\phi_i}{\partial t}$ are
all coupled with one another;
(2) It makes it possible to come up with an {\em explicit}
form for the mixing energy density coefficients $\lambda_{ij}$;
(3) Numerical solution of the coupled phase field equations becomes
more challenging.

\item
We have provided an {\em explicit} form for computing the mixing energy
density coefficients $\lambda_{ij}$ with general order parameters.
Note that the method
in \cite{Dong2014} 
requires the
solution of a linear algebraic system to determine $\lambda_{ij}$.
In contrast, $\lambda_{ij}$ in the current paper are
given in an explicit form, which applies to general order
parameters, including the special order-parameter set
employed in \cite{Dong2014}.

\item
We have developed an efficient algorithm for numerically
solving the  phase field equations
with general order parameters.
The algorithm transforms the ($N-1$) strongly-coupled
4-th order phase field equations for general order parameters
into $2(N-1)$ Helmholtz type equations that are
completely de-coupled from one another.
The computational complexity of the current algorithm 
for general order parameters is comparable to
that of \cite{Dong2014} for the special set
of order parameters.
The advantage with the special set of order parameters of
\cite{Dong2014} lies in that the phase field
equations have a simpler form. 
The current work shows that, 
even though the phase field equations have a more complicated 
form with general order parameters, 
the computational work involved in the numerical 
solutions of these equations is essentially the same as that
for the simpler phase field equations
with the special set of order parameters
of \cite{Dong2014}.

\end{itemize}



We have presented several example problems to 
demonstrate the accuracy and capability of 
the physical formulations and the numerical
algorithm developed herein.
%
These test problems involve multiple fluid phases,
large density contrasts, large viscosity contrasts, and multiple
pairwise surface tensions.
Several different sets of order parameters have
been employed in the simulations.
%
By comparing with the theory 
of Langmuir and de Gennes \cite{Langmuir1933,deGennesBQ2003},
we have shown that the formulations and the numerical
algorithm developed herein
have produced physically accurate results
for multiple fluid phases.
The simulation results also demonstrate
the complex dynamics induced by the interactions
among multiple types of fluid interfaces.


It is instructive to compare the general order parameters
discussed here for N-phase flows with the 
order parameter for two-phase flows.
There exists only one independent order parameter
for a two-phase system, and different order parameters have been
used by different researchers in the literature.
Several choices of 
the order parameter for two phases 
are touched on in e.g. \cite{AbelsGG2012},
and the resultant two-phase phase field equation would usually
only require a simple re-scaling for a different choice
of the order parameter.

On the other hand, for an N-phase system
there are ($N-1$) independent order
parameters. The possibilities for choosing the 
($N-1$) order parameters are much broader, and
different choices lead to varying degrees
of complexity in the resulting phase field
equations.
The class of order parameters considered
in the current work, equation \eqref{equ:order_param},
has a linear relation between 
($\rho_i-\rho_N$) and $\phi_j$.
One can readily imagine even broader classes of
order parameters, e.g.
\begin{equation}
\varphi_i(\vec{\phi}) = \rho_i(\vec{\phi}) - \rho_N(\vec{\phi})
= f_i(\vec{\phi}), 
\quad 1\leqslant i\leqslant N-1,
\end{equation}
where $f_i(\vec{\phi})$ are ($N-1$) given  functions 
and in general can be nonlinear.
This, however, will lead to even more complicated 
forms for the phase
field equations.
Note that, regardless of the set of order
parameters being employed, the phase field equations all
represent the mass balance relations for 
the $N$ individual fluid phases. 
We hold the view that the formulations 
with different order parameters
are merely different representations of the  N-phase
system, and that the different representations 
should be equivalent to one another.
This is an embodiment of the 
representation invariance principle \cite{MaW2014}.
The explicit expressions for 
the mixing energy density coefficients $\lambda_{ij}$
derived in the current work
are a direct result of this principle.

%




\section*{Acknowledgement}
The author gratefully 
acknowledges the support from NSF 
and ONR.

\section*{\underline{Appendix A:} Unique Solvability of $\lambda_{ij}$ 
Linear Algebraic System }

In this Appendix we prove that the linear algebraic
system about $\lambda_{ij}$ derived in \cite{Dong2014}
for the special set of order parameters defined
by \eqref{equ:special_order_param} has a unique solution
for any $N\geqslant 2$.
The unique solvability of that system for $N\geqslant 4$ 
is an un-settled
issue in \cite{Dong2014}.

The idea of the proof is as follows. We will show that
the system of equations about $\Lambda_{ij}$ 
(\eqref{equ:lambda_kk_expr} and \eqref{equ:lambda_kl_equation})
in the current work,
under a non-singular transform, 
is equivalent to the linear algebraic system
about $\lambda_{ij}$ in \cite{Dong2014} for 
the special set of order parameters 
defined by \eqref{equ:special_order_param}.
Since the system consisting of 
\eqref{equ:lambda_kk_expr} and \eqref{equ:lambda_kl_equation}
has a unique solution for any $N\geqslant 2$, the linear system 
from \cite{Dong2014}
must also have a unique solution.

The following is the linear algebraic system 
about $\lambda_{ij}$ from \cite{Dong2014},
which is based on the set of order parameters
defined by \eqref{equ:special_order_param},
\begin{equation}
\sum_{i,j=1}^{N-1}L_i^{kl}L_j^{kl}\lambda_{ij} 
= \frac{9}{2}\frac{\eta^2}{\beta^2}\sigma_{kl}^2,
\qquad 1\leqslant k<l\leqslant N,
\label{equ:special_lambda_ij_equation}
\end{equation}
where 
\begin{equation}
L_i^{kl} = \left\{
\begin{array}{ll}
\frac{\tilde{\rho}_k}{\tilde{\rho}_k+\tilde{\rho}_N}\delta_{ik}
- \frac{\tilde{\rho}_l}{\tilde{\rho}_l+\tilde{\rho}_N}\delta_{il},
& 
1\leqslant k<l\leqslant N-1, \
1\leqslant i\leqslant N-1, 
 \\
\frac{\tilde{\rho}_i}{\tilde{\rho}_i+\tilde{\rho}_N}\delta_{ik}
+ \frac{\tilde{\rho}_N}{\tilde{\rho}_i+\tilde{\rho}_N},
&
1\leqslant k<l= N, \
1\leqslant i\leqslant N-1.
\end{array}
\right.
\label{equ:L_ikl_expr}
\end{equation}
This is a system of $\frac{1}{2}N(N-1)$ equations about
$\frac{1}{2}N(N-1)$ unknowns, noting the
symmetry $\lambda_{ij}=\lambda_{ji}$ ($1\leqslant i,j\leqslant N-1$).

We next show that the linear system consisting of
\eqref{equ:lambda_kk_expr} and \eqref{equ:lambda_kl_equation}
about $\Lambda_{ij}$
can be transformed to the system \eqref{equ:special_lambda_ij_equation}
about $\lambda_{ij}$,
under the transform \eqref{equ:lambda_ij_spec_explicit}
between $\mathbf{A}$ and $\bm{\Lambda}$.
Re-write \eqref{equ:lambda_ij_spec_explicit} as
\begin{equation}
\begin{bmatrix} \Lambda_{ij} \end{bmatrix}_{(N-1)\times(N-1)}=
\bm{\Lambda} = 
\mathbf{Z}^{-1}\mathbf{L}^{-1}\mathbf{A}\mathbf{L}^{-1}\mathbf{Z}^{-1}
= \bm{\Lambda}_1\mathbf{L}^{-1}\mathbf{A}\mathbf{L}^{-1}\bm{\Lambda}_1
= 4\mathbf{R}^T \mathbf{AR}
\label{equ:lambda_transform}
\end{equation}
where we have used equation \eqref{equ:Z_mat_expr}
and $\bm{\Lambda}_1$ is defined in \eqref{equ:A1_volfrac_spec}, and
\begin{equation}
\mathbf{R} = \frac{1}{2}\mathbf{L}^{-1}\bm{\Lambda}_1
= \begin{bmatrix} r_{ij} \end{bmatrix}_{(N-1)\times(N-1)},
\qquad
r_{ij} = \frac{\tilde{\rho}_i}{\tilde{\rho}_i + \tilde{\rho}_N}\delta_{ij}
  + \frac{\tilde{\rho}_N}{\tilde{\rho}_i + \tilde{\rho}_N}
= L_i^{jN}.
\label{equ:R_expr}
\end{equation}
First consider the case $1\leqslant k<l\leqslant N-1$.
Equation \eqref{equ:lambda_kl_equation} becomes
\begin{equation}
\begin{split}
\frac{9}{2}\frac{\eta^2}{\beta^2}\sigma_{kl}^2
&
= \frac{1}{4}\left(\Lambda_{kk} + \Lambda_{ll} - 2\Lambda_{kl}  \right) \\
&
= \left( \frac{\tilde{\rho}_k}{\tilde{\rho}_k + \tilde{\rho}_N} \right)^2 
  \lambda_{kk} 
+ \left( \frac{\tilde{\rho}_l}{\tilde{\rho}_l + \tilde{\rho}_N} \right)^2 
  \lambda_{ll}
- 2\frac{\tilde{\rho}_k}{\tilde{\rho}_k + \tilde{\rho}_N} 
   \frac{\tilde{\rho}_l}{\tilde{\rho}_l + \tilde{\rho}_N} \lambda_{kl} \\
&
= \left(L_k^{kl}  \right)^2\lambda_{kk}
  + \left(L_l^{kl}  \right)^2\lambda_{ll}
  + 2L_k^{kl} L_l^{kl} \lambda_{kl} \\
&
=\sum_{i,j=1}^{N-1}L_i^{kl}L_j^{kl}\lambda_{ij},
\qquad
1\leqslant k<l \leqslant N-1
\end{split}
\end{equation}
where we have used
\eqref{equ:lambda_transform} and \eqref{equ:L_ikl_expr}. 
Therefore Equation \eqref{equ:lambda_kl_equation} is
transformed to \eqref{equ:special_lambda_ij_equation}
for the case $1\leqslant k<l \leqslant N-1$.
Next consider the case $1\leqslant k<l=N$.
Equation \eqref{equ:lambda_kk_expr} becomes
\begin{equation}
\begin{split}
\frac{9}{2}\frac{\eta^2}{\beta^2}\sigma_{kN}^2 
&
= \frac{1}{4}\Lambda_{kk} 
= \sum_{i,j=1}^{N-1} r_{ik} r_{jk} \lambda_{ij}
= \sum_{i,j=1}^{N-1} L_i^{kN}L_j^{kN}\lambda_{ij},
\qquad 1\leqslant k<l=N
\end{split}
\end{equation}
where we have used \eqref{equ:lambda_transform},
\eqref{equ:L_ikl_expr}, and \eqref{equ:R_expr}.
Therefore, equation \eqref{equ:lambda_kk_expr}
is transformed to \eqref{equ:special_lambda_ij_equation}
for the case $1\leqslant k<l=N$.

Since the transform \eqref{equ:lambda_ij_spec_explicit}
(or equivalently \eqref{equ:lambda_transform})
is non-singular,
we conclude that the linear system
consisting of \eqref{equ:lambda_kk_expr}
and \eqref{equ:lambda_kl_equation}
is equivalent to the linear system
given by \eqref{equ:special_lambda_ij_equation}.
We can then conclude that the linear system
\eqref{equ:special_lambda_ij_equation}
about $\lambda_{ij}$ has a unique solution
for any $N\geqslant 2$, and that
its solution is given by
\eqref{equ:lambda_ij_spec_explicit},
in which $\bm{\Lambda}$ is given by
\eqref{equ:Lambda_matrix_expr},
\eqref{equ:lambda_kk_expr}
and \eqref{equ:lambda_kl_expr}.

\section*{\underline{Appendix B:} Algorithm for N-Phase Momentum Equations}

In this Appendix we  present an algorithm 
for the N-phase momentum equations, 
\eqref{equ:nse} and \eqref{equ:continuity},
together with the velocity boundary condition,
\eqref{equ:vel_bc}.
While it is also based on a velocity correction-type
strategy to de-couple the pressure and velocity computations,
this scheme is different in formulation 
than that of \cite{Dong2014},
and it results in a smaller pressure error 
than the latter.
The algorithmic formulation  here 
for the N-phase momentum equations, however, can be traced to that
we developed in \cite{DongS2012} for two-phase flows.

We assume that the phase field variables $\phi_i$ 
($1\leqslant i\leqslant N-1$) are known, and our goal
is to compute the velocity and pressure from
\eqref{equ:nse} and \eqref{equ:continuity}.
Let 
\begin{equation}
P = p + 
\sum_{i,j=1}^{N-1} \frac{\lambda_{ij}}{2} \nabla\phi_i\cdot\nabla\phi_j
\label{equ:effective_P}
\end{equation}
denote an auxiliary pressure, which will also be 
loosely referred to as pressure hereafter.
We can then transform \eqref{equ:nse} into
\begin{equation}
\frac{\partial\mathbf{u}}{\partial t}
+ \mathbf{u}\cdot\nabla\mathbf{u}
= \frac{1}{\rho} \nabla P
+ \frac{\mu}{\rho} \nabla^2\mathbf{u}
+ \frac{1}{\rho}\nabla\mu\cdot\mathbf{D}(\mathbf{u})
- \frac{1}{\rho}\sum_{i,j=1}^{N-1}\lambda_{ij}\nabla^2\phi_j\nabla\phi_i
+ \frac{1}{\rho}\mathbf{f}.
\label{equ:nse_reform}
\end{equation}

Given ($\mathbf{u}^n$, $P^n$, $\phi_i^{n+1}$),
our algorithm for \eqref{equ:nse_reform} and \eqref{equ:continuity}
successively computes $P^{n+1}$ and $\mathbf{u}^{n+1}$
in a de-coupled fashion as follows: \\[0.1in]
\noindent\underline{For $P^{n+1}$:}
\begin{subequations}
\begin{equation}
\begin{split}
\frac{\gamma_0\tilde{\mathbf{u}}^{n+1}-\hat{\mathbf{u}}}{\Delta t}
+ \mathbf{u}^{*,n+1}\cdot\nabla\mathbf{u}^{*,n+1}
& + \frac{1}{\rho^{n+1}}\tilde{\mathbf{J}}^{n+1}\cdot\nabla\mathbf{u}^{*,n+1}
+ \frac{1}{\rho_0}\nabla P^{n+1}
= 
\left(\frac{1}{\rho_0}-\frac{1}{\rho^{n+1}}  \right)\nabla P^{*,n+1} \\
& - \frac{\mu^{n+1}}{\rho^{n+1}}\nabla\times\nabla\times\mathbf{u}^{*,n+1}
+ \frac{1}{\rho^{n+1}}\nabla\mu^{n+1}\cdot\mathbf{D}(\mathbf{u}^{*,n+1}) \\
& - \frac{1}{\rho^{n+1}}\sum_{i,j=1}^{N-1}\lambda_{ij}\nabla^2\phi_j^{n+1}\nabla\phi_i^{n+1}
+ \frac{1}{\rho^{n+1}}\mathbf{f}^{n+1},
\end{split}
\label{equ:pressure_1}
\end{equation}
\begin{equation}
\nabla\cdot\tilde{\mathbf{u}}^{n+1} = 0,
\label{equ:pressure_2}
\end{equation}
\begin{equation}
\left.\mathbf{n}\cdot\tilde{\mathbf{u}}^{n+1}\right|_{\partial\Omega}
= \mathbf{n}\cdot\mathbf{w}^{n+1}.
\label{equ:pressure_3}
\end{equation}
\end{subequations}
\\
\noindent\underline{For $\mathbf{u}^{n+1}$:}
\begin{subequations}
\begin{equation}
\frac{\gamma_0\mathbf{u}^{n+1}-\gamma_0\tilde{\mathbf{u}}^{n+1}}{\Delta t}
 - \nu_0 \nabla^2\mathbf{u}^{n+1}
= \nu_0 \nabla\times\nabla\times\mathbf{u}^{*,n+1},
\label{equ:velocity_1}
\end{equation}
\begin{equation}
\left.\mathbf{u}^{n+1}  \right|_{\partial\Omega} = \mathbf{w}^{n+1}.
\label{equ:velocity_2}
\end{equation}
\end{subequations}


In the above equations all the symbols follow
the notation outlined in Section \ref{sec:algorithm}.
$\mathbf{u}^{*,n+1}$ and $P^{*,n+1}$
are defined by \eqref{equ:def_nplus1_star}.
$\hat{\mathbf{u}}$ and $\gamma_0$ are defined
by \eqref{equ:def_hat_var}.
$\rho^{n+1}$ and $\mu^{n+1}$ are
given by \eqref{equ:rho_mu_discretized},
and also \eqref{equ:rho_mu_clamp} in case of large density
ratios among the N fluids.
$\tilde{\mathbf{J}}^{n+1}$ is 
given by \eqref{equ:J_tilde_discretized}.
$\mathbf{f}^{n+1}$ is the external body force
evaluated at time step ($n+1$).
$\mathbf{n}$ is the outward-pointing unit vector
normal to $\partial\Omega$.
$\tilde{\mathbf{u}}^{n+1}$ is an auxiliary
velocity that approximates $\mathbf{u}^{n+1}$.
$\rho_0$ is a chosen constant that must satisfy 
the condition 
\begin{equation}
0 < \rho_0 \leqslant \min(\tilde{\rho}_1,\tilde{\rho}_2,\dots,\tilde{\rho}_N).
\label{equ:rho_0_condition}
\end{equation}
$\nu_0$ in \eqref{equ:velocity_1} is a chosen positive
constant that is sufficiently large. A conservative
condition for $\nu_0$ is given in \cite{Dong2014}.
But in the current paper we will generally employ
the following value or larger,
\begin{equation}
\nu_0 = \max\left(
\frac{\tilde{\mu}_1}{\tilde{\rho}_1},
\frac{\tilde{\mu}_2}{\tilde{\rho}_2},
\cdots,
\frac{\tilde{\mu}_N}{\tilde{\rho}_N},
\right).
\label{equ:nu_0_condition}
\end{equation}


The above algorithm employs a velocity
correction-type idea \cite{GuermondS2003a,DongS2010,DongKC2014}  to de-couple
the computations for the pressure and the velocity.
The difference between this algorithm and
that of \cite{Dong2014} lies in that,
in the pressure substep (equation \eqref{equ:pressure_1})
all the terms in the momentum equations have been 
approximated at
the time step ($n+1$) in the current algorithm. 
In contrast, in the pressure substep of \cite{Dong2014}, 
while the time
derivative term is approximated at time step ($n+1$), all the
other terms are approximated at time step $n$ rather than ($n+1$).
In addition, the velocity substep of the scheme of \cite{Dong2014}
contains a number of correction terms to offset 
the effects caused by the less accurate approximations
using data from time step $n$ in the preceding pressure
substep. On the other hand,
the velocity substep of the current algorithm
(equation \eqref{equ:velocity_1}) does not contain
such correction terms.

It can also be noted that the variable density $\rho$
and the variable dynamic viscosity $\mu$ have been
treated with a reformulation of
the pressure term $\frac{1}{\rho}\nabla P$
and a reformulation of the 
viscous term $\frac{\mu}{\rho}\nabla^2\mathbf{u}$,
so that the linear
algebraic systems resulting from the discretization
involve only {\em constant} and {\em time-independent} 
coefficient matrices. 
The ideas for the reformulations stem from the original developments for
two-phase flows \cite{DongS2012,Dong2012,Dong2014obc}.


We next derive the weak forms for the pressure and
the velocity in order to facilitate the implementation
using $C^0$ spectral elements.
Let $q\in H^1(\Omega)$ denote the test function, and let
\begin{multline}
\mathbf{G}^{n+1} = 
\frac{1}{\rho^{n+1}}\mathbf{f}^{n+1}
- \left(
     \mathbf{u}^{*,n+1} 
     + \frac{1}{\rho^{n+1}} \tilde{\mathbf{J}}^{n+1}
  \right)\cdot\nabla\mathbf{u}^{*,n+1}
+ \frac{\hat{\mathbf{u}}}{\Delta t}
+ \left(\frac{1}{\rho_0} - \frac{1}{\rho^{n+1}}  \right)\nabla P^{*,n+1} \\
+ \frac{1}{\rho^{n+1}}\nabla\mu^{n+1}\cdot\mathbf{D}(\mathbf{u}^{*,n+1})
- \frac{1}{\rho^{n+1}}\sum_{i,j=1}^{N-1}\lambda_{ij}\nabla^2\phi_j^{n+1}\nabla\phi_i^{n+1}
+ \nabla\left( \frac{\mu^{n+1}}{\rho^{n+1}} \right) \times \bm{\omega}^{*,n+1},
\label{equ:G_expr}
\end{multline}
where $\bm{\omega} = \nabla\times \mathbf{u}$ is
the vorticity.
Take the $L^2$ inner product between equation \eqref{equ:pressure_1}
and $\nabla q$, and we get the weak form about $P^{n+1}$,
\begin{equation}
\int_{\Omega} \nabla P^{n+1} \cdot\nabla q
= \rho_0 \int_{\Omega} \mathbf{G}^{n+1}\cdot\nabla q
- \rho_0 \int_{\partial\Omega} \frac{\mu^{n+1}}{\rho^{n+1}} \mathbf{n}\times\bm{\omega}^{*,n+1}\cdot\nabla q
- \frac{\gamma_0\rho_0}{\Delta t}\int_{\partial\Omega}\mathbf{n}\cdot\mathbf{w}^{n+1} q,
\ \
\forall q\in H^1(\Omega)
\label{equ:p_weakform}
\end{equation}
where we have used integration by part,
equations \eqref{equ:pressure_2} and \eqref{equ:pressure_3},
the divergence theorem,
and the identity
$ 
\frac{\mu}{\rho}\nabla\times\bm{\omega}\cdot\nabla q
= \nabla\cdot\left(
  \frac{\mu}{\rho} \bm{\omega}\times\nabla q
  \right)
- \nabla\left( \frac{\mu}{\rho} \right)\times\bm{\omega}\cdot\nabla q.
$ 

Adding together the equations \eqref{equ:pressure_1} and
\eqref{equ:velocity_1}, we get
\begin{equation}
\frac{\gamma_0}{\Delta t}\mathbf{u}^{n+1} - \nu_0\nabla^2\mathbf{u}^{n+1}
= \mathbf{G}^{n+1} 
- \nabla\left( \frac{\mu^{n+1}}{\rho^{n+1}} \right) \times \bm{\omega}^{*,n+1}
- \frac{1}{\rho_0}\nabla P^{n+1}
- \left( \frac{\mu^{n+1}}{\rho^{n+1}} - \nu_0 \right) \nabla\times\bm{\omega}^{*,n+1}
\label{equ:velocity_1_reform}
\end{equation}
Let 
$
H^1_0(\Omega) = \left\{ \
v \in H^{\Omega} \ : \
v|_{\partial\Omega} = 0
\ \right\},
$
and $\varphi \in H_0^1(\Omega)$ denote
the test function.
Taking the $L^2$ inner product between
equation \eqref{equ:velocity_1_reform} and $\varphi$,
one can get the weak form about $\mathbf{u}^{n+1}$,
\begin{multline}
\int_{\Omega}\nabla\varphi\cdot\nabla\mathbf{u}^{n+1}
+ \frac{\gamma_0}{\nu_0\Delta t}\int_{\Omega}\varphi\mathbf{u}^{n+1}
= \frac{1}{\nu_0}\int_{\Omega}\left(
    \mathbf{G}^{n+1} - \frac{1}{\rho_0}\nabla P^{n+1}
  \right) \varphi \\
- \frac{1}{\nu_0}\int_{\Omega} \left(
    \frac{\mu^{n+1}}{\rho^{n+1}} - \nu_0
  \right) 
  \bm{\omega}^{*,n+1} \times \nabla\varphi,
\qquad
\forall \varphi \in H_0^1(\Omega),
\label{equ:vel_weakform}
\end{multline}
where we have used integration by part,
the divergence theorem, the identity ($\chi$ denoting
a scalar function)
\begin{equation*}
\int_{\Omega} \chi\nabla\times\bm{\omega}\varphi
= \int_{\partial\Omega} \chi\mathbf{n}\times\bm{\omega}\varphi
- \int_{\Omega} \nabla\chi\times\bm{\omega}\varphi
+ \int_{\Omega} \chi \bm{\omega}\times\nabla\varphi,
\end{equation*}
and the fact that the surface integrals of type 
$
\int_{\partial\Omega} \chi \varphi
$
vanish because $\varphi\in H^1_0(\Omega)$.

The weak forms for the pressure and the
velocity, \eqref{equ:p_weakform} and
\eqref{equ:vel_weakform}, can be
discretized in space using $C^0$ spectral 
elements 
in a straightforward
fashion. Note that the terms $\nabla^2\phi_i^{n+1}$ 
($1\leqslant i\leqslant N-1$) involved in
the $\mathbf{G}^{n+1}$ expression \eqref{equ:G_expr}
and in the $\tilde{\mathbf{J}}^{n+1}$ expression
(see \eqref{equ:J_tilde_discretized} and \eqref{equ:J_tilde_expr})
must be computed based on equation \eqref{equ:laplace_phi}.

Therefore, solving the N-phase momentum equations
amounts to the following two successive operations. First,
solve equation \eqref{equ:p_weakform} for
pressure $P^{n+1}$. Then, solve
equation \eqref{equ:vel_weakform}, together with
the Dirichlet condition \eqref{equ:velocity_2} on $\partial\Omega$,
for $\mathbf{u}^{n+1}$.



%
\bibliographystyle{plain}
\bibliography{nphase,obc,mypub,nse,sem,contact_line,interface}

\end{document}